\documentclass[longauth]{aa}
\usepackage{verbatim}
\usepackage{graphicx}
\usepackage{txfonts}
\usepackage{xcolor}
\usepackage{float}
\usepackage[caption = false]{subfig}

\usepackage{graphicx}	
\usepackage{amsmath}	
\usepackage{amssymb}	
\usepackage{lscape}	
\usepackage{multirow}
\usepackage{xcolor}
\usepackage{ulem}
\usepackage{subfig}
\usepackage{upgreek}
\usepackage{multirow}
\usepackage{multicol}
\usepackage{hyperref}
\hypersetup{
    citecolor=blue,
    colorlinks=true,
    linkcolor=blue,
    filecolor=magenta,      
    urlcolor=blue,
    pdfpagemode=FullScreen,
    }

\definecolor{amber}{rgb}{1.0, 0.49, 0.0}
\definecolor{violet}{rgb}{0.54, 0.17, 0.89}
\definecolor{gdago}{rgb}{0.7, 0.0, 0.7}

\usepackage{orcidlink}

\begin{document}

   \title{Targeting cluster galaxies for the 4MOST CHANCES Low-z sub-survey with photometric redshifts}

    \author{Hugo M\'endez-Hern\'andez\inst{1},
          Ciria Lima-Dias\inst{1},
          Antonela Monachesi \inst{1},
          Yara L. Jaffé\inst{2,4},
          Christopher P. Haines\inst{3,4},
          Gabriel S. M. Teixeira\inst{5},
          Elismar Lösch\inst{6},
          Raúl Baier-Soto\inst{2,4,13},
          Erik  V. R. Lima\inst{3},
          Amrutha B. M.\inst{3},
          C. R. Bom\inst{5},
          Giuseppe D'Ago\inst{7},
          Ricardo Demarco\inst{8},
          Alexis Finoguenov\inst{9},
          Rodrigo F. Haack\inst{10,11},
          Amanda R. Lopes\inst{10},
          C. Mendes de Oliveira\inst{6},
          Paola Merluzzi\inst{12},
          Franco Piraino-Cerda\inst{2,4,13},
          Anal\'ia V. Smith Castelli\inst{10,11},
          Cristobal Sif\'on\inst{13},
          Laerte Sodr\'e Jr\inst{6},
          Nicol\'as Tejos\inst{13},
          Sergio Torres-Flores\inst{1},
          Maria Argudo-Fern\'andez\inst{14,15},
          Jacob P. Crossett\inst{2,13},
          E. Ibar\inst{16,4},
          Ulrike Kuchner\inst{17},
          Ivan Lacerna\inst{3},
          Vitor H. Lopes-Silva\inst{18},
          Sebasti\'an Lopez\inst{19},
          Sean McGee \inst{20},
          Lorenzo Morelli\inst{3},
          Julie Nantais \inst{21},
          Patricio Olivares V. \inst{2},
          Diego Pallero\inst{2,4},
          Bianca M. Poggianti\inst{22},
          Emanuela Pompei\inst{23},
          V. M. Sampaio\inst{2,4},
          Benedetta Vulcani\inst{22},
          Alfredo Zenteno\inst{24},
          F. Almeida-Fernandes\inst{25},
          Maciej Bilicki \inst{26},
          M.S. Carvalho\inst{6},
          Cheng Cheng\inst{27,28},
          A. L., Figueiredo\inst{6},
          L. A. Guti\'{e}rrez-Soto\inst{10},
          F. R. Herpich\inst{29},
          A. Kanaan\inst{30},
          E A. D. Lacerda\inst{6},
          L. Nakazono\inst{31,32},
          G.B. Oliveira Schwarz\inst{33},
          T. Ribeiro\inst{34},
          Boudewijn F. Roukema \inst{35,36},
          Mar\'ilia J. Sartori\inst{6,37},
          Tha\'is Santos-Silva\inst{6,38},
          W. Schoenell\inst{39},
          }

   \institute{Departamento de Astronom\'ia, Universidad de La Serena, Avda. Ra\'ul Bitr\'an 1305, La Serena, Chile\\
              \email{hugo.mendez@userena.cl}
              \and
            Departamento de F\'isica, Universidad T\'ecnica Federico Santa Mar\'ia, Avenida Espa\~na 1680, Valpara\'iso, Chile\
            \and
            Instituto de Astronom\'ia y Ciencias Planetarias (INCT), Universidad de Atacama, Copayapu 485, Copiap\'o, Chile\
              \and
              Millennium Nucleus for Galaxies (MINGAL), Valpara\'iso, Chile\
             \and
             Centro Brasileiro de Pesquisas F\'isicas, Rua Dr. Xavier Sigaud 150, 22290-180 Rio de Janeiro, RJ, Brazil\
             \and
             Departamento de Astronomia, Instituto de Astronomia, Geof\'isica e Ciências Atmosféricas, Universidade de São Paulo, Rua do Matão 1226, Cidade Universit\'aria, São Paulo, 05508-090, Brazil\
             \and
             Institute of Astronomy, University of Cambridge, Madingley Road, Cambridge CB3 0HA, United Kingdom\
             \and
             Institute of Astrophysics, Facultad de Ciencias Exactas, Universidad Andr\'es Bello, Sede Concepci\'on, Talcahuano, Chile\
             \and
             Department of Physics, University of Helsinki, Gustaf H\"allstr\"ominkatu 2, 00560 Helsinki, Finland\
             \and
             Instituto de Astrof\'isica de La Plata, CONICET-UNLP, Paseo del Bosque s/n, B1900FWA, Argentina\
             \and
             Facultad de Ciencias Astron\'omicas y Geof\'isicas, Universidad Nacional de La Plata, Paseo del Bosque s/n, B1900FWA, Argentina\
             \and
             INAF-Osservatorio Astronomico di Capodimonte, Salita Moiariello 16, 80131 Napoli, Italy\
             \and
             Instituto de F\'isica, Pontificia Universidad Cat\'olica de Valpara\'iso, Casilla 4059, Valpara\'iso, Chile\
             \and
             Departamento de F\'isica Teórica y del Cosmos, Edificio Mecenas, Campus Fuentenueva, Universidad de Granada, E-18071 Granada, Spain
             \and 
             Instituto Universitario Carlos I de F\'isica Teórica y Computacional, Universidad de Granada, 18071 Granada, Spain
             \and
             Instituto de F\'isica y Astronom\'ia, Universidad de Valpara\'iso, Avda. Gran Bretaña 1111, Valpara\'iso, Chile\
             \and
             School of Physics and Astronomy, University of Nottingham, Nottingham NG7 2RD, UK             
             \and
             Valongo Observatory, Federal University of Rio de Janeiro, Ladeira do Pedro Antônio 43, Saúde, CEP 20080-090, Rio de Janeiro, RJ, Brazil\
             \and
             Departamento de Astronomía, Universidad de Chile, Casilla 36-D, Santiago, Chile\
             \and
             School of Physics and Astronomy, University of Birmingham, Birmingham B15 2TT, UK\
             \and
             Departamento de F\'isica y Astronom\'ia, Instituto de Astrof\'isica, Universidad Andres Bello, Fernandez Concha 700, Las Condes, Santiago 7591538, Chile\
             \and
             INAF-Osservatorio Astronomico di Padova, vicolo dell\'Osservatorio 5, 35136 Padova, Italy\
             \and
             European Southern Observatory, Science Operations, Alonso de Cordova 3107, Vitacura, 19001 Santiago, Chile\
             \and
             Cerro Tololo Inter-American Observatory, NSFs National Optical-Infrared Astronomy Research Laboratory, Casilla 603, La Serena,Chile\
             \and
             Instituto Nacional de Pesquisas Espaciais (INPE/MCTI), Av. dos Astronautas, 1758, S\~ao Jos\'e dos Campos, SP, Brazil\
             \and
             Center for Theoretical Physics, Polish Academy of Sciences, al. Lotnik\'ow 32/46, 02-668 Warsaw, Poland\
             \and             
             Chinese Academy of Sciences South America Center for Astronomy, National Astronomical Observatories, CAS, Beijing 100101, China\
             \and
             Key Laboratory of Optical Astronomy, NAOC, 20A Datun Road, Chaoyang District, Beijing 100101, China\
             \and
             Laborat\'orio Nacional de Astrof\'isica (LNA), Rua dos Estados Unidos 154, Bairro das Na\c{c}\~oes, Itajub\'a, 37504-364, MG, Brazil\
             \and
             Universidade Federal de Santa Catarina, Campus Universit\'ario Reitor Jo\~ao David Ferreira Lima, Florian\'opolis 88040-900, Brazil.\
             \and
             Observat\'orio Nacional / MCTIC, Rua General Jos\'e Cristino 77, Rio de Janeiro, RJ, 20921-400, Brazil\
             \and
             Departamento de F\'isica Matem\'atica, Instituto de F\'isica, Universidade de S\~{a}o Paulo, SP, Rua do Mat\~{a}o 1371, S\~{a}o Paulo, Brazil\
             \and
             Escola Politecnica, Universidade de S\~ao Paulo, Av. Prof. Luciano Gualberto, travessa do politecnico, 380, São Paulo, 05508-010, Brazil \
             \and
             Rubin Observatory Project Office, 950 N. Cherry Ave, Tucson 85719, USA\
             \and
             Institute of Astronomy, Faculty of Physics, Astronomy and Informatics, Nicolaus Copernicus University, Grudziadzka 5, PL-87-100 Toruń, Poland\
             \and
             Univ. Lyon1, Ens de Lyon, CNRS, Centre de Recherche Astrophysique de Lyon (CRAL) UMR5574, F-69230 Saint-Genis-Laval, France\
             \and
             Universidade do Vale do Para\'iba, Av. Shishima Hifumi, 2911, S\~ao Jos\'e dos Campos, 12244-000, SP, Brazil \
             \and
             Universidade Estadual de Feira de Santana, Departamento de F\'isica, Feira de Santana, BA 44.036-900, Brazil \
             \and 
             The Observatories of the Carnegie Institution for Science, 813 Santa Barbara St, Pasadena, CA 91101, USA\
             }
   \date{Received August 08, 2025}

 
\abstract
   {
    The evolution of galaxies is shaped by both internal processes and their external environments. Galaxy clusters and their surroundings provide ideal laboratories to study these effects, particularly mechanisms such as quenching and morphological transformation.
    The Chilean Cluster galaxy Evolution Survey (CHANCES) Low-z sub-survey is part of the 
    CHileAN Cluster galaxy Evolution Survey, a 4MOST community survey designed to uncover the relationship between the formation and evolution of galaxies and hierarchical structure 
    formation as it happens, through deep and wide multi-object spectroscopy. 
   }
   {
   We present the target selection strategy followed to select galaxy cluster candidate members for the CHANCES low-z sub-survey, in and around 50 clusters and two superclusters at z<0.07, out to \ ($\rm 5 \times R_{200}$) and down to $\rm m_{r}=20.4$.
   }
   {
   Combining public photometric redshift estimates from the DESI Legacy Imaging Survey and \mbox{T80S/S-PLUS iDR5}, 
   with custom photometric redshifts, we identify likely galaxy cluster candidate members whose photometric 
   redshifts are consistent with being at the known redshift of the cluster and measure the average deviations of their photometric redshifts with respect to the spectroscopic redshift measurements  $\sigma_{\mathrm{NMAD}}$.
   We test various selection parameters to maximize completeness while maintaining purity.
   }
   {
   We have successfully compiled our CHANCES-low-redshift catalogues, split into three different sub-surveys: 
   low-z bright ($\rm m_{r}<18.5$), low-z faint ($18.5\leq\rm m_{r}<20.4$) and low-z faint supplementary, 
   by selecting $\gtrsim 500,000$ galaxy cluster candidate members and including confirmed spectroscopic galaxy cluster members, from which we expect to obtain 4MOST low-resolution (R$\sim$6500) spectra for $\sim$320,000 galaxies.
   }
  {The CHANCES Low-z target catalogues form a statistically robust sample for spectroscopic follow-up, allowing studies of galaxy evolution and environmental effects in nearby cluster and supercluster environments.}


   \authorrunning{M\'endez-Hern\'andez et al.}
   \titlerunning{CHANCES Low-z Target Selection}
   \maketitle

\section{Introduction.}
Numerous studies have established that the evolution of galaxies is governed
by a complex interplay of internal processes and external environmental
influences. While massive galaxies are almost all quenched, regardless of
environment, the quenching of star formation in dwarf galaxies with masses
below 10$^{9\,}\mathrm{M_{\odot}}$ is largely driven by environment \citep
{Peng10,Geha12}, with quiescent dwarf galaxies almost always being found in
groups, clusters or as satellites of more massive galaxies \citep
{Haines07}. In particular, galaxy clusters are known to be very effective at
removing interstellar gas 
\citep{Gunn72,Larson80,Kennicutt83a,Byrd90}, altering the star formation of
 galaxies 
\citep{Dressler84,Lewis02} and even transforming them morphologically 
\citep{Dressler80,Postman84,Dressler97}. However, not all galaxies transform
 and quench inside of clusters, in fact, many arrive into clusters already
 quenched. The quenching of galaxies in lower-density environments such as
 groups and filaments prior entering the cluster is referred to as
 pre-processing 
\citep{Zabludoff98,Fujita04,Haines15,Bianconi18, Pallero2019, Kuchner2022,
 Lopes2024}.\\

Observational evidence already supports the significance of pre-processing in
various cosmic structures. For example, studies of the Fornax A group reveal
galaxies at different stages of pre-processing, with variations in their HI
content and molecular hydrogen conversion rates 
\citep{Loubser24}. Similarly, the Coma Supercluster demonstrates a progression
 of environmental quenching, where galaxies in filaments and groups exhibit
 systematically lower star formation rates as they approach the cluster
 core \citep{JimenezTeja25}. However, current studies are often limited in
 depth and sky coverage, emphasizing the need for broader and deeper
 surveys.

Despite significant progress in quantifying the importance of pre-processing
in galaxy evolution, several open questions remain. What are the dominant
mechanisms responsible for quenching star formation before cluster infall?
How does morphological transformation proceed in different environments?
Addressing these questions requires statistically significant datasets that
capture galaxies across a wide range of environments and mass regimes for
which deep and wide-area surveys are essential.

To fully characterize pre-processing and its impact on galaxy evolution, deep
and wide-area surveys are essential. Large-scale surveys covering extensive
portions of the cosmic web can provide the necessary statistical power to
disentangle environmental influences from intrinsic galaxy properties.
Furthermore, reaching lower stellar mass limits is critical for identifying
trends that might be masked when focusing only on massive galaxies. This is
precisely the objective of the CHileAN Cluster galaxy Evolution Survey \citep
[CHANCES;][]{Haines23,Sifon25}. CHANCES-Low-z will obtain $\sim$320,000
galaxy spectra in and around more than 100 clusters and 2 superclusters
(Horologium-Reticulum (HRS) and Shapley (SSC) superclusters) at $0<z<0.45$
with the 4MOST instrument \citep{deJong2019}, allowing us to study galaxy
pre-processing prior to cluster infall out to $\rm 5 \times R_{200}$ and down
to $\rm m_{r} = 20.4$~mag and $\rm log_{10}(M_{\star}/M_{\odot})\sim 8.2$. 

The CHANCES Low-z sub-survey is part of the CHileAN Cluster galaxy Evolution
Survey 
\footnote{\url{https://chances.uda.cl/}}, a 4MOST community survey designed to
uncover the relationship between the formation and evolution of galaxies and
hierarchical structure formation as it happens, through deep and wide
multi-object spectroscopy. CHANCES is split into 3 sub-surveys: CHANCES
Low-z, Evolution, and Circumgalactic Medium (CGM) \citep[see][for details]
{Haines23}.  The CHANCES-low-z sub-survey targets galaxies brighter than
$\rm m_{r}=20.4$ in 50 clusters out to $\rm 5\times R_{200}$ and 2 large
supercluster regions at $\rm z<0.07$. The Evolution sub-survey will observe
50 clusters at $0.07<z<0.45$ with the same magnitude limit and coverage out
to $\rm 5\times R_{200}$ (\textit{Haines et al. in prep.}) while CHANCES CGM
focuses on QSOs lying behind massive galaxy clusters at z>0.35, to detect
any Mg\,{\sc ii} absorption associated with galaxy clusters and their
surroundings (\textit{M\'endez-Hern\'andez et al. in prep.}). Cluster
selection, properties, and supercluster regions are detailed in \citep
{Sifon25}. 

One of the main tasks for the preparation of the survey has been a
careful target selection, which involves selecting the most likely
galaxy cluster candidate members based on photometric information
without biasing the galaxy sample or compromising completeness. This
paper explains in detail the target selection performed for CHANCES
Low-z, which was the most challenging of the 3 sub-surveys, for all
but one cluster in this sub-survey. We note that the strategy
followed to select galaxy cluster candidate members described here
does not include Antlia. The Antlia cluster presents some differences
to what we present here, since there is no Legacy Survey data
available for Antlia which we use for all clusters to select their
targets, and it is too close(z $\leq$ 0.012) for obtaining reliable
photometric redshift(hereafter photo-z or $\rm z_
{phot}$) determinations. The selection for the Antlia cluster will be
presented in \citep{LimaDias25}.  This paper is organized as
follows. Section \ref{sec:data} presents the data used for the
target selection, Section \ref{sec:CMTS} explains the strategy
followed for selecting the galaxy cluster candidate members,
Section  \ref{sec:Validation} presents the environmental analysis of
one of our cluster sample using the final CHANCES-Low-z catalogues.
Finally our conclusions are presented in Section \ref
{sec:Conclusions}.

We assume a flat $\mathrm{\uplambda CDM}$ cosmology with cosmological
parameters corresponding to the central values inferred by the Planck
CMB observations \citep{Planck20}, of which the most relevant are the
current expansion rate, $\mathrm{H_{0}=67.4\,kms^{-1}\,Mpc^
{-1}}$, and the present-day matter density parameter, $\mathrm
{\Omega_{m}}$=0.315.

\section{Data.} \label{sec:data}

\begin{figure*}
    \begin{center}
    \includegraphics[width=0.9\linewidth]{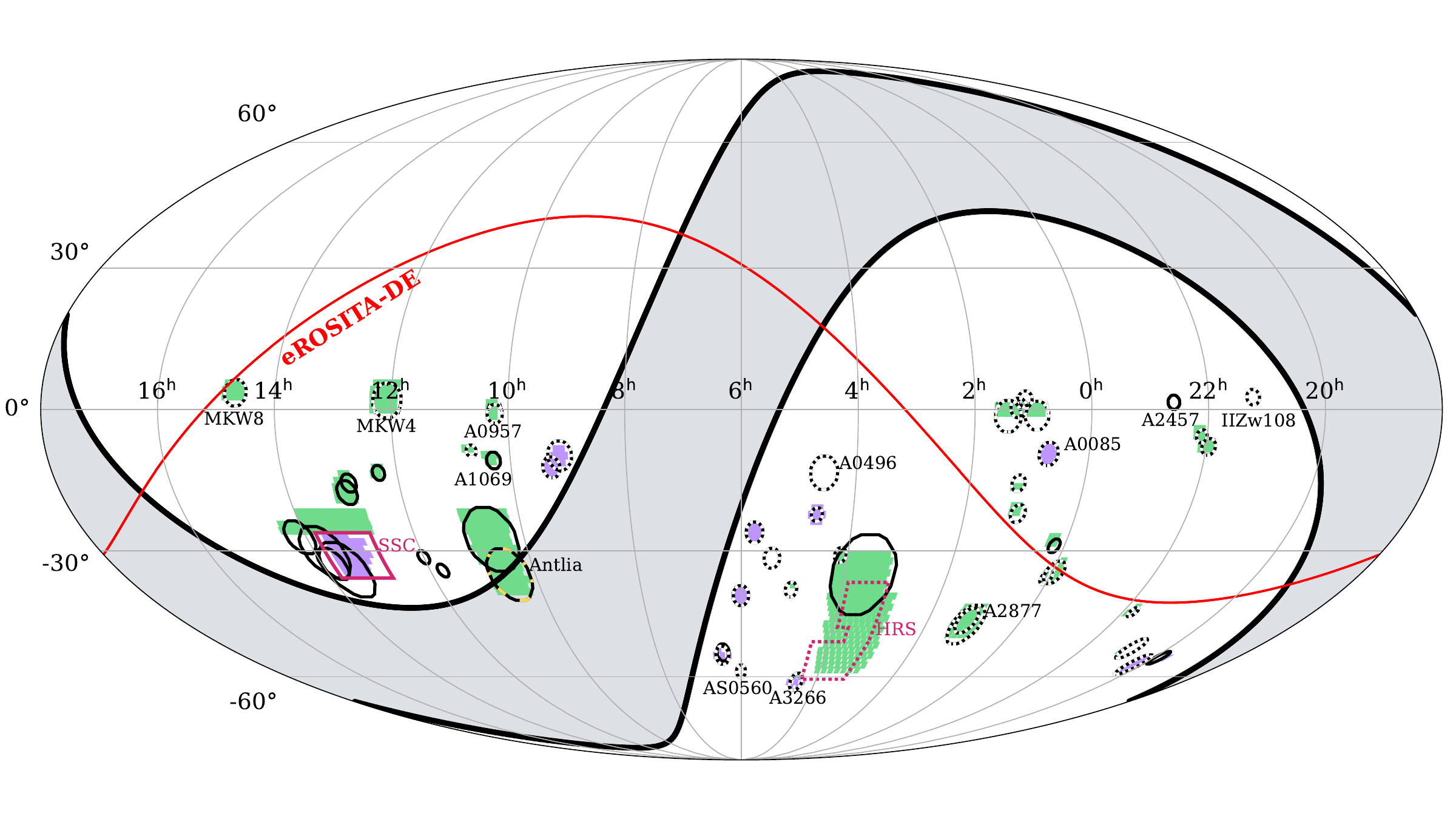}
    \end{center}
    \caption{Photometric data coverage for the CHANCES Low-z sub-survey.
    Black circles show the CHANCES-low-z cluster  location with their 
    corresponding $\rm 5 \times R_{200}$ regions. The Horologium-Reticulum (HRS) 
    and Shapley (SSC) superclusters regions are shown in magenta dashed boxes. 
    The \mbox{S-PLUS} -DR4 and -iDR5 footprint coverage around our clusters 
    are shown in green, 
    while the additional CHANCES-T80S campaigns are shown in purple. The clusters
    with available LS-DR9 photometric redshifts are showed with dotted circles, 
    while 49 of 50 clusters (except from Antlia, yellow dash-dotted circle) 
    from our sample are covered by \mbox{LS-DR10} and \mbox{LS-DR10-CBPF} photometric redshifts.
    The grey shaded region bounded by black solid lines shows Galactic latitudes $|b|\leqslant20^{\circ}$, 
    while the region below the red curve corresponds to the eROSITA-DE survey \citep{Merloni24}.}
    \label{fig:skycov}
\end{figure*}

CHANCES Low-z will target 50 clusters and 2 supercluster regions at $z<0.07$.
The selection criteria for the clusters are explained in detail in \citet
{Sifon25}. In short,  we used literature catalogues based on X-ray emission
and the Sunyaev-Zel’dovich effect to define the cluster sample in a
homogeneous way, considering cluster mass and redshift, as well as the
availability of ancillary data such as the DESI Legacy Imaging Survey
(LS, \citealt{Day19}, hereafter LS) and \mbox{S-PLUS} \citep
{MendesO19} photometry to facilitate the selection of target galaxies.  As a
result, we are sampling known clusters with an essentially uniform
distribution in mass at $M_{200}\gtrsim10^{14} M_{\odot}$, plus a few
well-known lower-mass systems. The distribution of the clusters in the sky
and the data coverage is shown in Figure ~\ref{fig:skycov}. 

Wide-field photometric observations in multiple narrow- and broad-band filters
allow us to derive accurate photometric redshifts of galaxies from which we
can compute cluster membership and characterize the environment. At the same
time, the photometry allows the study of galaxy transformations through their
morphologies and spatially-resolved stellar populations. 

To construct the CHANCES low-z cluster galaxy target catalogues, we made use
of LS and \mbox{S-PLUS} photometric datasets from which different photometric
redshifts are available, (e.g., from DR9 \& DR10 Legacy Survey and from S-
PLUS-DR4), to select galaxy cluster candidate members.

\subsection{DESI Legacy Imaging Survey.}\label{sec:pz_pls}

The DESI Legacy Imaging Surveys are a combination of three public projects:
the Dark Energy Camera Legacy Survey (DECaLS), the Beijing–Arizona Sky
Survey (BASS), and the Mayall z-band Legacy Survey (MzLS). LS provides deep
optical and infrared imaging across over 14,000 deg$^2$, covering both the
Northern and Southern Galactic Caps, utilizing the $g$, $r$, $z$ bands from
DECaLS, BASS, and MzLS, along with W1 and W2 bands from WISE \citep
{Day19}. The DECaLS imaging is acquired with the Dark Energy Camera
(DECam, \citealt{DePoy08}) on the Blanco 4-meter telescope at CTIO, Chile,
reaching 5$\sigma$ depths in extinction-corrected magnitudes of 24.39, 23.82,
22.95, in the $g$, $r$, and $z$ bands, respectively \citep{Day19}. The images
have been processed through the NOAO community pipeline, \texttt
{Legacy-pipeline}, optimized for precise source extraction and photometry.
The photometric processing employs the \texttt{Tractor} algorithm, which
models source fluxes using parametric profiles convolved with the point
spread function (PSF), ensuring consistent photometry across varying
observational conditions, and produces probabilistic inference on source
properties. 

Photometric redshifts for LS are derived using machine learning techniques. 
\citet{Zhou21,Zhou23} implemented a random forest algorithm trained on
spectroscopic redshifts(hereafter spec-z or $\rm z_{spec}$) from
approximately 2.2 million galaxies, achieving a normalized redshift bias of
$2.4\times 10^{-4}$ and a scatter $\rm \sigma_{NMAD}$ of 0.02, with an
outlier rate near 5.1\%. Photometric redshifts have been calculated for all
galaxies with at least one exposure in each of the $grz$-bands, i.e. \texttt
{NOBS\_G, NOBS\_R, NOBS\_Z} $\ge 1$, and have been incorporated into the LS
DR9 and DR10 data releases. These photometric redshifts are expected to be
reliable for galaxies with $z$-band magnitudes brighter than 21 \citep
{Zhou21,Zhou23}. The resulting LS catalogues offer comprehensive photometric
and redshift information, facilitating studies of galaxy evolution and
large-scale structure. We note that 49 of 50 clusters (excluding Antlia) and
the HRS and SSC  superclusters are covered by \mbox{LS-DR10}.

CHANCES low-z targets are taken from \mbox{LS-DR10}. This latest data release
includes additional DECam data from the Dark Energy Survey \citep
{DESC16,deJong15,Shanks15}, the DELVE Survey \citep{Drlica21}, and the DECam
eROSITA Survey (DeROSITAS; \citealt{Zenteno25}), including the $i$-band
imaging from those surveys. \textcolor{black}{We make use of the 
\mbox{LS-DR10} bricks covering a $15^\prime\times15^\prime$ sky patch to
create our \mbox{LS-DR10} parent catalogues from which we selected the
galaxy cluster candidate members by adopting the following criteria. We
select unique objects and ensure they match with GAIA \citep
{DeAngeli23,GAIADR3} by using the following conditions in
combination: \texttt{(cat['type'] != 'DUP') \& (cat['brick\_primary'] ==
True)}. The first condition ensures excluding Gaia sources that are
coincident with sources modelled with extended fit by the Tractor fitter.
Such sources were retained and do not have optical flux assigned in the
individual tractor tables. The second condition allowed us to select the
best observation of a source when it appears in multiple bricks. We remove
star-like  point sources by imposing \texttt{(cat
['type'] != 'PSF')}. Finally, we convert \mbox{LS-DR10} flux measurements in
units of nanomaggies into magnitudes by making use of the equation $m_{X} =
22.5 - 2.5\times\mathrm{log10(f_X)}$, where $\mathrm{f_X}$ represents the
flux of a given passband in nanomaggies. Errors on magnitudes have been then
retrieved from the \texttt{flux\_ivar} column with the equation
$|2.5\times0.434\frac{\mathrm{f_X}}{\sqrt{\mathrm{f_X\_ivar}}}|$.}

We use \mbox{LS-DR10} as our parent photometric catalogues throughout CHANCES,
ensuring uniform photometric and astrometric calibrations. Since CHANCES
targets have to share the focal plane with other 4MOST surveys, it is
mandatory that all surveys share the same astrometric solution. \mbox
{LS-DR10} is widely used as the parent photometric catalogues of the 4MOST
extragalactic surveys, and has its astrometry tied to GAIA DR2.  Figure \ref
{fig:skycov} shows the CHANCES-low-z sample distribution. Each circle
corresponds to the $\rm 5\times R_{200}$ radius, dotted circles indicate
those clusters also covered by 
\mbox{LS-DR9} $\mathrm{z_{phot}}$. Out of the 50 low-z clusters, 34 are
covered by \mbox{LS-DR9} $\rm z_{phot}$ while all clusters, except for
Antlia, of the CHANCES-low-z sample are covered by \mbox{LS-DR10} $\rm z_
{phot}$. 

Nevertheless, we noted that the faint (18.5<$\rm m_{r}$<20.5) low-mass cluster
galaxies at z<0.07 are largely absent from the training sets used by \cite
{Zhou21,Zhou23}. This is mainly caused by redshift and colour cuts applied to
some of the main deep surveys (e.g. SDSS-DR14 
\citealt{Abolfathi18, Strauss02} BOSS \citealt
{Dawson13,Dawson16}, VIPERS \citealt{Scodeggio18}, DEEP2 \citealt
{Newman13}, DESI \citealt{DESI1,DESI2,DESI3}, SDSS-DR16 \citealt
{Ahumada20}) used in the \mbox{LS-DR9} and \mbox{LS-DR10} photo-z training
samples, that exclude low-z galaxies(i.e. BOSS target galaxies at
0.15<z<0.70 while VIPERS and DEEP2 exclude galaxies at $z\lesssim0.5$). For
this reason we urge the need to develop our own photo-z based on \mbox
{LS-DR10} photometric datasets, as is explained in the following section.

\subsection{CHANCES Legacy Survey $\rm z_{phot}$ estimations.}\label{sec:pz_CBPF}

The publicly available photometric redshifts from \mbox{LS-DR10} provides a
valuable initial resource for our target selection. However, we identified a
need to re-estimate these $z_{\rm phot}$ to better suit our scientific
objectives specially at the faint-end. To this end, we developed a custom
photometric redshift model named \mbox{CBPF-$\rm z_{phot}$}\footnote
{CBPF stands for Centro Brasileiro de Pesquisas Físicas.}. This novel
approach allowed tailored estimation and greater control over the redshift
modelling process described as follows.

Our photometric redshifts re-estimations were obtained using Mixture Density
Networks(MDNs; \citealt{Bishop_1994}), which are neural networks combined
with a parametric mixture model \citep{McLachlan_2019, goushen_2010}, which
in our case outputs a Gaussian mixture distribution. The network architecture
was developed based on methods proven effective in previous photo-z
studies \citep{teixeira2024, Lima22}. Motivated by the absence of faint
(18.5<$\rm m_{r}$<20.5) low-mass cluster galaxies at z<0.07 in the training
sets used by \mbox{LS-DR9} and \mbox{LS-DR10}, resulting in an observed
galaxy cluster candidate members under-density in the red sequence faint-end
as shown by the \mbox{LS-DR9} and \mbox{LS-DR10} photo-z selection
(see Section~\ref{sec:CMTS}),
we modified the network architecture to separately process
colours and magnitudes using dedicated dense layers. These were then combined
through a late concatenation step, followed by additional dense layers and a
Gaussian mixture output (illustrative diagram provided in Figure \ref
{fig:mdn-CBPF}). This architectural improvement led to a significantly more
populated faint red sequence, consistent with results from \mbox{S-PLUS} $\rm
z_{phot}$, known for its enhanced redshift accuracy due to richer photometric
information (see Section~\ref{sec:CMTS} for a comparison between our
internal photo-z estimations (\mbox{LS-DR10-CBPF}) with \mbox{LS-DR9}
and \mbox{LS-DR10} public photo-z).

The model was trained using magnitudes and colours from the \mbox
{LS-DR10} dataset, while the reference spectroscopic redshifts used to train
and evaluate the model were obtained from a cross-match between \mbox
{LS-DR10} and the Southern-Hemisphere Spectroscopic Redshift
Compilation \citep{erik_specz2024}; further details regarding the spec-z
compilation are provided in Section~\ref{sec:spcz}. To construct a reliable
training sample, we selected galaxies based on the star–galaxy separation
( \texttt{(cat['type'] != 'PSF')}) already described in Section \ref
{sec:pz_pls}. We then applied the following criteria to ensure high-quality
photometry and remove unphysical colours:

\begin{tabular}{p{0.4\columnwidth}p{0.6\columnwidth}}
\begin{itemize}
\renewcommand\labelitemi{}
\setlength\itemsep{0.3em}
    \item $\bullet \quad \rm m_{g} < 24.0$
    \item $\bullet \quad \rm m_{r} < 23.0$
    \item $\bullet \quad \rm m_{i} < 23.0$
    \item $\bullet \quad \rm m_{z} < 22.5$
    \item $\bullet \quad \rm m_{W1} < 25$
    \item $\bullet \quad \rm m_{W2} < 25$
  \end{itemize} &
\begin{itemize}
\renewcommand\labelitemi{}
\setlength\itemsep{0.3em}
    \item 
    \item $\bullet \quad -1 < \rm m_{g} - m_{r} < 4$
    \item $\bullet \quad -1 < \rm m_{r} - m_{i} < 4$
    \item $\bullet \quad -1 < \rm m_{i} - m_{z} < 4$
    \item $\bullet \quad \rm 0.01 < z_{spec} < 1$
\end{itemize} \\
\end{tabular}

The global magnitude cuts exclude very faint galaxies, low-significances detections, 
and spurious sources, reflecting the intrinsic
limitations of the instrument. Although these thresholds are relatively 
generous, only a small number of objects lie close to these limits because we 
also applied a signal-to-noise criterion (SNR > 2 in the griz bands) 
when constructing the training sample. In addition, color cuts-defined in an 
agnostic way to remove sources with unphysical colours \citep{Drlica-Wagner_2018}.
After these cuts, we reserved 10\% of the resulting catalogue as a test sample
to validate the performance of the model on unseen data. From the remaining
dataset, we constructed our training sample by randomly selecting objects to
achieve a uniform spectroscopic redshift distribution up to $z_{\rm spec} =
1$, mitigating potential biases towards over-represented redshift values in
the training set. This upper limit does not affect the performance at 
lower redshifts, but it helps reduce systematic biases at higher redshifts
\citep[as demonstrated in][]{teixeira2024}.

The inputs to our MDN model included magnitudes from the \textit{g, r, i, z,
W1}, and \textit{W2} bands, along with all possible derived colours. The
network was trained to output probability density functions with peaks
aligned to the spectroscopic redshift values.

To ensure maximal coverage, we trained independent models for different
combinations of missing bands. Most clusters contain at least 80\% of objects
with complete photometric coverage, that is, reliable measurements in all six
bands: 
\textit{g, r, i, z, W1}, and \textit{W2}. The model trained using this full
set of bands is labelled \texttt{GRIZW1W2}. To handle sources with incomplete
data, we also trained additional models by omitting selected bands(using the
same architecture and training set, but with different magnitude inputs),
resulting in the \texttt{GRIZ}, \texttt{GRI}, and \texttt{GRZ} labelled
models. A similar approach was adopted in \citet{teixeira2024}. As expected,
in our case, the best results were obtained with the \texttt
{GRIZW1W2} model. Since this configuration covers the vast majority of
objects in our sample, all reported metrics throughout this work refer to
this model unless stated otherwise. 

Our improved photometric redshift model achieved a median bias, 
$(\delta z)^{\rm CBPF}{=}{+}0.008$, scatter $\rm \sigma_{NMAD}^{\rm CBPF}=0.028$, 
and $\rm \eta^{CBPF} = 0.05$, specifically for test sample objects within 
$0<\rm z_{spec}<0.3$ and $18.5<\rm m_{r}<20.5$ (mag), Here, $\rm \delta z = z_{phot}-z_{spec}$ 
is the bias of the photometric redshift estimate, $\rm \sigma_{NMAD}$ 
is the normalized median absolute deviation (see Section \ref{sec:CMTS} for details) 
and $\eta$ is the outlier fraction of the $\rm z_{phot}$ estimates (i.e. the fraction 
of galaxies with ${\mid}\delta{\rm z}{\mid}/(1+\rm z_{spec}) > 0.15$). 
For comparison, the corresponding values for the public \mbox{LS-DR10} 
photometric redshifts give a median bias of 
$(\delta z)^{\rm {LS-DR10}}=0.014$, $\rm \sigma_{NMAD}^{{LS-DR10}}=0.032$ for the scatter, and $\rm \eta ^{{LS-DR10}}= 0.19$. 

The systematic uncertainties $\sigma_z$ of the photo-z arise from dataset 
limitations (e.g., signal quality, depth) and can be estimated from the 68\% 
credible intervals of the PDFs for each object. These intervals were released 
along with the public \mbox{LS-DR10} photometric redshifts while  
the \mbox{LS-DR10-CBPF} intervals were obtained from our photo-z estimations. 
We find an averaged 
$\sigma_z^{\mathrm{CBPF}} = 0.044$ and $\sigma_z^{\mathrm{LS-DR10}} = 0.145$, 
demonstrating the improved precision of our novel estimates.

The overall accuracy metrics and uncertainties for the complete test set up to 
magnitude 20.5 are shown in Table \ref{tab:photoz_metrics}. 
Thus, the novel CBPF-$\rm z_{phot}$ estimates demonstrate comparable 
accuracies to the public \mbox{LS-DR10} values while notably enhancing the 
physical interpretability of colour-magnitude diagrams, particularly in the context 
of our target selection, and significantly improving photo-z quality for fainter 
objects at low-redshifts. Additional details on the full \mbox{LS-DR10-CBPF} 
photometric redshift estimations can be found in (\textit{Teixeira et al. in prep.}). 

\subsection{T80S/S-PLUS.}\label{sec:pz_S-PLUS}

The Southern Photometric Local Universe Survey (\mbox{S-PLUS}) footprint
covers around 9300 square degrees of the sky, utilising a robotic 0.83 m
telescope (T80-South, hereafter T80S) based at the Cerro Tololo
Inter-American Observatory (CTIO) in Chile. This telescope is equipped with
the T80Cam, featuring a detector of 9232$\times$9216 10$\mu$m-pixels,
providing a field of view of approximately 2 square degrees and a pixel scale
of 0.55 arcseconds.

\mbox{S-PLUS} employs the Javalambre 12-filter photometric system \citep
{Cenarro19}, which consists of five broad-band filters ($u,g,r,i,z$) and
seven narrow-band filters(J0378, J0395, J0410, J0430, J0515, J0660, J0861),
designed to coincide with, respectively, the [O\,{\sc ii}], Ca H + K, H\,
{\sc $\delta$}, G band, Mg\,{\sc b} triplet, H\,{\sc $\alpha$}, and Ca
triplet features. While originally tailored for stellar classification, the
system’s spectral coverage also enables detailed analyses of galaxies and
planetary nebulae. The J0660 narrow-band filter is centered at
$\uplambda$6614\AA, allowing it to capture both  H\,{\sc $\alpha$} and the
[N\,{\sc ii})] $\uplambda\uplambda$6548-6584\AA\, emission lines for objects
with redshifts up to $\sim$0.02.

We gather our \mbox{T80S/S-PLUS} datasets from the DR4 \& iDR5 \mbox
{S-PLUS} datasets and our own CHANCES-T80S observational programmes
CN2020B-30 and CN2022B-77 (PIs Y. Jaffé and C. Sif\'on), hereafter we will
refer to them as T80S/S-PLUS. The \mbox{CHANCES-S-PLUS} datasets followed the
DR4 \& iDR5 photometric data reduction and photometric redshift estimations,
fully described in 
\citet{Herpich24} and \citep{LimaDias25}, respectively, following
the standard \mbox{S-PLUS} data reduction and calibration \citep
{Oliveira25,AlmeidaF22}. In particular, the \mbox{S-PLUS} photometric
redshift estimations include a single-point estimations (SPEs) and
probability distribution functions (PDFs) based on galaxy spectroscopy for
all objects. These estimates were obtained using a supervised machine
learning algorithm based on a Bayesian Mixture Density Network model. More
details can be found in \cite{Lima22}. A global overview of the \mbox
{S-PLUS} main survey including the instrumentation, strategies and goals can
be found in \citet{MendesO19}. We crossmatch \mbox{S-PLUS} catalogues with
their \mbox{LS-DR10} counterparts to produce homogenous tables with GAIA
astrometric information and \mbox{LS-DR10} photometry, only keeping
the \mbox{S-PLUS} $\rm z_{phot}$ from which we select our galaxy cluster
candidate members. 

The green regions in Figure \ref{fig:skycov} indicate the clusters covered by
the \mbox{S-PLUS} DR4 main survey footprint, while purple ones correspond our
dedicated CHANCES-T80S observational campaigns. Table \ref
{tab:S-PLUSCoverage} summarizes the \mbox{T80S/S-PLUS} coverage of our low-z
cluster sample. The \mbox{T80S/S-PLUS} data represents a 55\% of our CHANCES
low-z data, with 35 out of the 50 clusters (${\sim}67\%$) having
partial \mbox{T80S/S-PLUS} coverage. Only 9 clusters, ${\sim}17\%$ of the
entire low-z sample, are fully covered out to $\rm 5\times R_
{200}$ with \mbox{T80S/S-PLUS} data. 

To compare the galaxy cluster candidate members selected by adopting different
$\mathrm{z_{phot}}$ estimates coming from \mbox{T80S/S-PLUS} and LS-DR9,
LS-DR10, and LS-DR10-CBPF we considered 6 clusters that are completely
covered out to $\rm 5\times R_{200}$ by \mbox{T80S/S-PLUS} and have a good
spectroscopic coverage down to $\rm{m_{r}}{=}20.4$ (A0500, A0957, A1631,
A2717, A3223, A3809), and complemented by clusters(A2399, A3490, A0780,
A1644, A3266, A3376, A3667, A0085, A3716, A2415) with a \mbox
{T80S/S-PLUS} coverage higher than 80\% alongside a good $\mathrm{z_
{spec}}$ coverage in the faint-end ($18.5<\rm m_{r}<20.4$). The complete list
of targets used for the target selection in this work (including both S-PLUS
data and CHANCES-T80S observations) is available at the archival section of
the S-PLUS database splus.cloud \footnote{\url{https://www.splus.cloud}}

\subsection{Spectroscopic redshifts.}\label{sec:spcz}

Many of the regions targeted have been observed by spectroscopic surveys
before. We use spectroscopic information from the literature to discard
galaxies outside the redshift range we are considering and also to assess the
completeness and purity of our photometric target selection. For this task,
we used the Southern Hemisphere Spectroscopic Redshift Compilation \citep
{Erik24_compilation}. This compilation was created with the goal of
establishing a basis for photometric redshift models \citep{Lima22} to be
used in the \mbox{S-PLUS} survey \citep{MendesO19}, covering the entire sky
below the declination of +10$^{\circ}$.

The current version of this compilation contains over 5000 catalogues from
different sources, such as VizieR\footnote{\url
{https://vizier.cds.unistra.fr/}}, High Energy Astrophysics Science Archive
Research Center(HEASARC)\footnote{\url
{https://heasarc.gsfc.nasa.gov/}}, NASA/IPAC Extragalactic Database
(NED)\footnote{\url{http://ned.ipac.caltech.edu/}}, the Sloan Digital Sky
Survey's eighteenth data-release 
\citep{Almeida23_SDSSDR18}, and others. Since some catalogues contain
duplicated information, we apply a duplicate removal procedure and the
number of catalogues used for the objects that remain is 1852. The total
number of objects is 8437460, of which 2548065 are galaxies, 4812282 are
stars, and 507894 are quasars. The catalogue is publicly available \footnote
{\url{https://github.com/ErikVini/specz_compilation}} and includes the full
reference list of catalogues used.\\

\begin{figure*}
    \centering
    \includegraphics[width=0.96\linewidth]{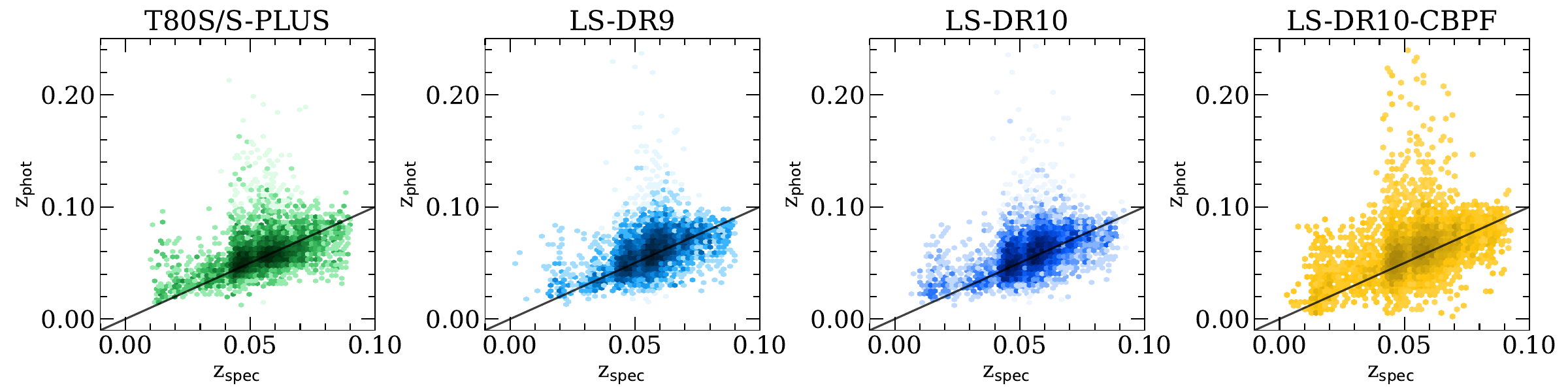}
    \caption{Comparison between photometric (z$_{phot}$) and spectroscopic (z$_{spec}$) 
    redshifts for each of the photo-z sets used in this work, from left to rigth \mbox{T80S/S-PLUS}, 
    \mbox{LS-DR9}, \mbox{LS-DR10} and \mbox{LS-DR10-CBPF}. Coloured symbols correspond to the  
    bright ($\rm m_{r}<18.5$) galaxy cluster candidate members
    belonging to 16 clusters with good spatial and z$_{\mathrm{spec}}$ coverage for all surveys.}
    \label{fig:photoz-specz}
\end{figure*}

A direct comparison of the $\mathrm{z_{phot}}$ coming from \mbox
{T80S/S-PLUS}, \mbox{LS-DR9}, \mbox{LS-DR10-} and \mbox{LS-DR10-CBPF}-$\rm z_
{phot}$ against $\mathrm{z_{spec}}$, by considering 16 clusters well covered
by $\rm z_{phot}$ \mbox{T80S/S-PLUS}, \mbox{LS-DR10} and \mbox
{LS-DR10-CBPF} is shown in Figure \ref{fig:photoz-specz}. We find that, the
mean spectroscopic and photometric redshift difference:
$\rm \langle \delta_z \rangle = \langle|z_{spec} - z_{phot}|\rangle$ is
$\langle \delta_{z}\rangle=0.013 \pm 0.019$ if we consider all the combined
target selection adopted for bright galaxies ($\rm m_{r} < 18.5$, see
Sec \ref{sec:CMTS}) using $\rm z_{phot}$ estimations from \mbox
{T80S/S-PLUS}, \mbox{LS-DR9}, \mbox{LS-DR10} and \mbox{LS-DR10-CBPF}. If we
then inspect this difference for each $\rm z_{phot}$ independently, we find
that $\rm \delta_z$ is 0.009, 0.011, 0.012, and 0.013  with Spearman
correlation coefficients ($\rho$) of 0.57, 0.53, 0.58, and  0.53 for \mbox
{T80S/S-PLUS},
\mbox{LS-DR9}, \mbox{LS-DR10} and \mbox{LS-DR10-CBPF} respectively. 

\section{Cluster membership and target selection strategy.}\label{sec:CMTS}

\begin{figure*}
\subfloat{\includegraphics[width = 0.25\linewidth]{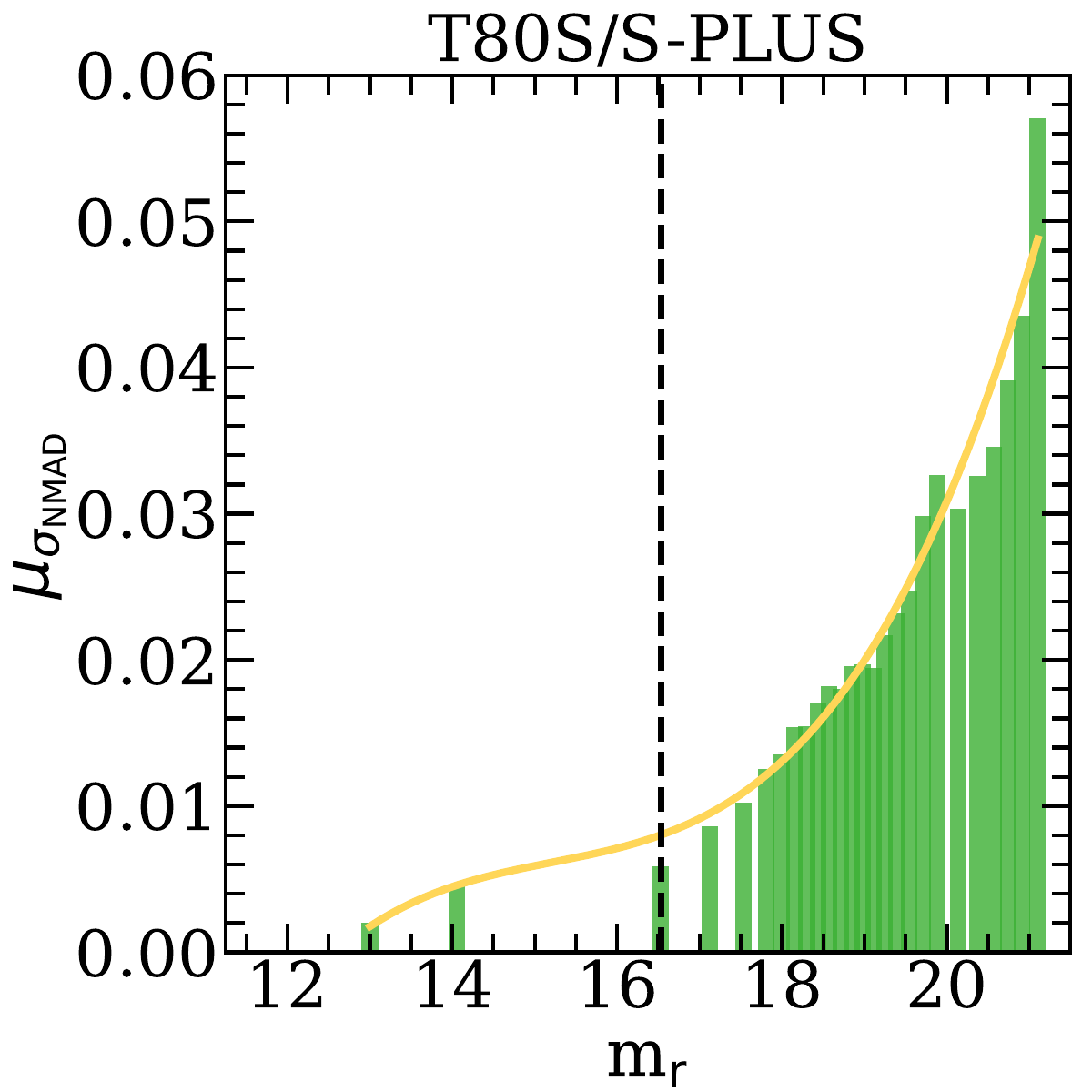}} 
\subfloat{\includegraphics[width = 0.25\linewidth]{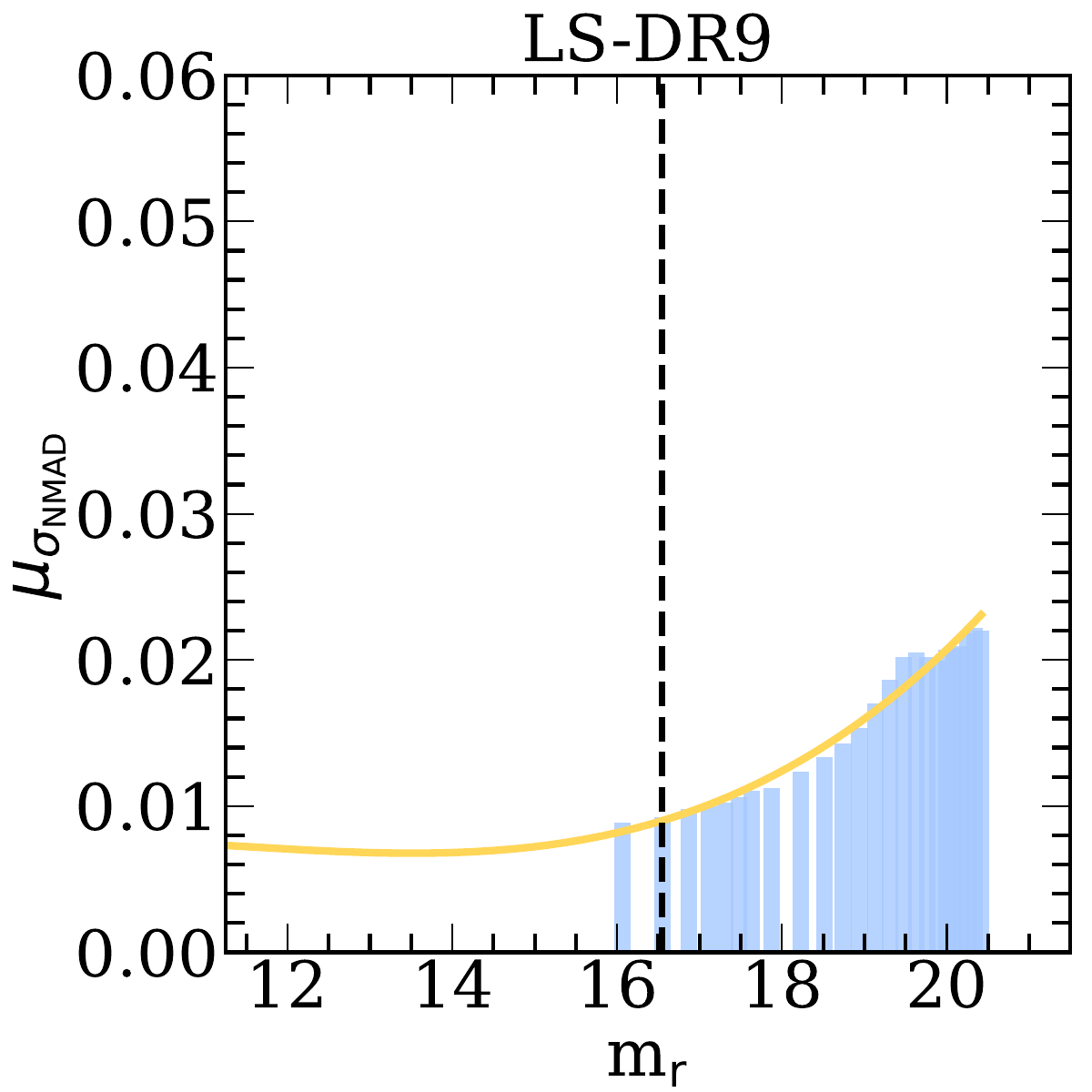}}
\subfloat{\includegraphics[width = 0.25\linewidth]{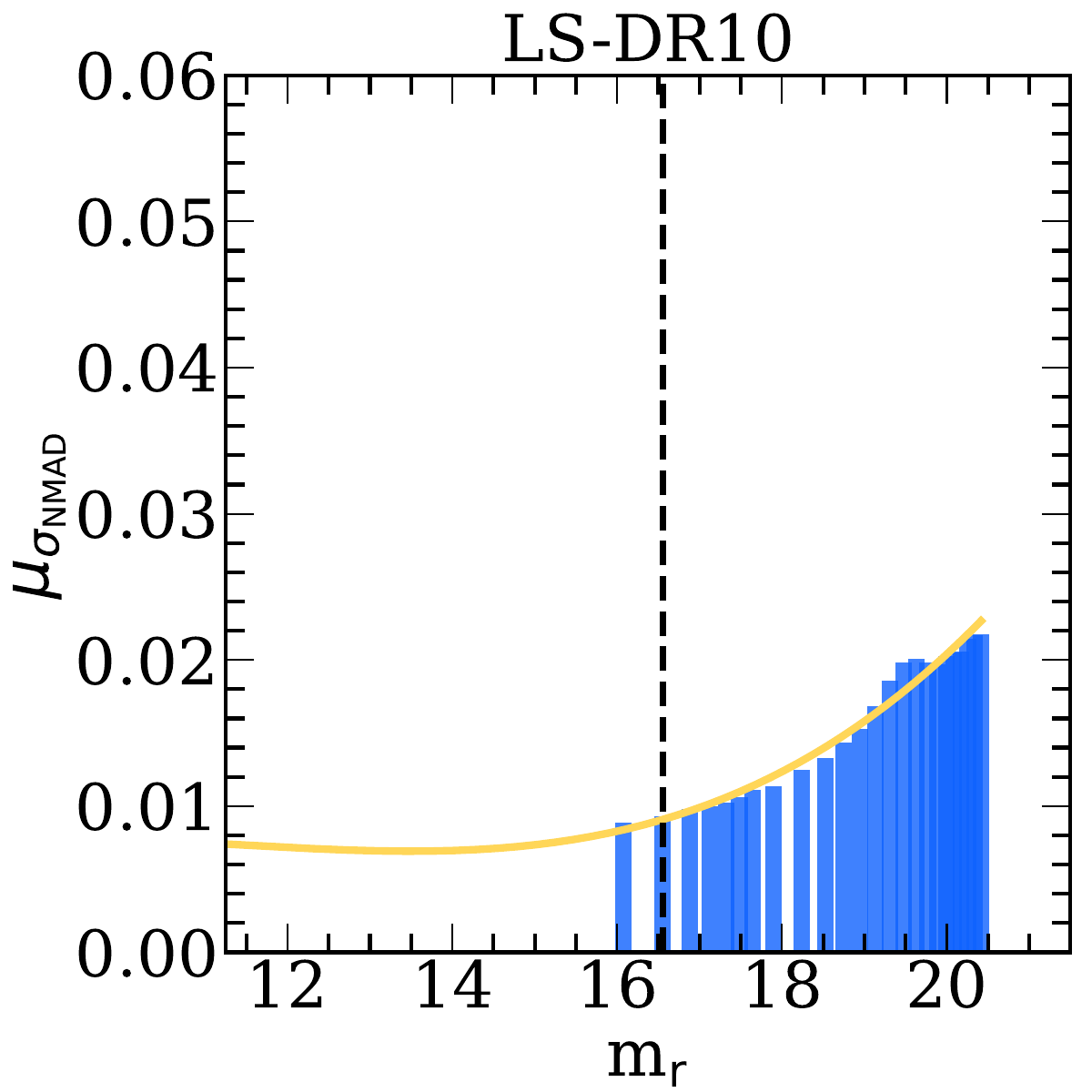}} 
\subfloat{\includegraphics[width = 0.25\linewidth]{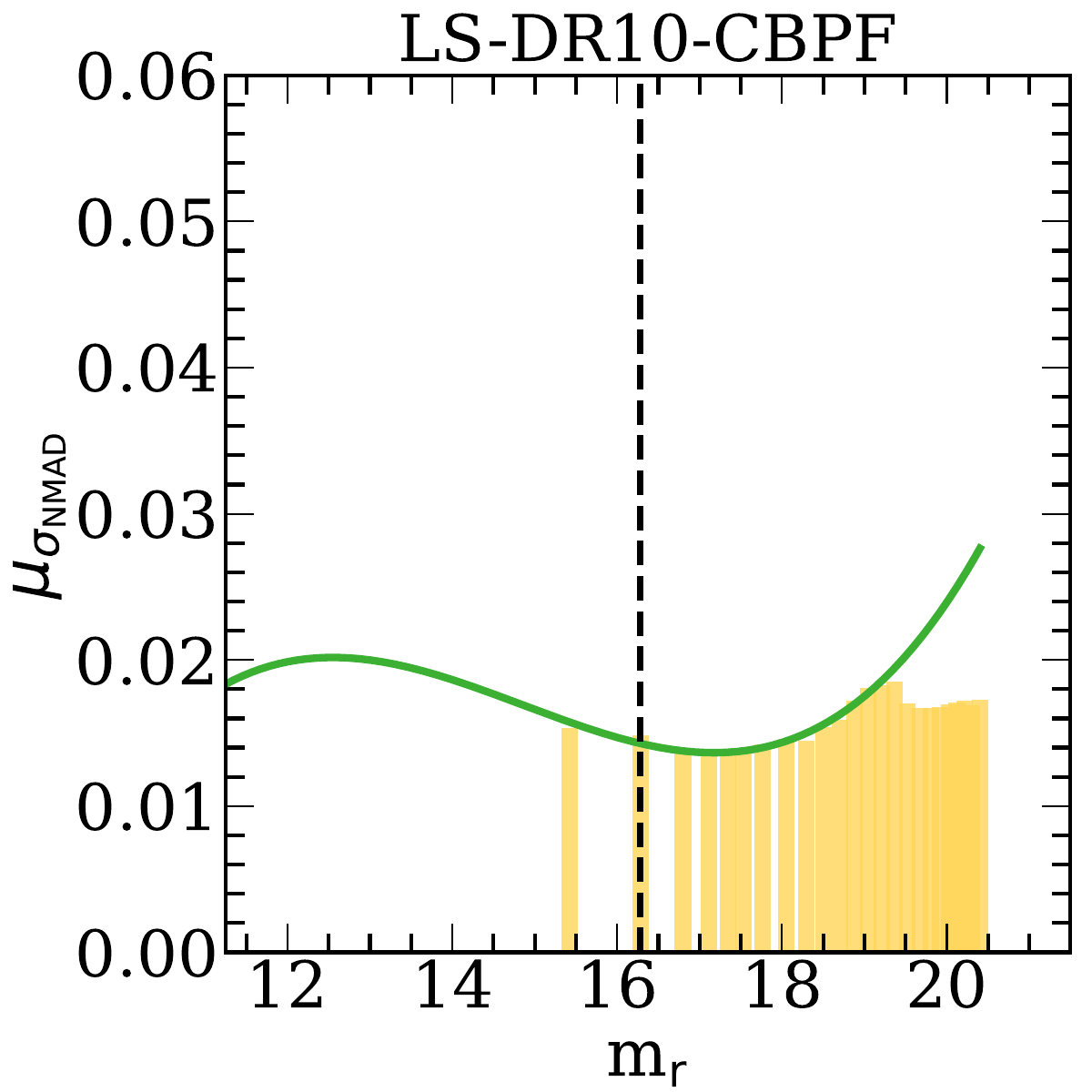}}
\caption{$\sigma_{\rm NMAD}$ curves used to select the most likely galaxy
cluster candidate members for the low-z-CHANCES target catalogues for each
set of photometric redshifts: 
\mbox{T80S/S-PLUS} (green), \mbox{LS-DR9} (light-blue), \mbox{LS-DR10}
(blue) and \mbox{LS-DR10-CBPF} (yellow). The curves were obtained through a
polynomial fit on the $\sigma_{\mathrm{NMAD}}$ histogram with equally $\rm
log(m_{r})$ sized bins for each of the $\rm z_{phot}$ training sample
(see Section \ref{sec:CMTS} for details), dashed lines indicate the $\rm m_
{r}$-limit at which a constant value was adopted for selecting the cluster
galaxy candidate members. These curves evidence the differences in $\rm m_
{r}$ distributions of the different training samples used to determine $\rm
z_{phot}$.}
\label{fig:SNMADCURVES}
\end{figure*}

We use all the photometric information and $\rm z_{phot}$ available, described
in the previous section, to select candidate cluster members whose
photometric redshifts are consistent with being at the known redshift of the
cluster. To do so, we first, measure the average deviations of the
photometric redshifts with respect to the spectroscopic redshift measurements
$\langle\sigma_{\mathrm{NMAD}}\rangle$ in 0.2 mag bins of $r$-band magnitude,
using the normalized median absolute deviation as a robust measure of the
uncertainties, where $\sigma_{\mathrm{NMAD}}$ is defined as: 

\begin{equation}
\sigma_{\mathrm{NMAD}} =1.48 \times \mathrm{median}\Bigg( \left| \frac{\delta z  - \mathrm{median}(\delta z)}{1+z_{\mathrm{spec}}} \right| \Bigg) 
\end{equation}

\noindent where $z_{\mathrm{phot}}$ is the photometric redshift and $\delta z
= (z_{\mathrm{phot}}-z_{\mathrm{spec}})$ is its bias. \citep
{Brammer08} shows that $\sigma_{\mathrm{NMAD}}$ is robust, with low
sensitivity to outliers, gradually increasing for fainter magnitudes. We
then, fit a 3rd-order polynomial function to the average normalized median
absolute deviation(refer to Section \ref{sec:appendix} to see the best-fit
values), which assess the quality of the $\rm z_{phot}$ estimates
$\langle\sigma_{\mathrm{NMAD}}\rangle$) per $\mathrm{m_{r}}$-bin for \mbox
{T80S/S-PLUS} and LS photometric redshifts individually, as shown in
Figure \ref{fig:SNMADCURVES}.

Using these curves, we can now select galaxy cluster candidate members
considering galaxies with photometric redshifts within the mean cluster
redshift ($\rm z_{cl}$) considering a given error range. In other words: 
$\rm | z_{phot} - z_{cl} | < N \times \sigma_{\rm NMAD}$, where $\rm N$ can be
adjusted according to the data set to maximise completeness. Note that we
assume a constant value of $\rm N \times \sigma_{NMAD}$ for objects brighter
than $\rm m_{r}\lesssim$16.5 to include spectroscopic confirmed bright galaxy
cluster members from the \mbox{T80S/S-PLUS} datasets. Correspondingly, for
consistency we assumed a similar constant value for the adopted LS selection
functions.

\begin{figure*}
\subfloat{\includegraphics[width = 0.25\linewidth]{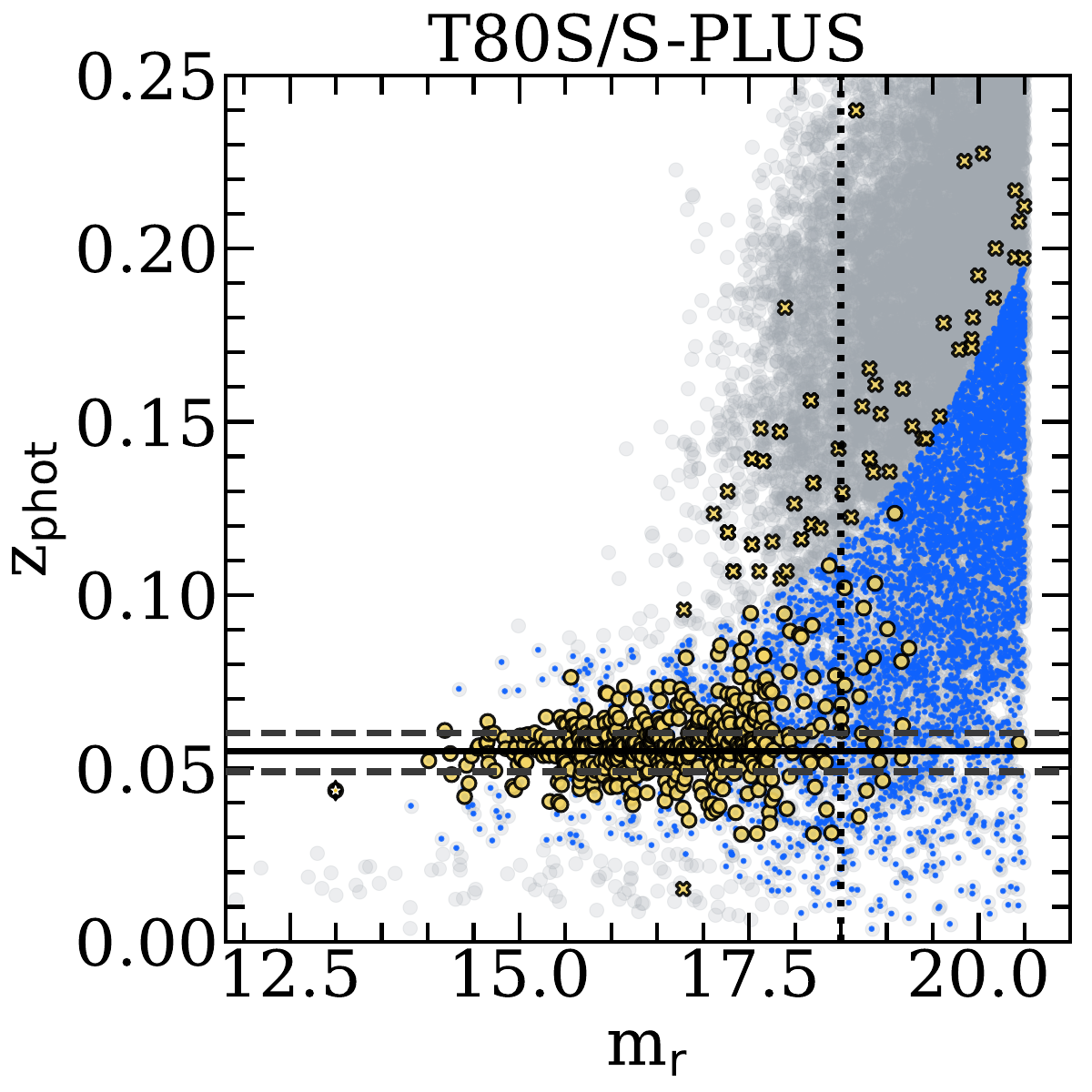}} 
\subfloat{\includegraphics[width = 0.25\linewidth]{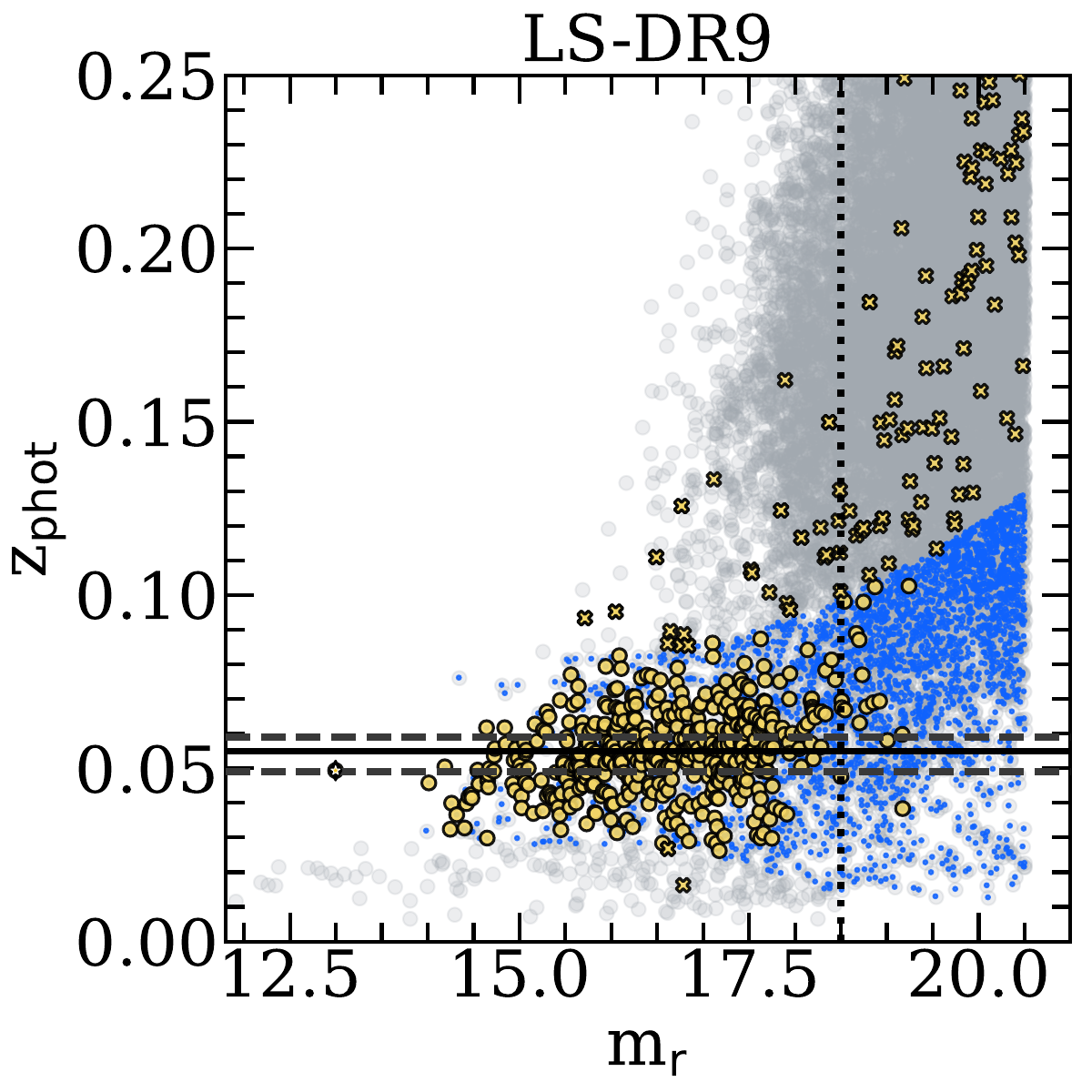}}
\subfloat{\includegraphics[width = 0.25\linewidth]{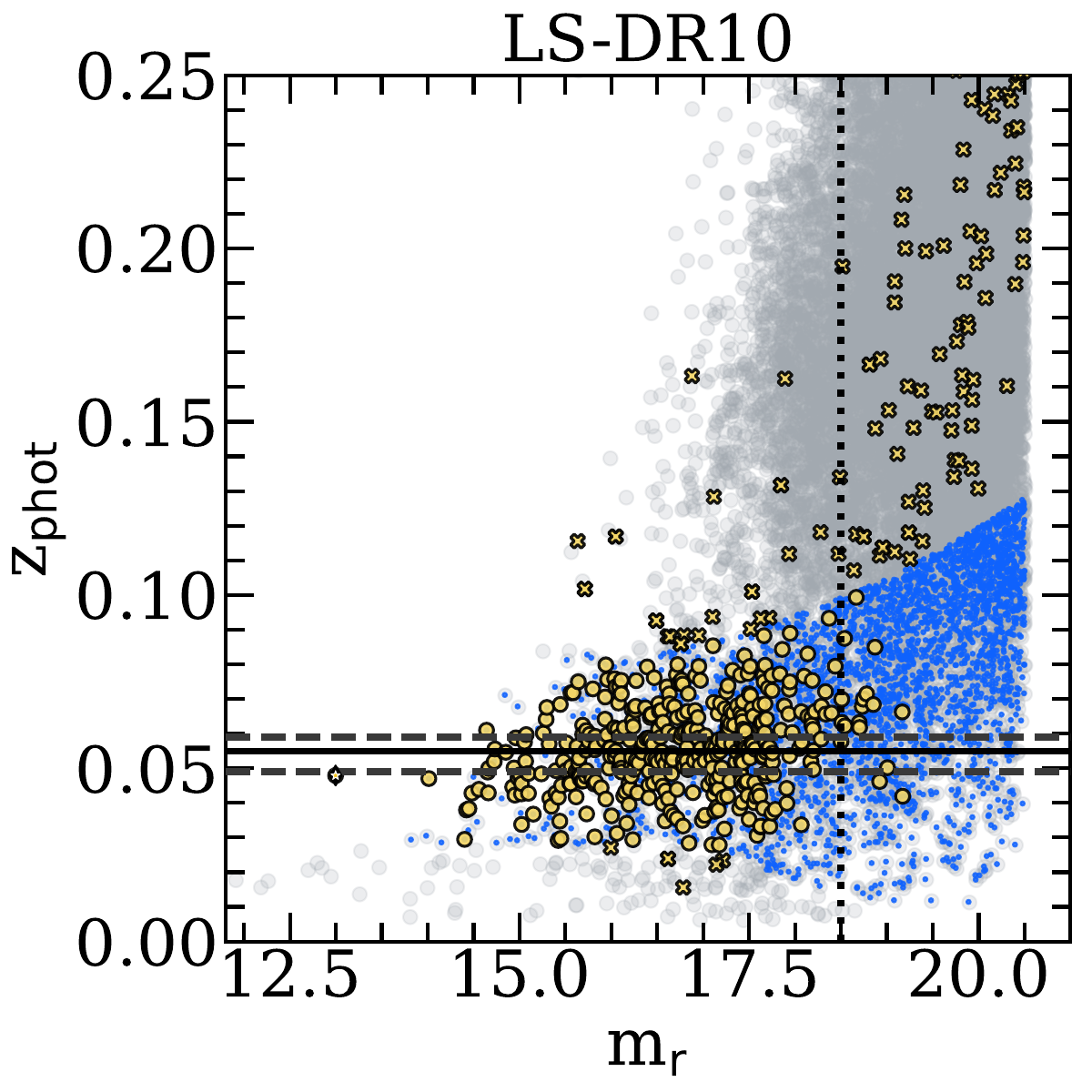}}
\subfloat{\includegraphics[width = 0.25\linewidth]{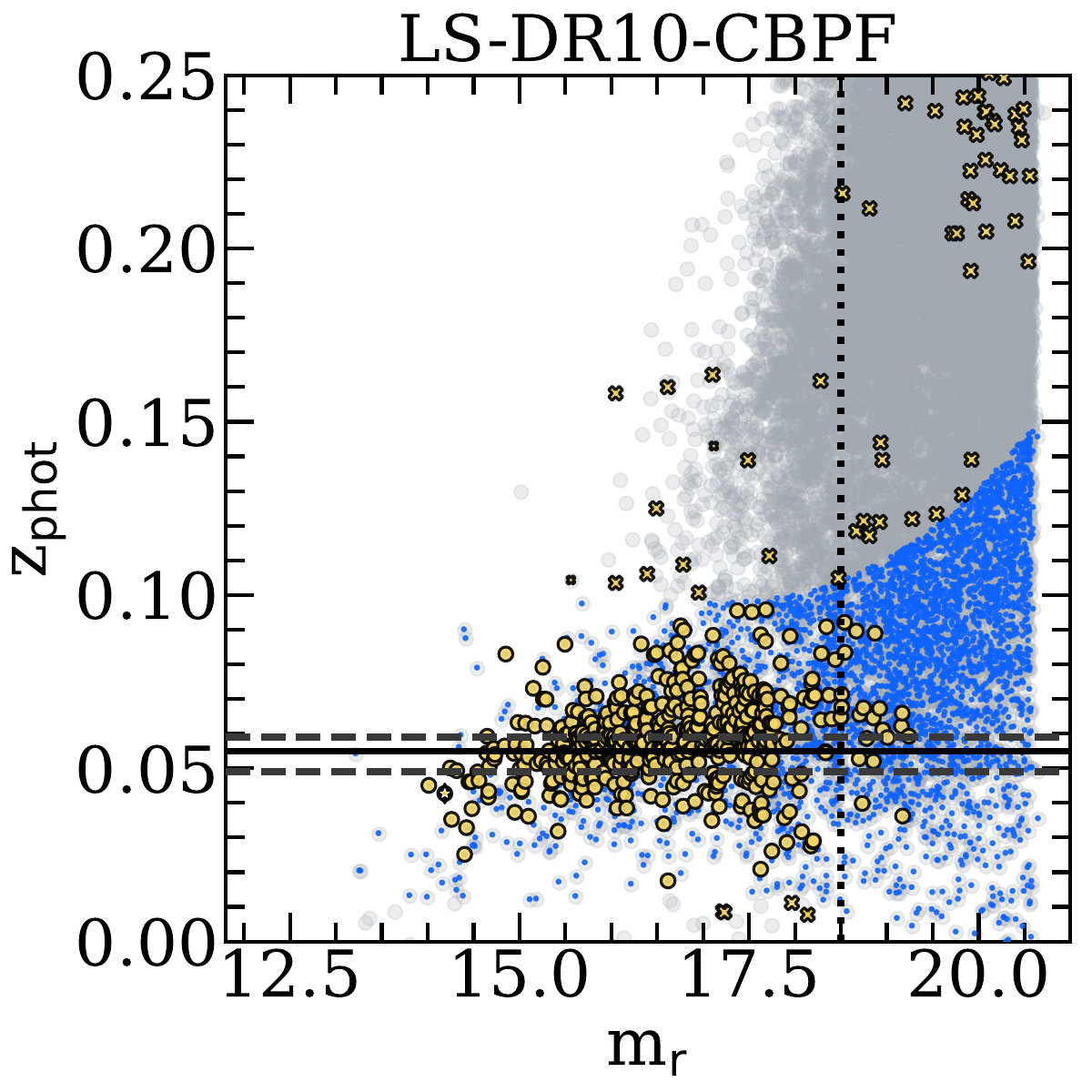}}
\caption{Photometric redshift as a function of $r$-band magnitude for each set
of photometric redshifts, from left to right: \mbox{T80S/S-PLUS}, \mbox
{LS-DR9}, \mbox{LS-DR10} and \mbox{LS-DR10-CBPF}. We include the
corresponding parent sample (grey circles), the A0085 galaxy cluster
candidate members selected from our photometric method (blue), the
spectroscopically confirmed members (yellow circles), and the confirmed
spectroscopic cluster members and spectroscopic objects within the cluster
redshift range (yellow crosses) that were not selected by our method. The
solid black line indicates the cluster redshift obtained from the mean
spectroscopic redshift distribution, while the dashed lines indicate the
cluster spectroscopic redshift lower and upper limits. The  bright/faint
boundary at $\rm m_{r}=18.5$ is indicated by a vertical dotted-line.
Adopting different $\rm z_{phot}$ with their corresponding  $\rm \sigma_
{NMAD}$ curves lead to a different cluster galaxy candidate members
selection, specially at  fainter magnitudes ($\rm m_
{r}\geqslant18.5$), where S-PLUS selects more candidates as compared with
LS-$\rm z_{phot}$.}
\label{fig:mrzphot}
\end{figure*}

To illustrate our galaxy cluster candidate member selection process, we make
use of Abell 85  
\mbox{T80S/S-PLUS}, \mbox{LS-DR9}, LS-DR10 and our  LS-DR10-CBPF $\mathrm{z_
{phot}}$ available. Figure~\ref{fig:mrzphot} shows the corresponding parent
samples in grey symbols, blue symbols correspond to the photometrically
selected objects, while yellow open symbols correspond to photometric
selected objects with known spectroscopically redshift determinations
($\rm z_{spec}$). Figure~\ref{fig:mrzphot} shows the striking differences
among the $\rm z_{phot}$ used in this work, while T80/SPLUS assigns photo-z
values as high as $\rm z_{phot}\lesssim0.2$, LS-DR9 -DR10 and -DR10-CBPF
$\rm z_{phot}$ only reach $\rm z_{phot}\lesssim0.15$, caused by the
different selection functions shown in \ref{fig:SNMAD}. These variations are
more evident at lower $\rm m_{r}\geqslant18.5$ where spectroscopically
confirmed cluster members and cluster candidate members with known
spectroscopic redshift lying within the $\rm |z_{cls}|<\sigma$ range are
missed by our z-phot selection method. LS-DR9 and LS-DR10 photo-z
estimations miss twice the spectroscopic objects missed by S-PLUS, while our
LS-DR10-CBPF $\rm z_{phot}$ recovers most of them.

\begin{figure*}
\subfloat{\includegraphics[width = 0.25\linewidth]{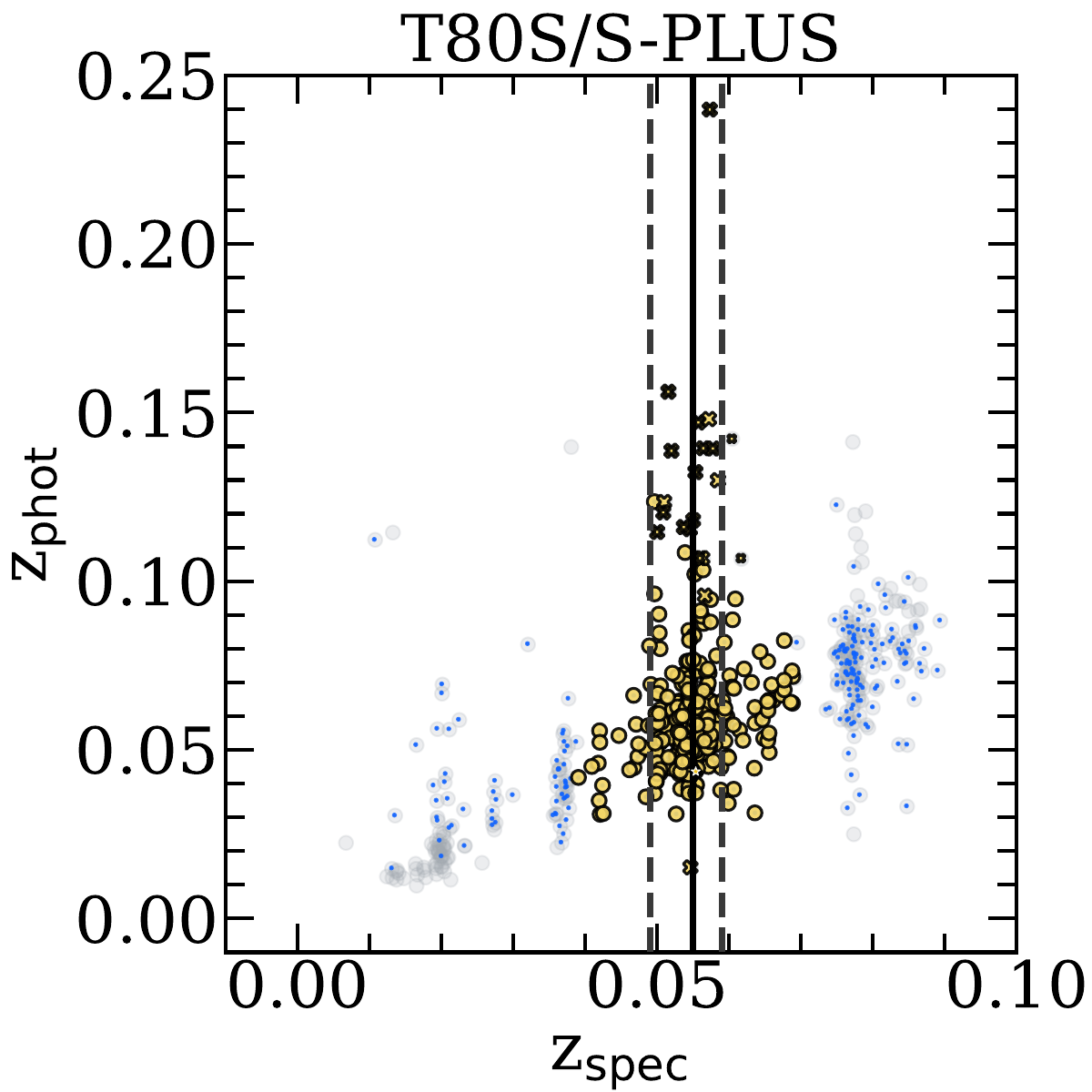}} 
\subfloat{\includegraphics[width = 0.25\linewidth]{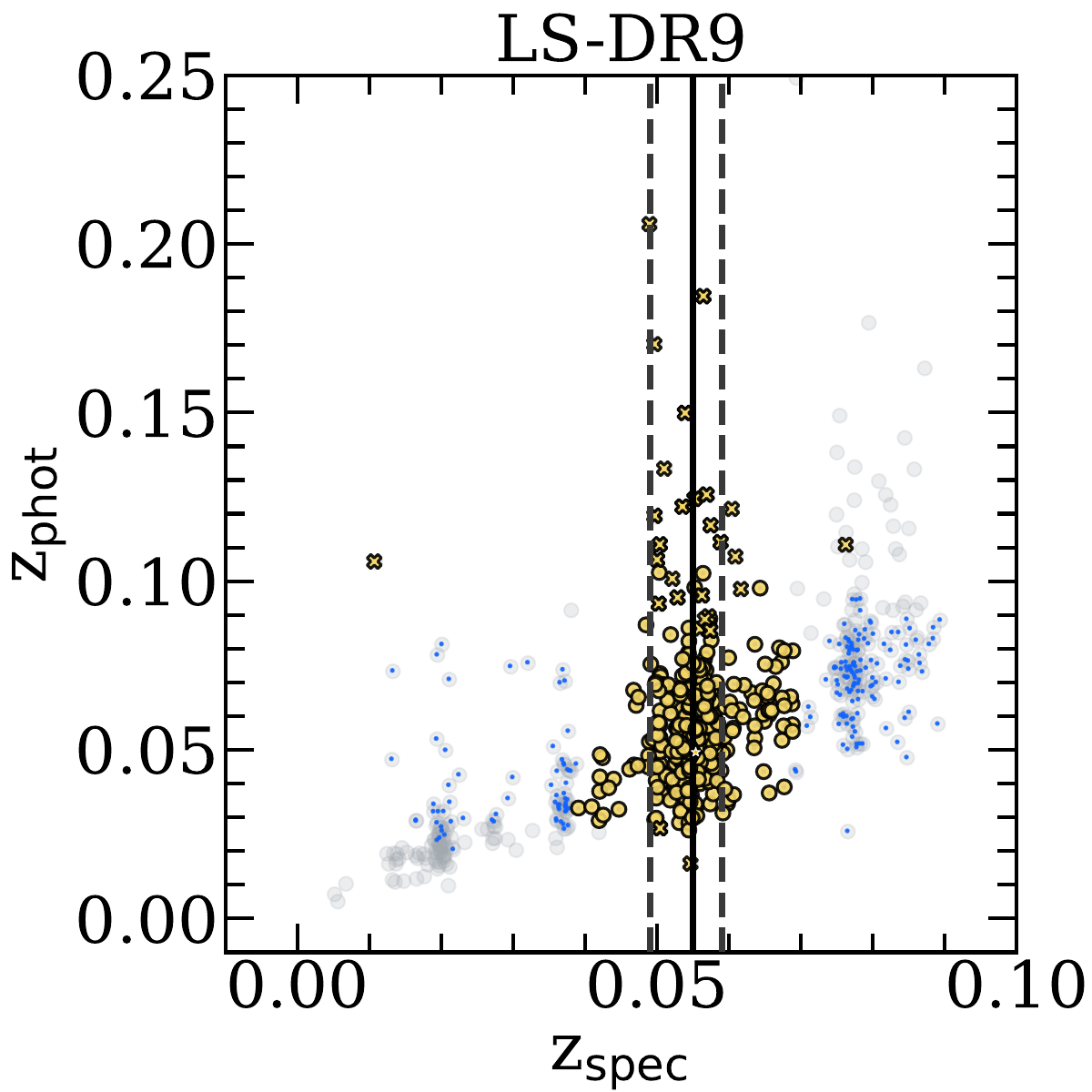}} 
\subfloat{\includegraphics[width = 0.25\linewidth]{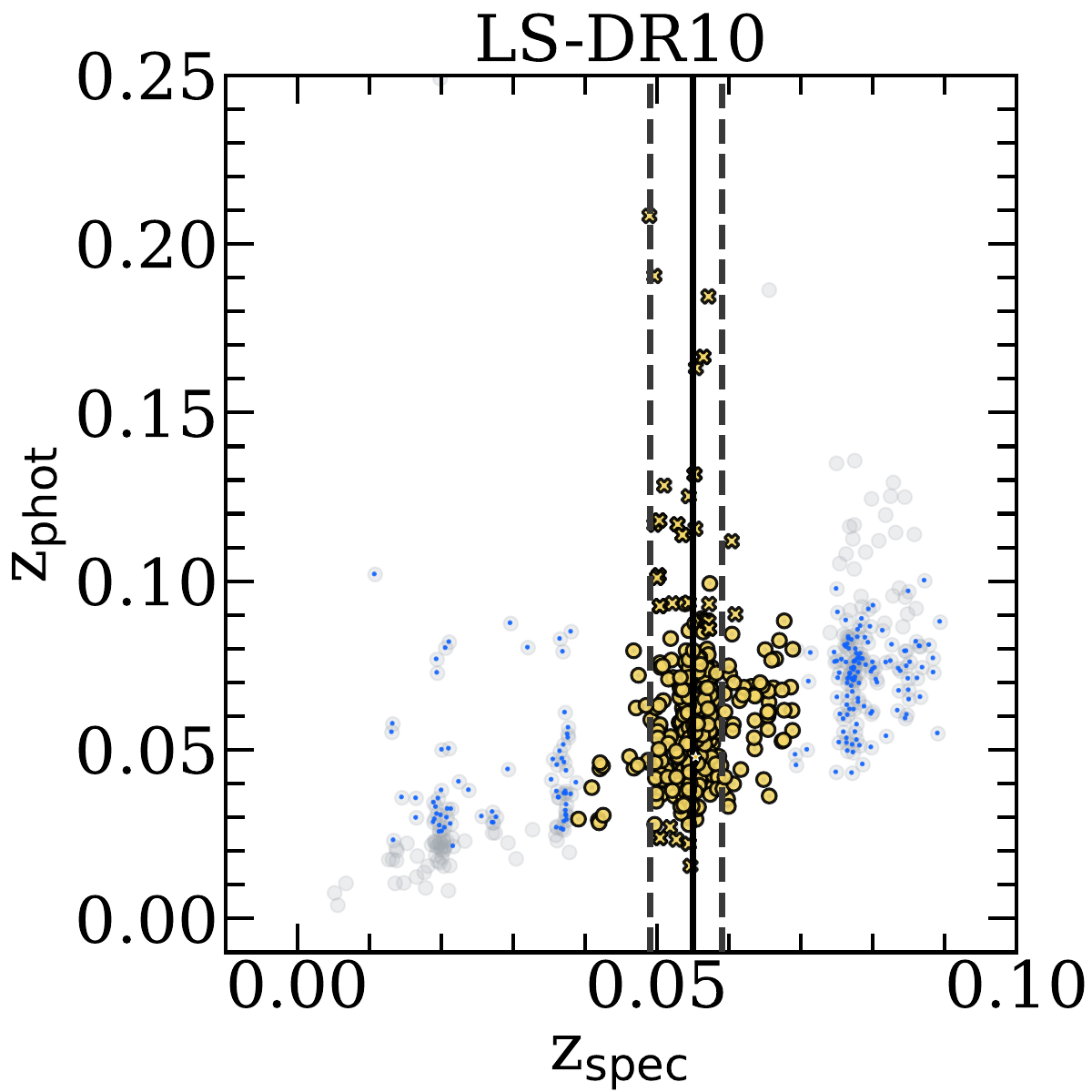}}
\subfloat{\includegraphics[width = 0.25\linewidth]{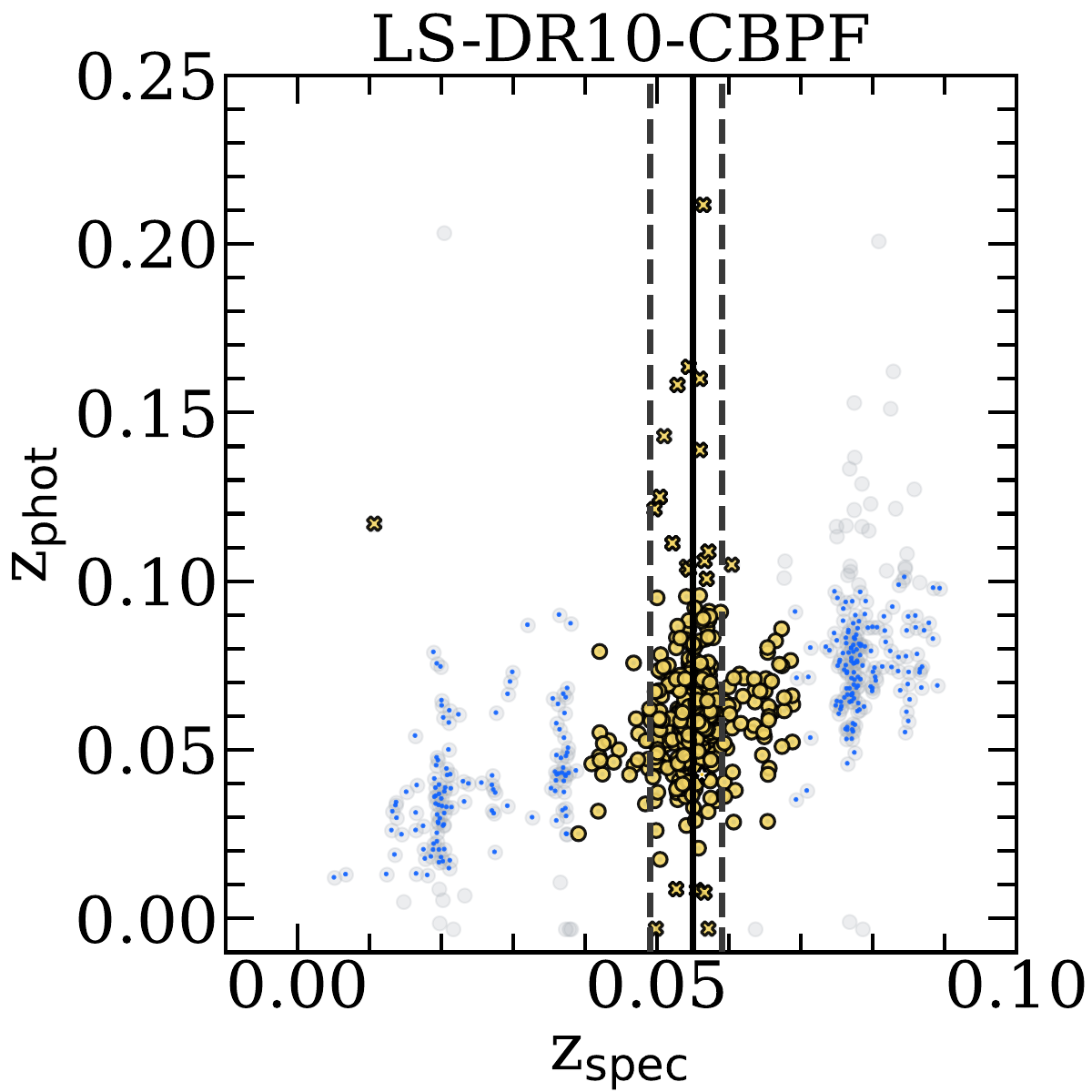}}
\caption{Photometric redshift versus spectroscopic redshift for each set of
photometric redshifts, from left to right: \mbox{T80S/S-PLUS}, \mbox
{LS-DR9}, \mbox{LS-DR10} and \mbox{LS-DR10-CBPF}. We include the
corresponding parent sample (grey circles), the A0085 galaxy cluster
candidate members selected from our photometric method (blue), the
spectroscopically confirmed members (yellow circles), and the confirmed
spectroscopic cluster members and spectroscopic objects within the cluster
redshift range (yellow crosses) that were not selected by our method. The
solid black vertical-line indicates the cluster redshift obtained from the
mean spectroscopic redshift distribution, while the dashed vertical-lines
indicate the cluster lower and upper limits.}
\label{fig:zphotzspec}
\end{figure*}

\begin{figure*}
\subfloat{\includegraphics[width = 0.25\linewidth]{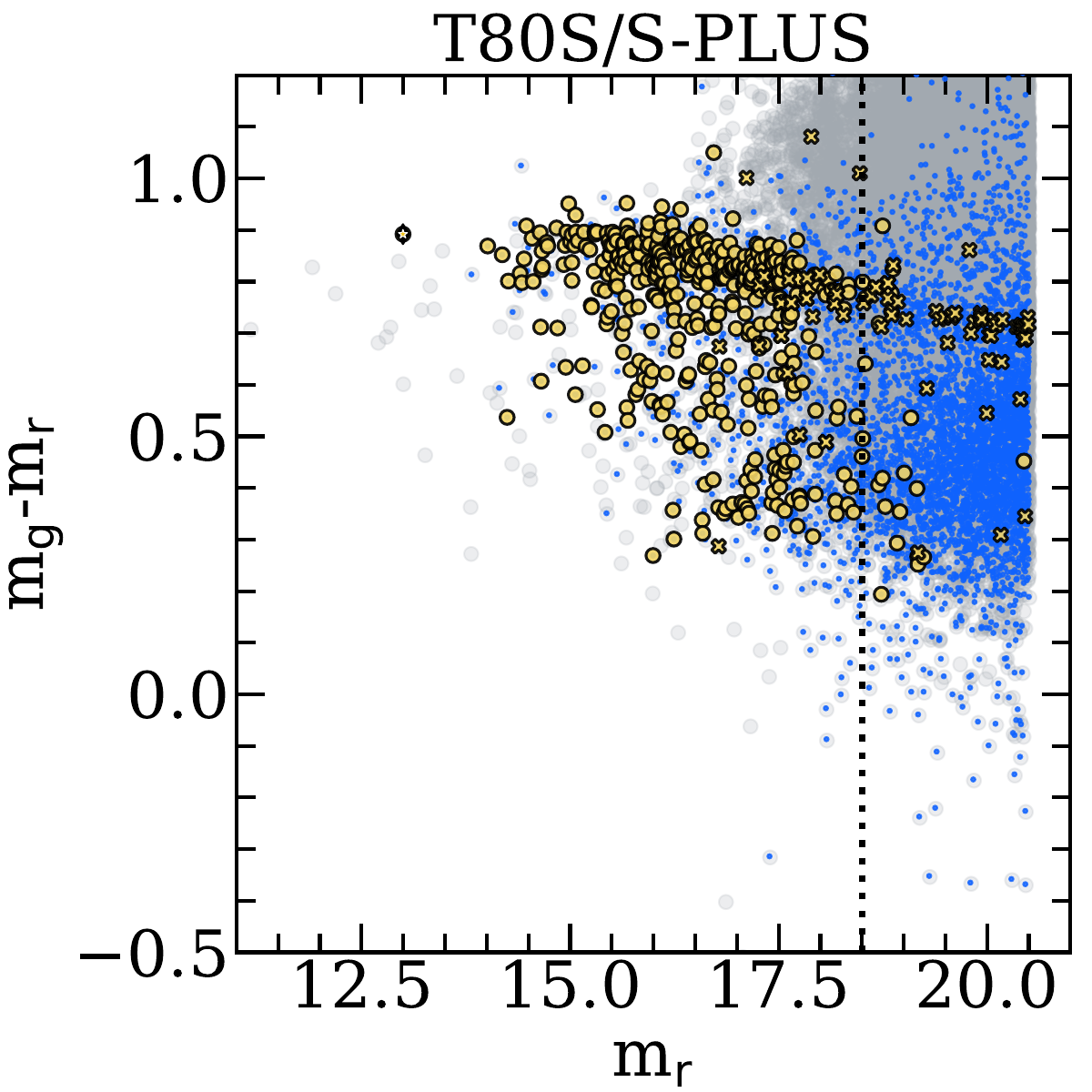}} 
\subfloat{\includegraphics[width = 0.25\linewidth]{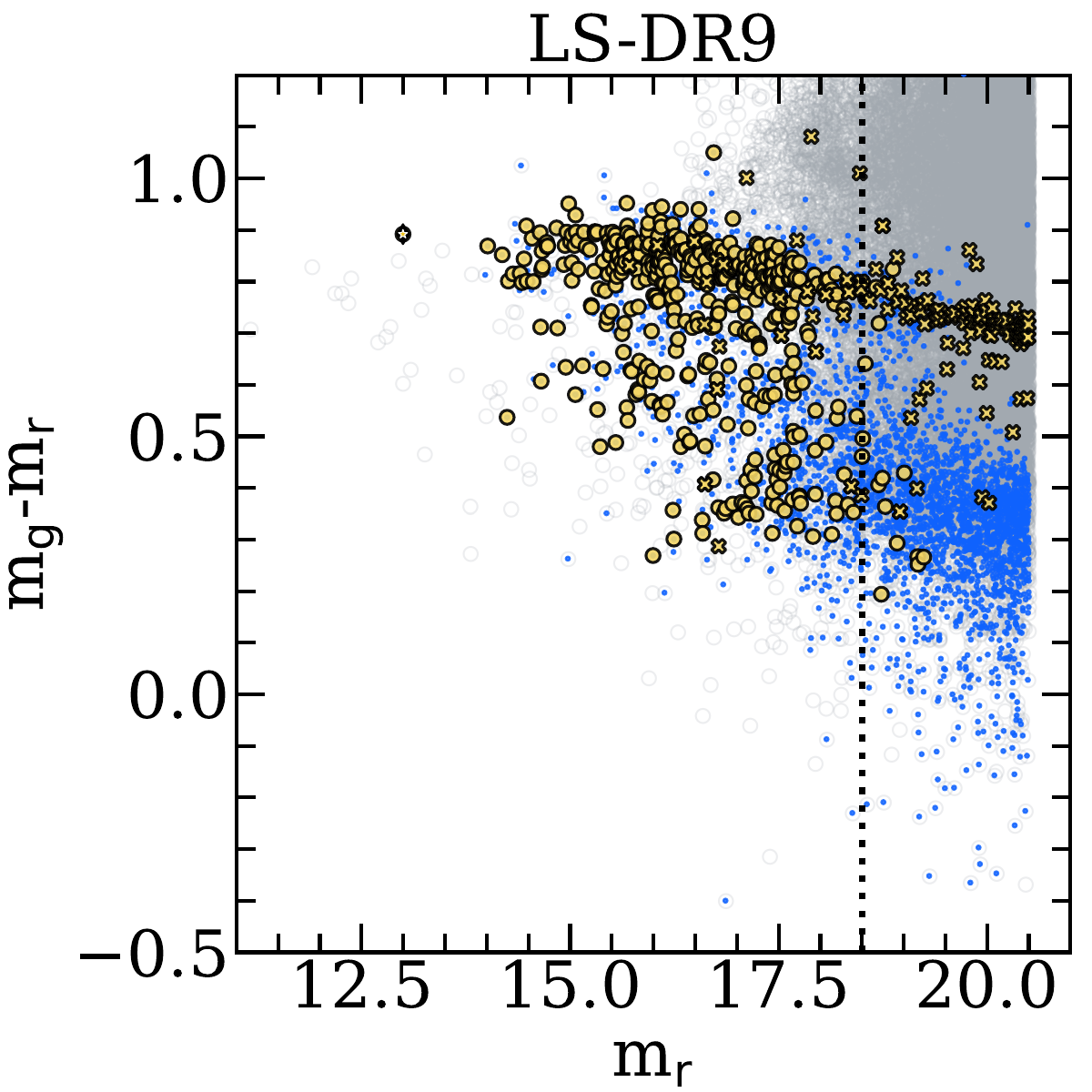}} 
\subfloat{\includegraphics[width = 0.25\linewidth]{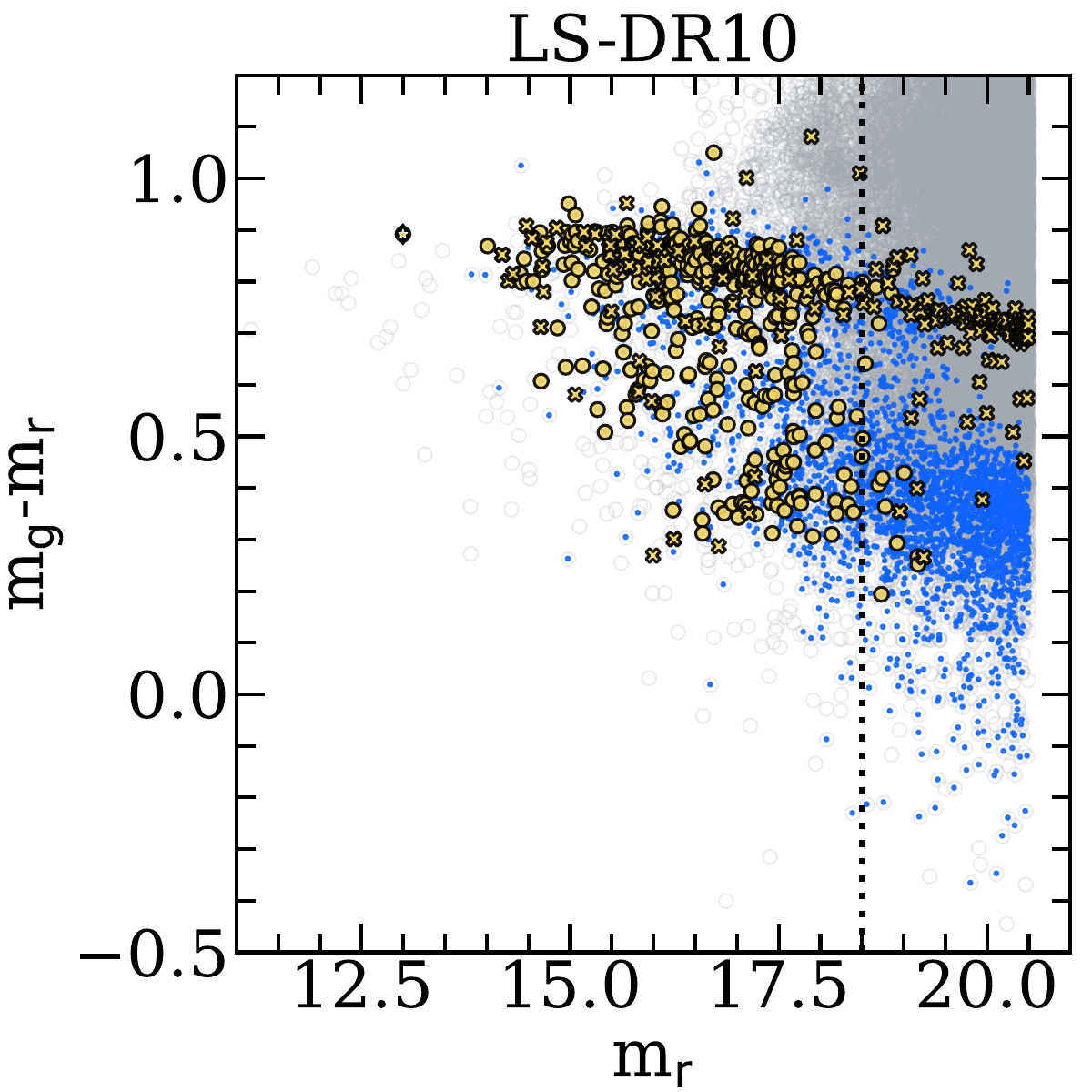}}
\subfloat{\includegraphics[width = 0.25\linewidth]{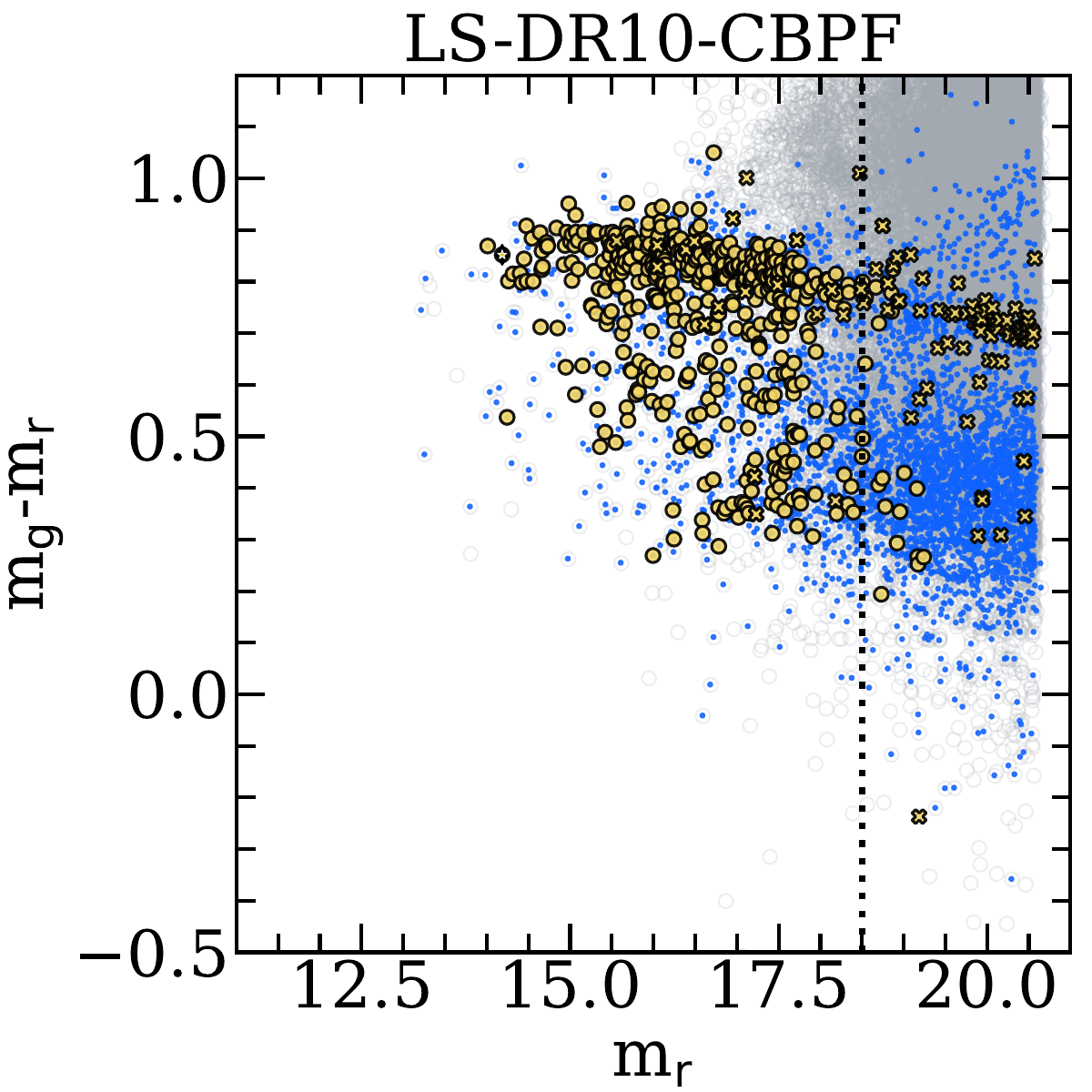}}
\caption{Colour-magnitude diagram for each set of photometric redshifts from
left to right: \mbox{T80S/S-PLUS}, \mbox{LS-DR9}, \mbox{LS-DR10} and \mbox
{LS-DR10-CBPF}. We include the corresponding parent sample (grey circles),
the A0085 galaxy cluster candidate members selected from our photometric
method (blue),the spectroscopically confirmed members (yellow circles), and
the confirmed spectroscopic cluster members and spectroscopic objects within
the cluster redshift range (yellow crosses) that were not selected by our
method. The  bright/faint boundary at $\rm m_{r}=18.5$ is indicated by a
vertical dotted-line. The different CMDs evidence the out-striking
differences of the selected cluster galaxy candidate members, specially at
fainter magnitudes ($\rm m_{r}\geqslant18.5$), where only T80S/S-PLUS- and
LS-DR10-CBPF-$\rm z_{phot}$ effectively populate the red faint regime
($\rm m_{g}-m_{r}>0.5$).}
\label{fig:ColMagDia}
\end{figure*}

\begin{figure*}
\subfloat{\includegraphics[width = 0.25\linewidth]{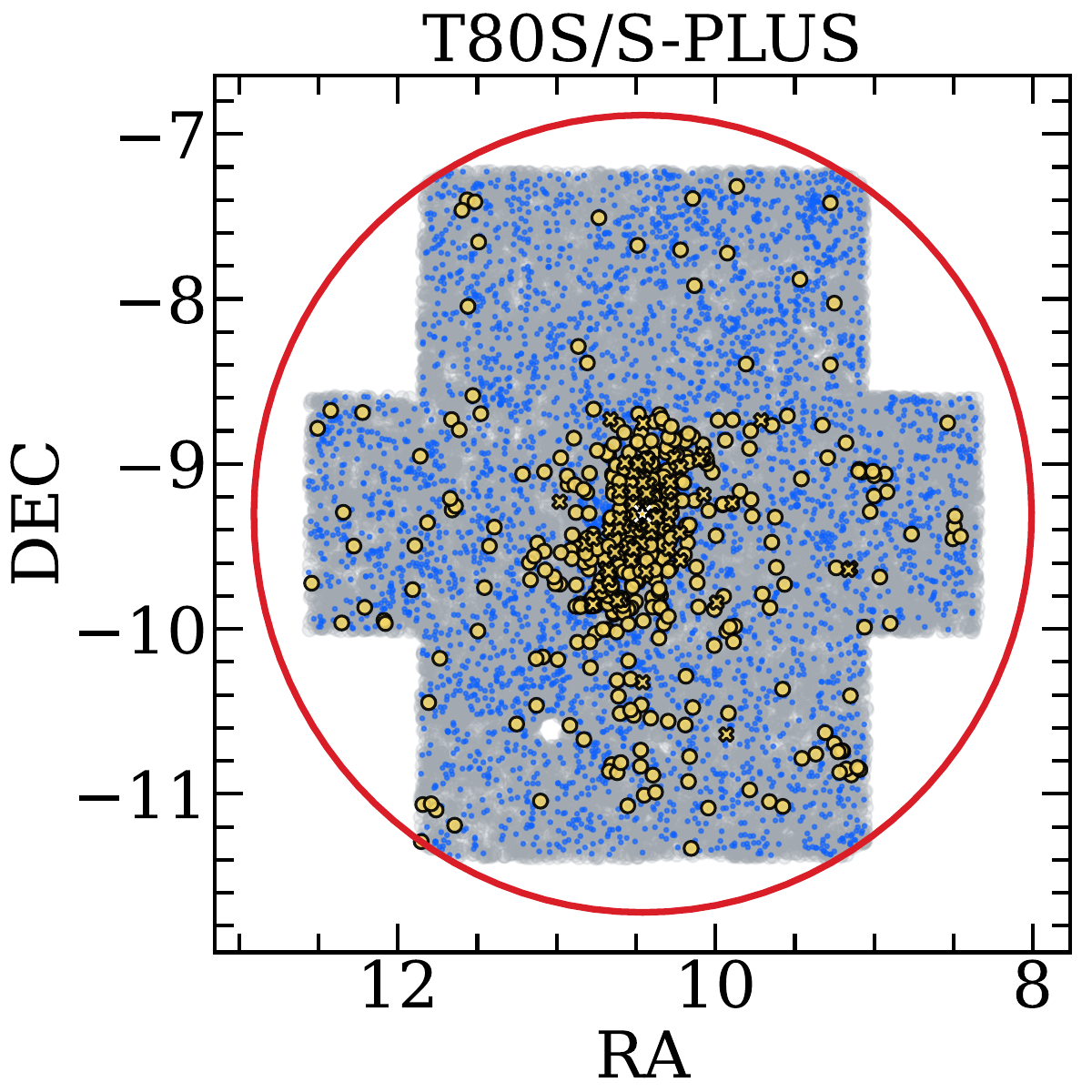}} 
\subfloat{\includegraphics[width = 0.25\linewidth]{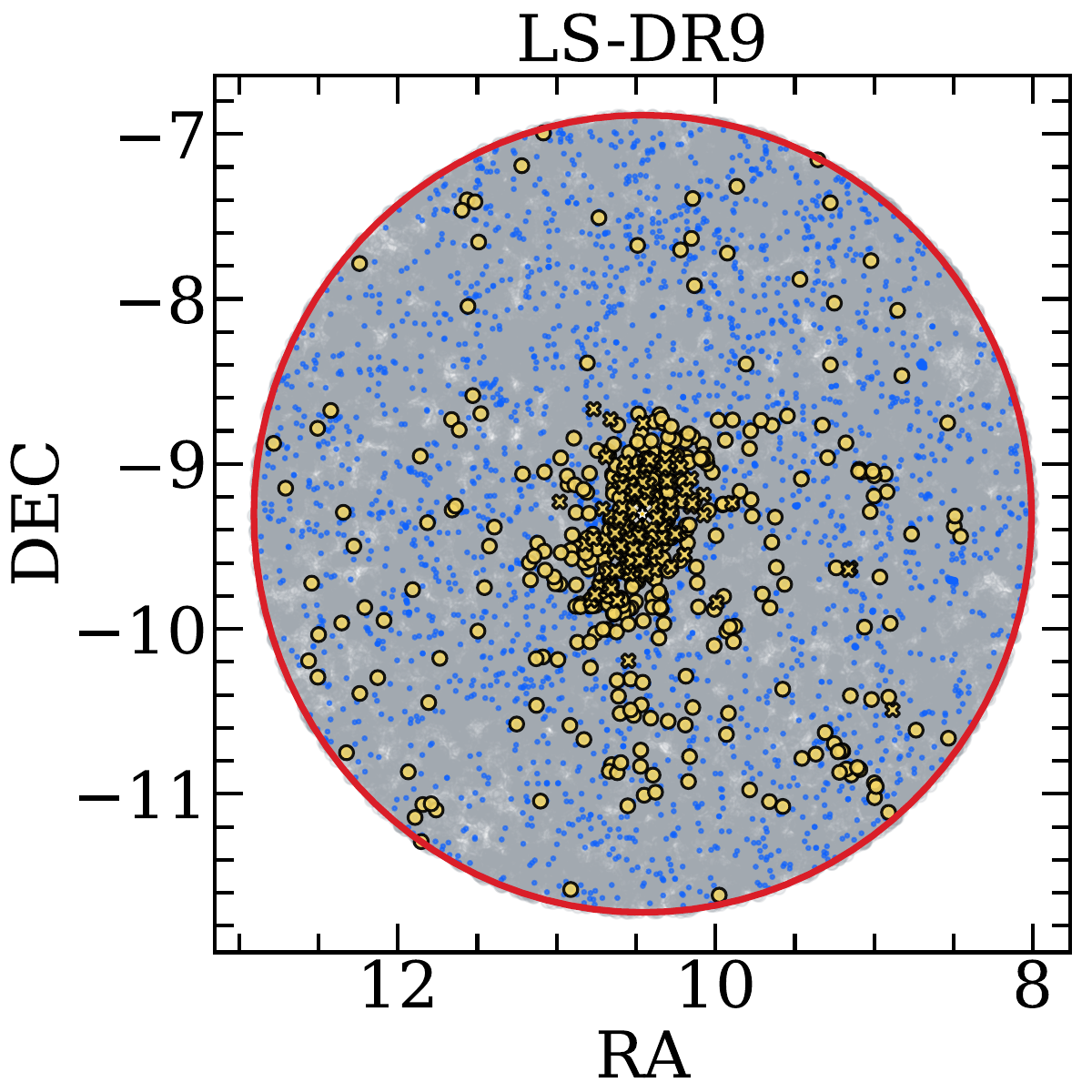}} 
\subfloat{\includegraphics[width = 0.25\linewidth]{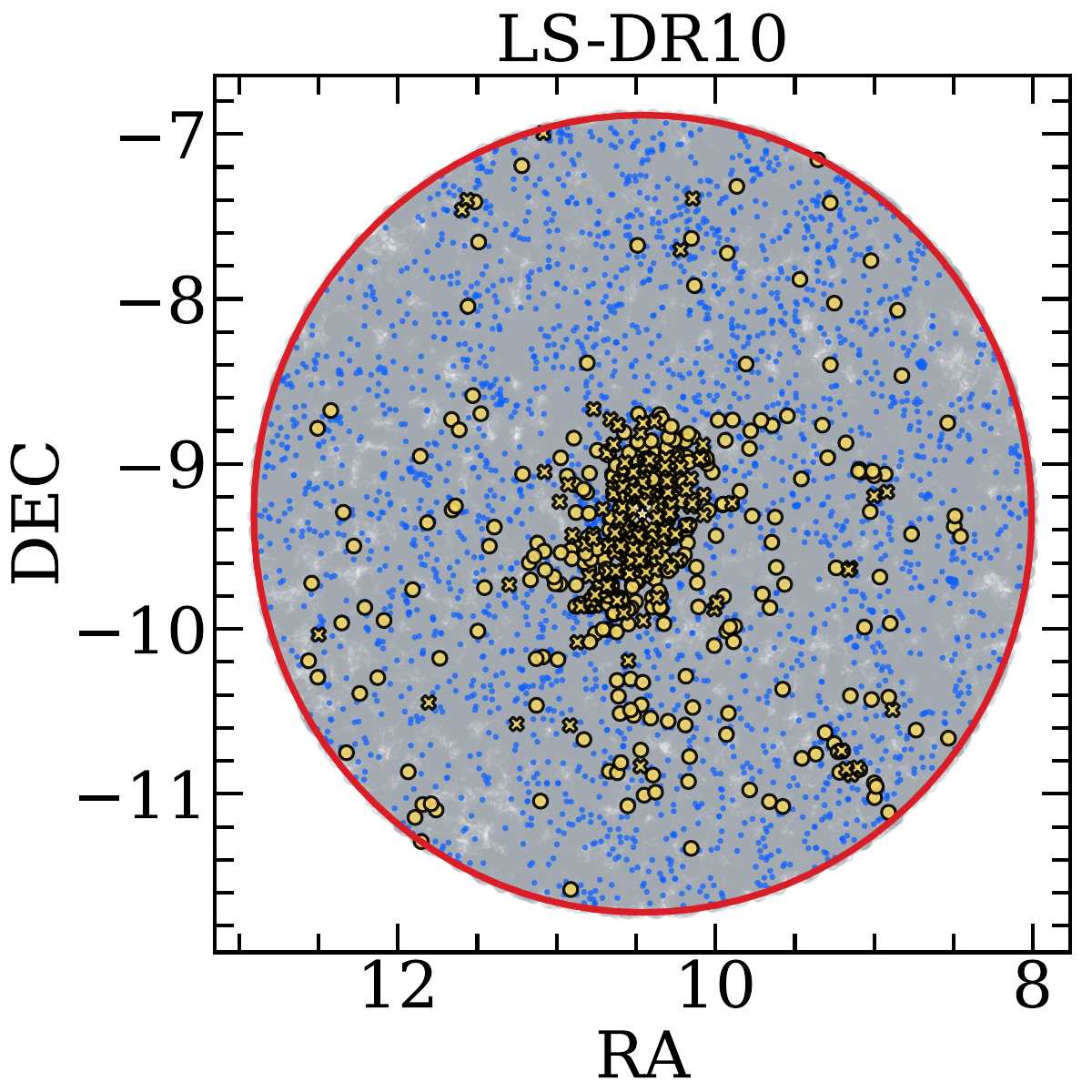}}
\subfloat{\includegraphics[width = 0.25\linewidth]{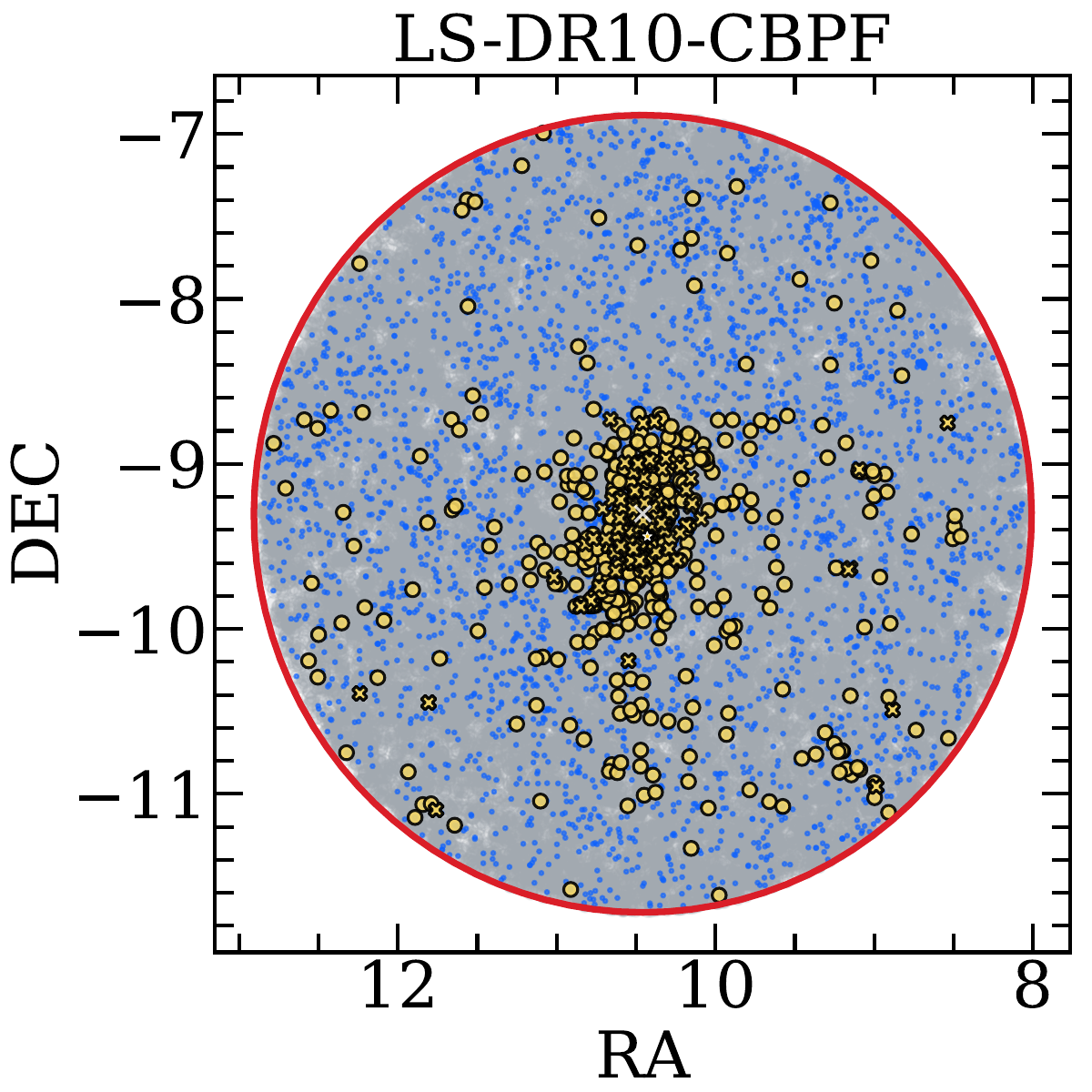}}
\caption{Spatial distribution of the A0085 galaxy cluster candidate members
using our target selection process for the four different parent catalogues
used, from left to right: \mbox{T80S/S-PLUS}, \mbox{LS-DR9}, \mbox
{LS-DR10} and 
\mbox{LS-DR10-CBPF}. We include the corresponding parent sample(grey circles),
the A0085 galaxy cluster candidate members selected from our photometric
method (blue),the spectroscopically confirmed members (yellow circles), and
the confirmed spectroscopic cluster members and spectroscopic objects within
the cluster redshift range (yellow crosses) that were not selected by our
method.
The cluster galaxy candidate members of the four different $\rm z_
{phot}$ are homogeneously distributed and any significant differences are
appreciated.}
\label{fig:SpatialDist}
\end{figure*}

Figures \ref{fig:zphotzspec} and \ref{fig:ColMagDia} show the corresponding
$\mathrm{z_{phot}}$-$\mathrm{z_{spec}}$ and the colour-magnitude
(CMD) diagrams respectively. All four datasets show a similar redshift
distribution, ranging values between $\mathrm{z_{spec}}$ = 0.02. and 0.07,
where most of the selected objects are located around the cluster redshift at
z=0.056. Spectroscopic objects out of this range correspond to selected
objects from the $\rm \sigma_{NMAD}$ curve-fit and strictly depend on the
$\rm z_{phot}$ adopted. However, when we consider the CMDs, we notice some
striking differences in the faint-end regime at magnitudes fainter than $\rm
m_{r}\geqslant18.5$. While \mbox{T80S/S-PLUS} selects objects that completely
cover the faint-end including objects redder than $\rm  m_{g}-m_
{r}\geqslant1.0$, the \mbox{LS-DR9}- and  \mbox{LS-DR10-}$\mathrm{z_
{phot}}$ show an underdensity of faint-red objects ($\rm m_{g}-m_
{r}=0.5-0.9$ and $\rm m_{r}\geqslant18.5$), regardless of the excellent
coverage of confirmed spectroscopic objects. This motivated our internal CBPF
$\mathrm{z_{phot}}$ determinations from the \mbox{LS-DR10} photometric
datasets. Figure \ref{fig:SpatialDist} shows that adopting the different
available $\mathrm{z_{phot}}$ result in an target selection homogenously
distributed within $\rm 5\times R_{200}$. We note that 
although Figures \ref{fig:zphotzspec} and \ref{fig:ColMagDia} shows a 
good coverage of spectroscopic objects in all the magntiude range and
specially in the faint-red-end ($\rm mag_{r}\geqslant18.5$) for Abell-85, 
this is not the case for all the sample of the 50+2  CHANES-low-z clusters.
Most of our clusters have a scarce spectroscopic coverage at fainter magnitudes, 
which directly impacts the training datasets to determine the photozs. 
The low spectroscopic coverage at fainter magnitudes is one of the 
main motivations  for proposing the CHANCES survey: to explore this regime 
and populate it with good-quality spectra.

\begin{figure*}
    \centering
    \includegraphics[width=0.9\linewidth]{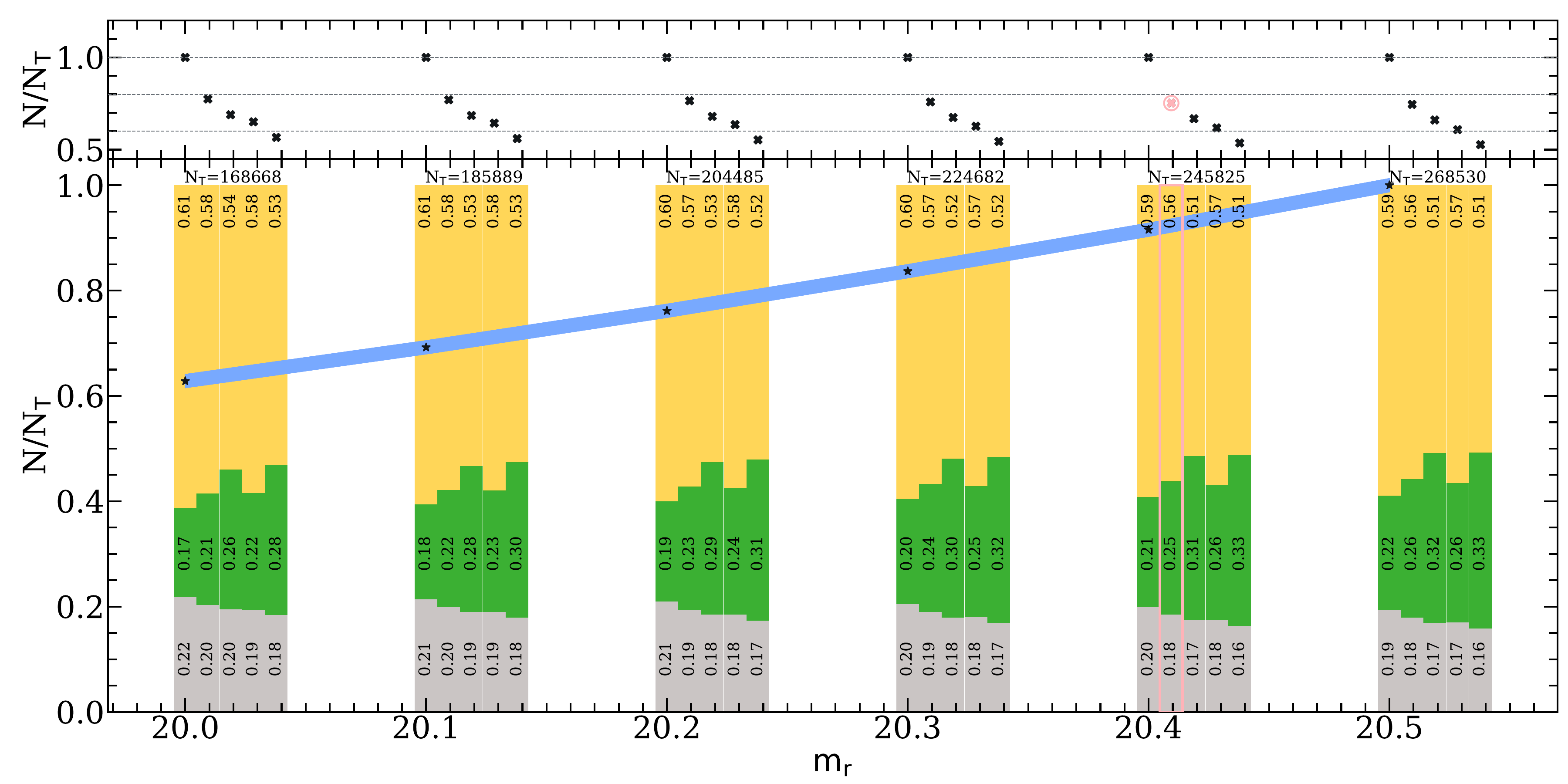}
    \caption{Total number of selected objects ($\rm N_{T}$) for the faint-end
     ($\rm m_{r}\geq18.5$) regime by adopting different $\rm N \times \sigma_
     {NMAD}$ combinations for \mbox{T80S/S-PLUS} and 
    \mbox{LS-DR10-CBPF} galaxy cluster candidate members selection and
     imposing different $\rm m_{r}$-limit values (20, 20.1, 20.2, 20.3, 20.4
     and 20.5). The lower panel show a set of 5 different $\rm
     N \times \sigma_{NMAD}$ combinations within the $\rm m_
     {r}$-range 20-20.5. The leftmost column of each set corresponds to
     target selections adopting $\rm N \times \sigma_{NMAD}=4$  for
     both \mbox{T80S/S-PLUS} and \mbox{LS-DR10-CBPF}, followed by the $\rm
     N \times \sigma_{NMAD}$ combinations (from left to right): 3.5/3.0,
     3.5/2.5, 3.0/2.5; and 3.0/2.0 for \mbox{T80S/S-PLUS} and \mbox
     {LS-DR10-CBPF} respectively. As a reference $\rm N_{T}$ 
     number ($\rm N_{T}$) of selected objects is included on top of each $\rm
     m_{r}$-limited set considering $\rm N \times \sigma_{NMAD}=4$ for
     both \mbox{T80S/S-PLUS} and CBPF surveys. The blue line highlights how
     $\rm N_{T}$ varies as a function of $\rm m_{r}$, if $\rm m_{r}$=20.5
     ($\rm N/N_{T}=1$) is considered $\rm N_{T}$ is reduced to 60\% at $\rm
     m_{r}$-limit = 20. Each bar is decomposed by their corresponding parent
     survey catalogue: \mbox{T80S/S-PLUS} (green), CBPF (yellow) and common
     objects selected by both surveys(grey). The upper panel shows how the
     different $\rm N \times \sigma_{NMAD}$ adopted combinations for
    \mbox{T80S/S-PLUS} and \mbox{LS-DR10-CBPF} affect $\rm N_{T}$ at each $\rm
     m_{r}$-limit bin, considering $\rm N \times \sigma_{NMAD}=4$ as a
     reference ($\rm N/N_{T}=1$). At all $\rm m_{r}$ bins the number of
     objects is reduced for the different $\rm N \times \sigma_
     {NMAD}$ combinations, dropping down to a $\lesssim$60\% of the selected
     objects at $\rm N \times \sigma_{NMAD}$-combination: 3.0, 2.0 for \mbox
     {T80S/S-PLUS} and \mbox{LS-DR10-CBPF} respectively. The final $\rm m_
     {r}$-limit at =20.4 and $\rm N \times \sigma_
     {NMAD}$-combination 3.5/\mbox{T80S/S-PLUS} and 3.0/\mbox
     {LS-DR10-CBPF} adopted to compile the Low-z faint (S1505) and Low-z
     faint suplementary (S1506) catalogues is highlighted in red.    
      }
    \label{fig:1505SNMAD-Test}
\end{figure*}

To compile the final versions of the CHANCES low-redshift catalogues, we
tested different values of $\rm N$ for \mbox{T80S/S-PLUS} and LS $\rm z_
{phot}$. In particular we tested the effects of adopting different $\rm m_
{r}$-limits in the range 20.1-20.5 and adopting $\rm N \times \sigma_
{NMAD}$ combinations: 3.5/3.0, 3.5/2.5, 3.0/2.5, and 3.0/2.0 for \mbox
{T80S/S-PLUS} and \mbox{LS-DR10-CBPF} (see Figure \ref
{fig:1505SNMAD-Test}) We find that the total number of selected objects
($\rm N_{T}$) declines as a function of $\rm m_{r}$, if $\rm m_{r}$=20.5 is
considered $\rm N_{T}$ is reduced to 60\% at $\rm m_{r}$-limit = 20.
Moreover, independent of $\rm m_{r}$-limit adopted, $\rm N_{T}$ declines from
assuming $\rm N\times \sigma_{NMAD}=4$ for both $\rm{z_{phot}}$ 
\mbox{-T80S/S-PLUS} and \mbox{-LS-DR10-CBPF} to adopting different $\rm
N\times \sigma_{NMAD}$ combinations, by dropping down to a $\lesssim$60\% of
the selected objects at $\rm N\times \sigma_{NMAD}$-combination: 3.0, 2.0
for \mbox{T80S/S-PLUS} and \mbox{LS-DR10-CBPF} respectively.

\begin{figure*}
\subfloat[LS-DR10-CBPF]{\includegraphics[width = 0.333\linewidth]{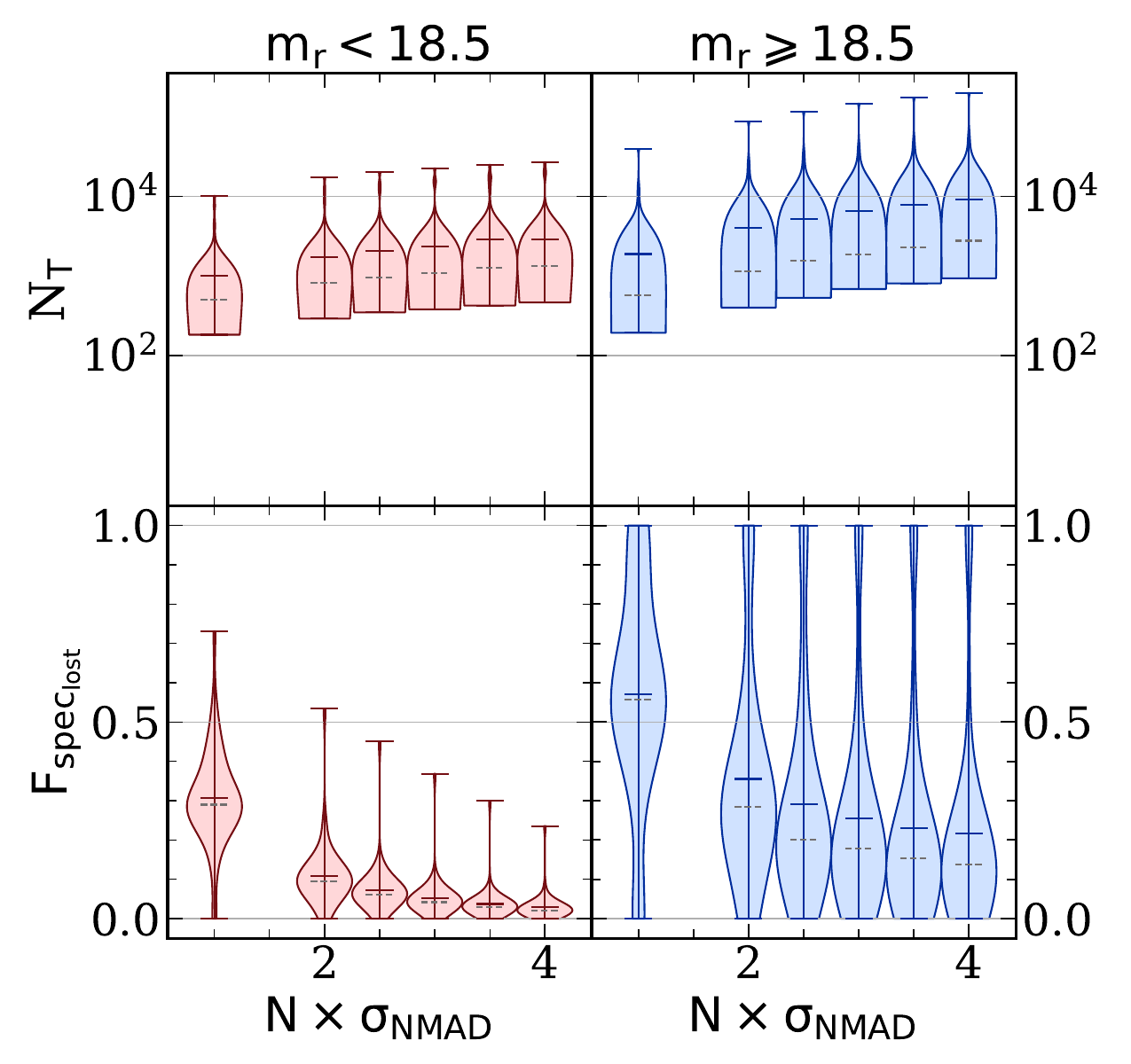}} 
\subfloat[LS-DR10-CBPF-S-PLUS N=10]{\includegraphics[width = 0.333\linewidth]
{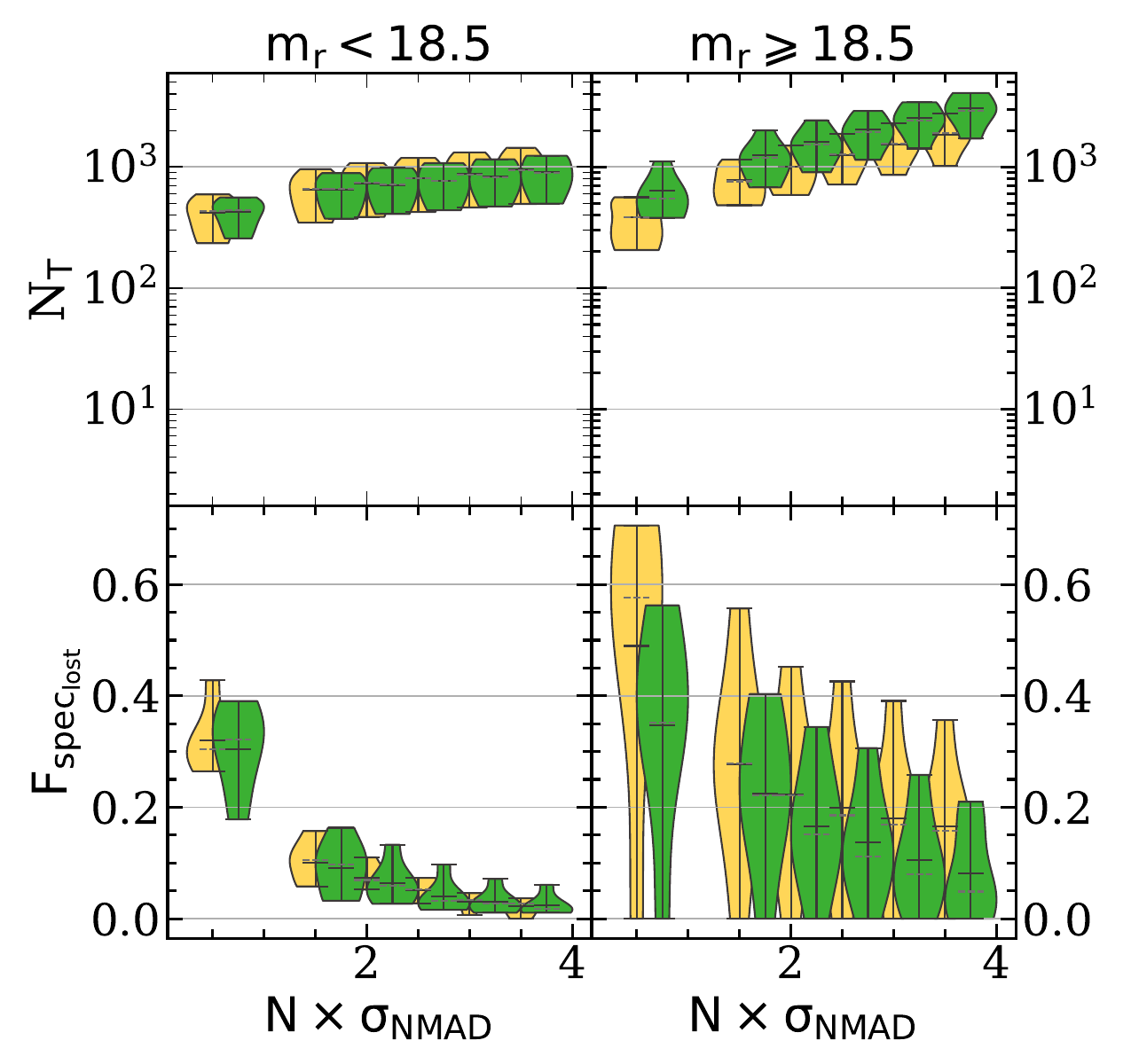}}
\subfloat[LS-DR10-CBPF-S-PLUS N=16]{\includegraphics[width = 0.333\linewidth]
{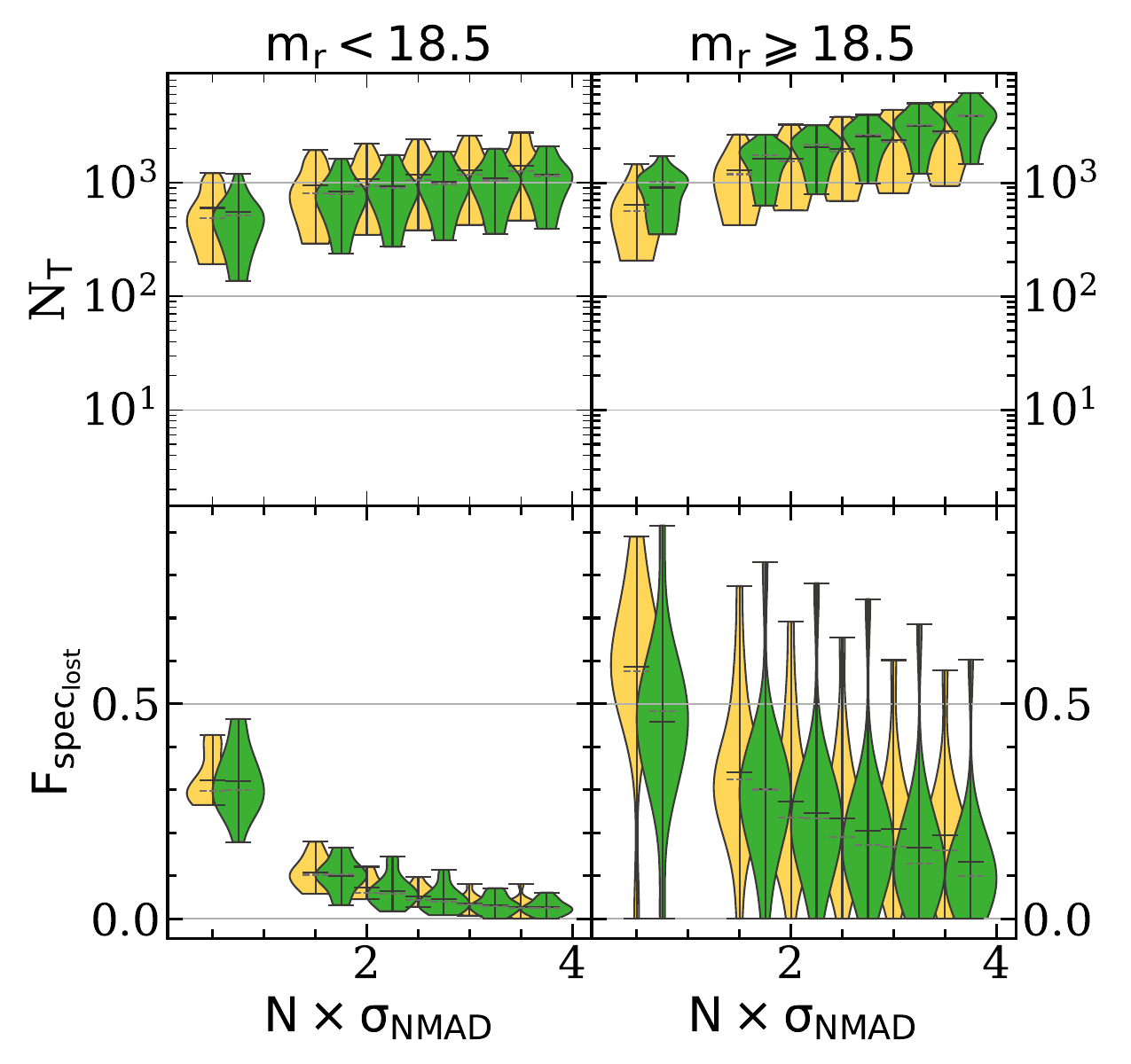}}
\caption{Number of galaxy cluster candidate members selected photometrically
 (top) and the fraction of lost spectroscopically confirmed members
 normalized to $\rm N \times \sigma_{NMAD}=1$ (bottom) as a function of $\rm
 N\times\sigma_\mathrm{NMAD}$ considering the CHANCES low-redshift
 magnitude-limited regimes: bright-end ($\rm m_{r}<18.5$, red) and faint-end
 ($\rm m_{r}\geqslant18.5$, blue). Assuming $\rm N\times\sigma_\mathrm
 {NMAD}=4$ our target selection process selects the largest number of  galaxy
 cluster candidate members with the lowest fraction loss of spectroscopic
 members, while assuming $\rm N\times\sigma_{NMAD}=1$ we select the lowest
 number of galaxy cluster candidate members, with the highest fraction of
 lost spectroscopically confirmed members. The first panel show this by
 considering only our internal \mbox{LS-DR10-CBPF}-$\rm z_
 {phot}$ estimations, the middle and right panels show a direct comparison
 between Ls-DR10-CBPF- and S-PLUS-$\rm z_{phot}$ considering 10 clusters with
 good \mbox{T80S/S-PLUS} coverage (middle panel) and an extended sample
 considering 16 clusters with \mbox{T80S/S-PLUS} coverage >80\%
 (rightmost panel), see Section \ref{sec:pz_S-PLUS} for details. Assuming
 $\rm N \times \sigma_{NMAD}=$ 3.5 and 3 for S-PLUS and LD-DR10-CBPF-$\rm z_
 {phot}$ respectively successfully reduces the total number of the galaxy
 cluster candidate members selected with a spectroscopic members loss of
 $\sim2-3\%$.}
\label{fig:FractionViolin}
\end{figure*}

Given that  we will obtain higher S/N for galaxies with $\rm m_{r}<18.5$, and
that in that regime the photo-zs for selecting targets perform best, we split
our low-z sub-survey at this magnitude limit. Additionally, we further divide
the faint-end(galaxies with $\rm m_{r}\geqslant18.5$)  in two sub-surveys. We
end up with three low-z sub-surveys: S1501, S1505, and S1506 each with
different membership selection strategy \footnote{S1502 and S1503/04
correspond to CHANCES-evolution and CHANCES-CGM sub-surveys respectively.}. 

In particular, we have adopted the following restrictions:

\begin{itemize}
\item Low-z bright (S1501): bright ($\rm m_{r} < 18.5$) photometric
members according to \mbox{T80S/S-PLUS} photometric redshifts, considering a
threshold of $\rm N \times \sigma_{NMAD}=$3.5, and/or photometric members
according to public photometric redshifts from \mbox{LS-DR9} and/or \mbox
{LS-DR10} ($\rm N \times \sigma_{NMAD}=3$), and in addition, galaxies that
are considered members according to our custom made photometric redshifts
CBPF ($\sigma_{\rm NMAD}=$3) and $\rm z_{spec}<0.1$. 

\item Low-z faint (S1505): Faint ($18.5\leqslant\rm m_
{r}<20.4$) photometric members considering photometric redshifts from \mbox
{T80S/S-PLUS} when available ($\rm N \times \sigma_{NMAD}=$3.5) and/or our
custom-made photometric redshifts CBPF ($\rm N \times \sigma_{NMAD}=$3)
and $\rm z_{spec}<0.1$. 

\item Low-z faint suplementary (S1506): To ensure a high completeness
of faint red galaxies, we also create a subset of galaxies equal to the
Low-z faint sample with added red-sequence galaxies not selected by our
photometric redshifts (see Section \ref{sec:redseq}) 
from Legacy (CBPF) meaning: S-PLUS + Legacy-CBPF (+ Red
Sequence)
\end{itemize}

The CHANCES-low-redshift sub-surveys presented in this work have been compiled
using the 
\texttt{EECHOz} \footnote{Efficiently Extracting Cluster candidate members
with Homogeneity using Optical colour and photometric-z available at 
\url
{https://github.com/4MOST-CHANCES/CHANCES-EECHOz-Low-redshift-TargetSelection}}
code \citep{EECHOz}. Figure \ref{fig:1505SNMAD-Test} shows a detailed comparison of the
faint-end regime ($\rm m_{r}\geq18.5$) by combining different $\rm
N \times \sigma_{NMAD}$ values for \mbox{T80S/S-PLUS-} and CBPF-$\rm z_
{phot}$ and different $\rm m_{r}$ upper-limits within the $\rm m_
{r}$-range 20-20.5. After comparing the total number of objects selected for
different values of  $\rm N$, we decided to adopt  $\rm N=3.5$ for 
\mbox{T80S/S-PLUS} and $\rm N=3$ for all LS photometric redshifts. This
combination of values provides the best compromise between the
(minimum) number of objects selected and a high spectroscopic redshift
completeness. Figure \ref{fig:FractionViolin} shows the total number of
selected objects (upper panels) and the fraction of the selected
spectroscopic confirmed members (lower panels) by adopting different $\rm
N \times \sigma_{NMAD}$ values. Larger $\rm N \times \sigma_{NMAD}$ values
results in a larger number of photometric selected objects and a smaller
fraction of lost spectroscopic confirmed members that are not being
selected, while smaller $\rm N \times \sigma_{NMAD}$ values results in a
smaller number of photometric selected objects with a larger fraction of
lost spectroscopic confirmed members that are not being selected. The first
panel shows how this affects our bright($\rm m_{r}<18.5$) in blue) and
faint ($\rm m_{r}\geq18.5$) in red) catalogues, the middle and right panels
show a direct comparison between \mbox{T80S/S-PLUS-} and CBPF-$\rm z_
{phot}$ estimations by considering all the clusters with any \mbox
{T80S/S-PLUS} coverage (middle) and by considering those clusters(N=16) with
good \mbox{T80S/S-PLUS} coverage ($>80\%$). For all the $\rm
N \times \sigma_{NMAD}$ adopted values,
\mbox{T80S/S-PLUS} selects more objects than CBPF. The former allow us to
define the final $\rm N \times \sigma_{NMAD}$ adopted values to fit the
total number of selected objects with the 4MOST fiber/hours allocated to
CHANCES.\\

\begin{figure}
    \begin{center}
\includegraphics[width=1\linewidth]{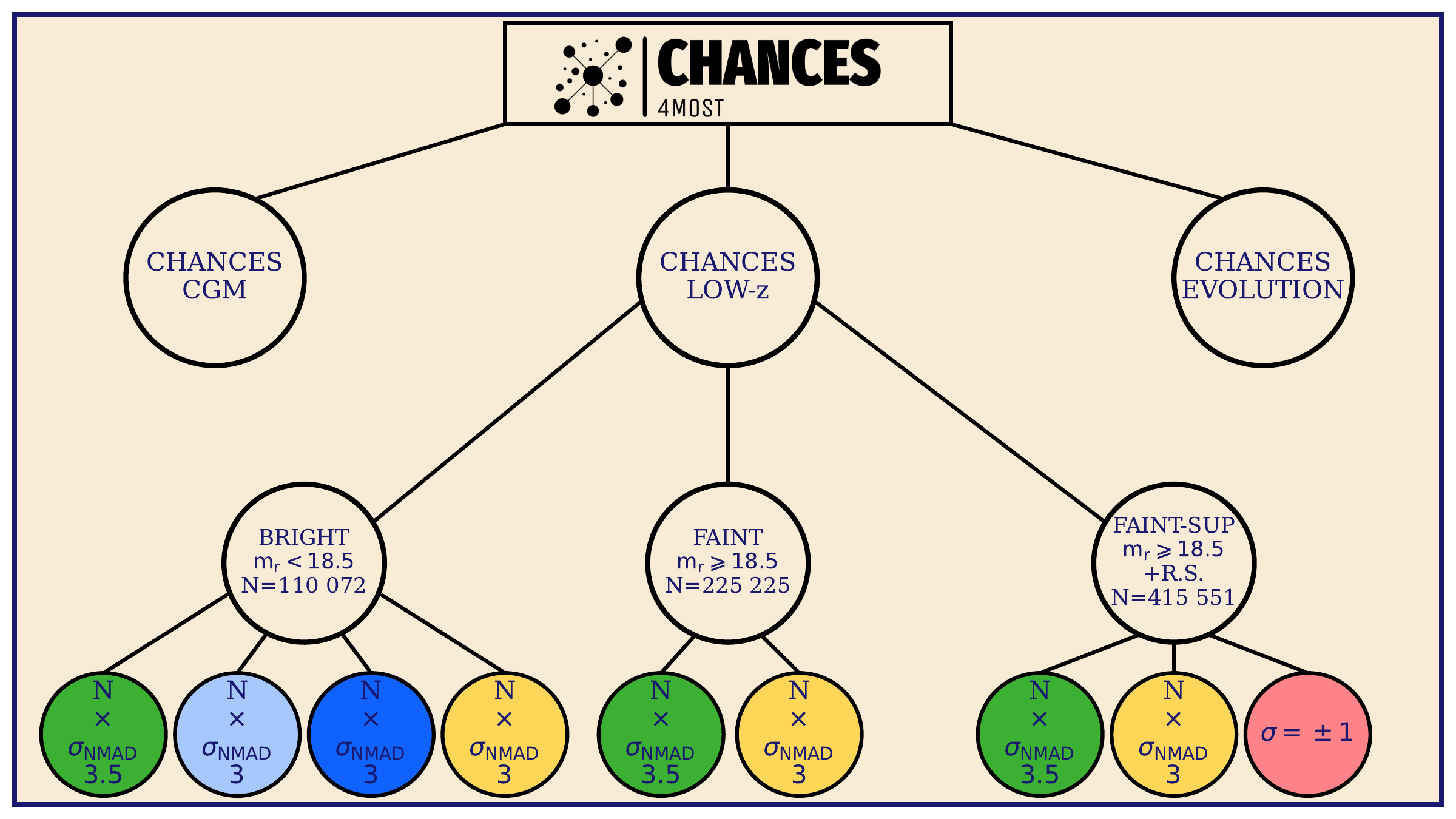}
    \end{center}
    \caption{CHANCES-low-z photometric selection diagram showing the different
     survey datasets used to compile the Low-z bright (S1501), Low-z faint
     (S1505) and Low-z faint supplementary (S1506). The \mbox
     {T80S/S-PLUS}, \mbox{LS-DR9}, LS-DR10  and LS-DR10-CBPF datasets are
     shown in green, cyan, blue and yellow, the red circle represents the
     $\pm1\sigma$ red-sequence selected objects.}
    \label{fig:CDiagram}
\end{figure}

In all cases, galaxies with known spectroscopic redshifts (Section~\ref
{sec:spcz}) outside the range $\pm3\sigma$ of the CHANCES-low-z subsurvey
redshift were excluded. Figure \ref{fig:CDiagram} summarizes the different
restrictions adopted to build our three CHANCES-low-z catalogues, as well as
the final number of target members selected for each of the three low-z
catalogues. We note that assuming a magnitude limit of $\rm m_{r}$=20.4, the
4MOST $f.o.m.$ (0.5) and the corresponding completenesses for  S1501
(84.1\%), S1505 (65\%) and S1506 (10\%), we expect to observe ~320,000
objects within the CHANCES-low-redshift sub-survey. We describe each
sub-survey selection strategy in what follows.

\subsection{Low-z bright (S1501).}\label{sec:bright}

For galaxies with $\rm m_{r}<18.5$ all the available photometric redshifts
perform well. Thus, to maximize completeness and to preserve homogeneity
(particularly with the selection on the faint-end, see Section \ref
{sec:faint}), we consider galaxies to be galaxy cluster candidate members if
any of their photometric redshifts coincide with the cluster redshift within
errors. The errors considered are the normalized median average deviation
($\sigma_{\mathrm{NMAD}}$) which increases at fainter magnitudes. 

\begin{figure*}
    \centering
    \includegraphics[width=0.9\linewidth]{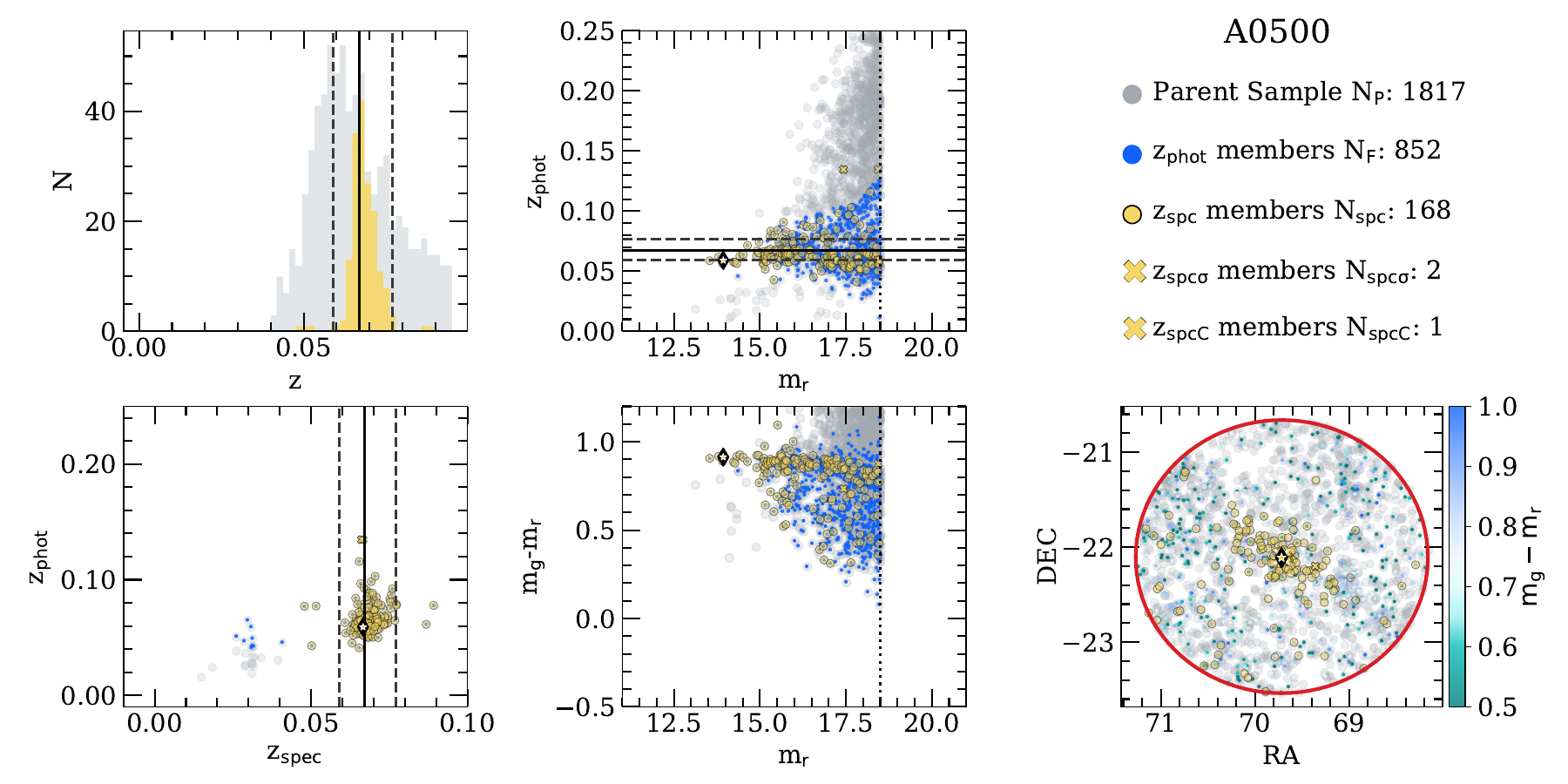}
    \caption{Photometric target selection of Abell 500 for bright objects
     ($\rm m_{r}<18.5$) using \mbox{T80S/S-PLUS} $\rm{z_{phot}}$. grey
     symbols correspond to the \mbox{T80S/S-PLUS} photometric parent sample,
     yellow symbols correspond to the spectroscopic objects, while blue
     symbols correspond to the photometric selected Abell 500 galaxy cluster
     candidate members. The upper panels show from left to right, the
     spectroscopic redshift distribution($\rm{z_{spec}}$), the $\rm m_{r}-z_
     {phot}$ diagram and relevant cluster information including the parent
     sample, total number of galaxy cluster selected objects, the
     spectroscopically confirmed cluster members (yellow circles), the
     confirmed spectroscopic cluster members (open circles), and the
     spectroscopically confirmed cluster members and spectroscopic objects
     within the cluster redshift range (yellow crosses) that were not
     selected by our method. The lower panels show from left to right show
     the $\rm{z_{spec}}-\rm{z_{phot}}$, colour-magnitude and spatial
     distribution diagrams.}
    \label{fig:ExBright}
\end{figure*}

As an example, Figure \ref{fig:ExBright} shows the photometric selection
procedure of Abell 500 where we select 1020 galaxies with \mbox
{T80S/S-PLUS}, 934 with LS DR9, 874 with LS DR10 and 1032 with CBPF. The
corresponding Figures from \mbox{LS-DR9}, LS-DR10 and LS-DR10-CBPF for Abell
500 can be found in the Appendix \ref{sec:appendix}.

To assess the completeness of each selection, we consider the fraction of
known spectroscopic members in a given cluster (from the literature) that are
recovered through the photometric target selection by adopting each of the
different $\rm z_{phot}$. On the other hand we evaluate the purity of our
selection by considering the fraction of photometric selected objects that
are spectroscopically confirmed as members. Table \ref{tbl:PC} shows the
average completeness and purity of 6 clusters completely covered by 
\mbox{T80S/S-PLUS} and Legacy surveys with a good spectroscopic coverage down
to the faint regime ($\rm m_{r}=20.4$). We also include the same values by
extending our sample to those clusters (N=16) with a good \mbox
{T80S/S-PLUS} coverage larger than 80\% (see Section \ref
{sec:pz_S-PLUS}). When we combine all photometric redshift selections, their
combined completenesses and purities are 0.93, 0.89 and 0.80, 0.83 for S1501
and S1505 respectively.

Finally, Figure \ref{fig:S1501_Decomp} shows the decomposition of the 34 clusters with
any \mbox{T80S/S-PLUS} coverage from S1501 selection, clusters completely
covered ($\rm 5\times R_{200}$) by \mbox{T80S/S-PLUS} are delimited by a
blue-solid line, while those clusters with good \mbox{T80S/S-PLUS} coverage
($\geqslant80\%$) are indicated by a blue-dashed line. The green region
corresponds to the fraction of objects selected by adopting \mbox
{T80S/S-PLUS-} $\rm z_{phot}$, the light-blue region corresponds to 
objects selected by adopting \mbox{LSDR9} $\rm z_{phot}$,
the light-blue region corresponds to  objects selected by adopting 
\mbox{LSDR10} $\rm z_{phot}$,
the yellow region corresponds to the fraction
of objects selected by adopting CBPF- $\rm z_{phot}$ and the grey region
corresponds to the fraction of selected objects in common. 
This decomposition shows that the highest fraction ($\geqslant48\%$) of selected objects 
corresponds to those who are selected by two or more surveys, 
followed by those objects selected independently by adopting 
\mbox{T80S/S-PLUS} ($\leqslant43\%$) and \mbox{LS-DR10-CBPF} ($\leqslant20\%$).  
Objects exclusively selected by adopting \mbox{LS-DR9} and \mbox{LS-DR10} 
$\rm z_{phot}$ show the lowest contribution ($\leqslant10\%$) to our S1501 sub-survey.

\begin{table*}
    \caption{Completeness and purity values for low- and bright-z sub-surveys.}
    \centering
    \begin{tabular}{c|cc|cc||cccccccc}
        \multirow{2}{*}{ID} & \multicolumn{2}{c|}{C$_{\rm A}$} & \multicolumn{2}{c||}{C$_{6}$} & \multicolumn{2}{c}{T80S/S-PLUS} &\multicolumn{2}{c}{LS-DR9} & \multicolumn{2}{c}{LS-DR10}  & \multicolumn{2}{c}{LS-DR10-CBPF} \\
        \cline{2-13}
        & C & P & C & P & C & P & C & P & C & P & C & P\\
        \hline\hline
        S1501 & 0.853 & 0.801 & 0.868 & 0.857 & 0.933 & 0.871 & 0.668 & 0.902 & 0.813 & 0.859 & 0.760 & 0.855 \\
        S1505 & 0.760 & 0.813 & 0.859 & 0.834 & 0.833 & 0.835 & -     & -     & -     & -     & 0.532 & 0.781 \\
        \hline
        \\
        \multirow{2}{*}{ID} & \multicolumn{2}{c|}{C$_{\rm A}$} & \multicolumn{2}{c||}{C$_{16}$} & \multicolumn{2}{c}{T80S/S-PLUS} &\multicolumn{2}{c}{LS-DR9} & \multicolumn{2}{c}{LS-DR10}  & \multicolumn{2}{c}{LS-DR10-CBPF} \\
        \cline{2-13}
        & C & P & C & P & C & P & C & P & C & P & C & P\\
        \hline
        S1501 & 0.853 & 0.801 & 0.934 & 0.807 & 0.860 & 0.849 & 0.751 & 0.854 & 0.777 & 0.839 & 0.862 & 0.807 \\
        S1505 & 0.760 & 0.813 & 0.889 & 0.829 & 0.858 & 0.829 & -     & -     & -     & -     & 0.669 & 0.823 \\

    \end{tabular}
    \tablefoot{Completeness and purity values for low- and bright-z sub-surveys
     considering each $\rm z_{phot}$ selections individually and the
     corresponding combined values. The upper panel considers (C$_
     {6}$) considers clusters A0500, A0957, A1631, A2717, A3223, A3809, which
     have good $\rm z_{spec}$ coverage throughout all the $r$-band magnitude
     range m$_{r}$<20.4, while the lower cells correspond to the extended
     sample (C$_{16}$) including clusters with \mbox{T80S/S-PLUS} coverage
     larger than 80\%: A0500, A1631, A1644, A2399, A3266, A3376, A3490,
     A3667, A3716, A3809. C$_{\rm_{A}}$ considers all the CHANCES-low-z
     clusters regardless their \mbox{T80S/S-PLUS} or $\rm z_
     {spec}$ coverages.} 
    \label{tbl:PC}
\end{table*}

\subsection{Low-z faint (S1505).}\label{sec:faint}

For galaxies with $\rm 18.5\leqslant m_{r}<20.4$~mag the photometric redshifts
generally have more difficulties selecting potential galaxy cluster candidate
members. As discussed before in Section \ref{sec:CMTS} and shown in
Figure \ref{fig:ColMagDia}, the public \mbox{LS-DR10} $\rm z_
{phot}$ under-selects faint red galaxies in clusters. Instead, \mbox
{T80S/S-PLUS} photometric selection performs better in this regime, however
only 6 clusters of CHANCES-low-z cluster sample are completely covered out to
$\rm 5\times R_{200}$, as mentioned in Section \ref{sec:pz_S-PLUS}. 

For homogeneity, we developed our own $\rm z_{phot}$ estimations (CBPF) using
the available \mbox{LS-DR10} photometric information (see Section \ref
{sec:pz_CBPF}) which alleviate in great part the red-faint-end
under-selection issue(see Figure \ref{fig:ColMagDia}). As a result, for the
faint-end we only use a combination of our CBPF- and 
\mbox{T80S/S-PLUS}-$\rm z_{phot}$ when available to select targets, using the
same strategy and $N$ values as for the bright end for homogeneity.
Figure \ref{fig:S1505_Decomp} shows the decomposition of the 34 clusters with
any \mbox{T80S/S-PLUS} coverage from S1505 selection, clusters completely
covered ($\rm 5\times R_{200}$) by \mbox{T80S/S-PLUS} are delimited by a
blue-solid line, while those clusters with good \mbox{T80S/S-PLUS} coverage
($\geqslant80\%$) are indicated by a blue-dashed line. The green region
corresponds to the fraction of objects selected by adopting \mbox
{T80S/S-PLUS-} $\rm z_{phot}$, the yellow region corresponds to the fraction
of objects selected by adopting CBPF- $\rm z_{phot}$ and the grey region
corresponds to the fraction of selected objects in common. If we consider
only the 6 clusters with spatial and spectroscopical coverages, we find that
$\sim45\%$ and $\sim28\%$ where selected exclusively by \mbox
{T80S/S-PLUS} and CBPF respectively, while the remaining $\sim27\%$ of the
targets were selected by both of them. Finally, the individual completeness
reached by both $\rm z_{phot}$ is shown in second row's right columns of
Table \ref{tbl:PC}.

\subsection{Low-z faint suplementary (S1506): Adding the red
sequence.}\label{sec:redseq}

\begin{figure*}[htbp]
    \centering
    \begin{minipage}{\linewidth}
        \centering
        \includegraphics[width=\linewidth]{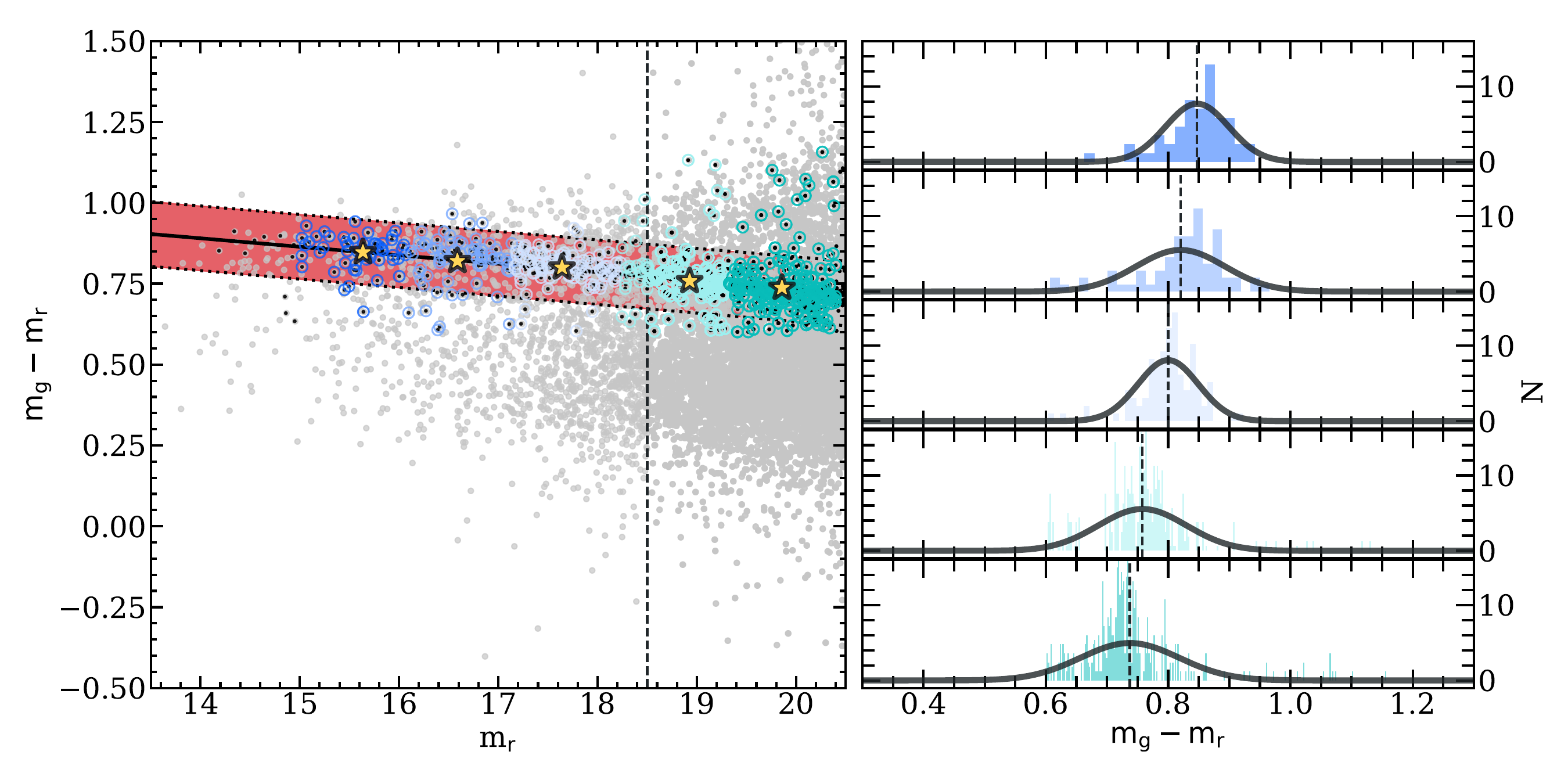}
    \end{minipage}
    \caption{Red sequence fit from the colour-magnitude diagram (CMD) on Abell
     85. The left panel shows the CMD, the grey points correspond to the
     galaxy cluster candidate members selected by our photometric method, the
     open coloured symbols show the colour-range bins used to fit the
     red-sequence, yellow stars correspond to the mean $\rm m_{g} - m_
     {r}$  colour in each bin obtained from a gaussian fit, while the best
     fit is shown with a black solid line including the  $\pm1\sigma$ region
     in red and delimeted by dotted lines. The dashed line indicates the
     bright/faint boundary at m$_{r}$=18.5. The right panel shows the $\rm m_
     {g} - m_{r}$ bins distributions, overlaid in black lines show the best
     guassian fit for each distribution while the dashed lines indicated the
     mean average $\rm m_{g} - m_{r}$ colour. }
    \label{fig:CMD_red-sequence_fit}
\end{figure*}

While CBPF-$\rm z_{phot}$ improved the target selection in the red-faint-end
of the CMD significantly compared to the \mbox{LS-DR10}-$\rm z_{phot}$, the
very faint-end ($\rm m_{r} \gtrsim 19.5$) of the red sequence still presents
lower target densities. To overcome this issue, we fitted the red sequence
using five equally sized $\rm m_{r}$ bins ($\Delta \rm m_{r}=0.9$), spanning
the full magnitude range of the CMD ($\rm 15.5 \leqslant m_
{r} \leqslant 20.5$). To perform the fit, we considered all previously
selected photometric members located within $\rm 0.75 \times R_{200}$ and
within the colour range $\rm 0.6 \leqslant m_{g} - m_{r} \leqslant 1.2$, in
order to ensure a proper fit based on the most likely galaxy cluster
candidate members. We then applied a supervised one-component Gaussian
Mixture Model (GMM, \citealt{scikit-learn}) to the colour distribution to
determine the mean $\rm m_{g}-m_{r}$ value in each magnitude bin(see right
panles of Figure \ref{fig:CMD_red-sequence_fit}). Using the five mean values
obtained from the GMM, we performed a linear regression to fit the red
sequence(see left panel of  Figure \ref{fig:CMD_red-sequence_fit}), to
include galaxies within $\pm 1\times\sigma$ of the linear fit. The left panel
of Figure \ref{fig:CMD_red-sequence_fit} shows the red-sequence fit on the
CMD for Abell 85, indicating in different colours the magnitude bins
considered to do the red-sequence fit. The right panels of Figure \ref
{fig:CMD_red-sequence_fit} show the colour distribution for each bin, along
with the corresponding Gaussian fit and its mean value. For clusters with a
spectroscopic sample that spans the full magnitude range of the CMD, from the
bright to the faint-end, and/or includes more than 120 spectroscopically
confirmed members, we applied a similar red sequence fitting procedure,
considering only the spectroscopic members. We fitted two gaussians in four
magnitude bins to separate the blue cloud from the red sequence, and
performed the fit using only the mean of the red sequence galaxies.\\

\begin{figure*}
    \centering
    \includegraphics[width=0.9\linewidth]{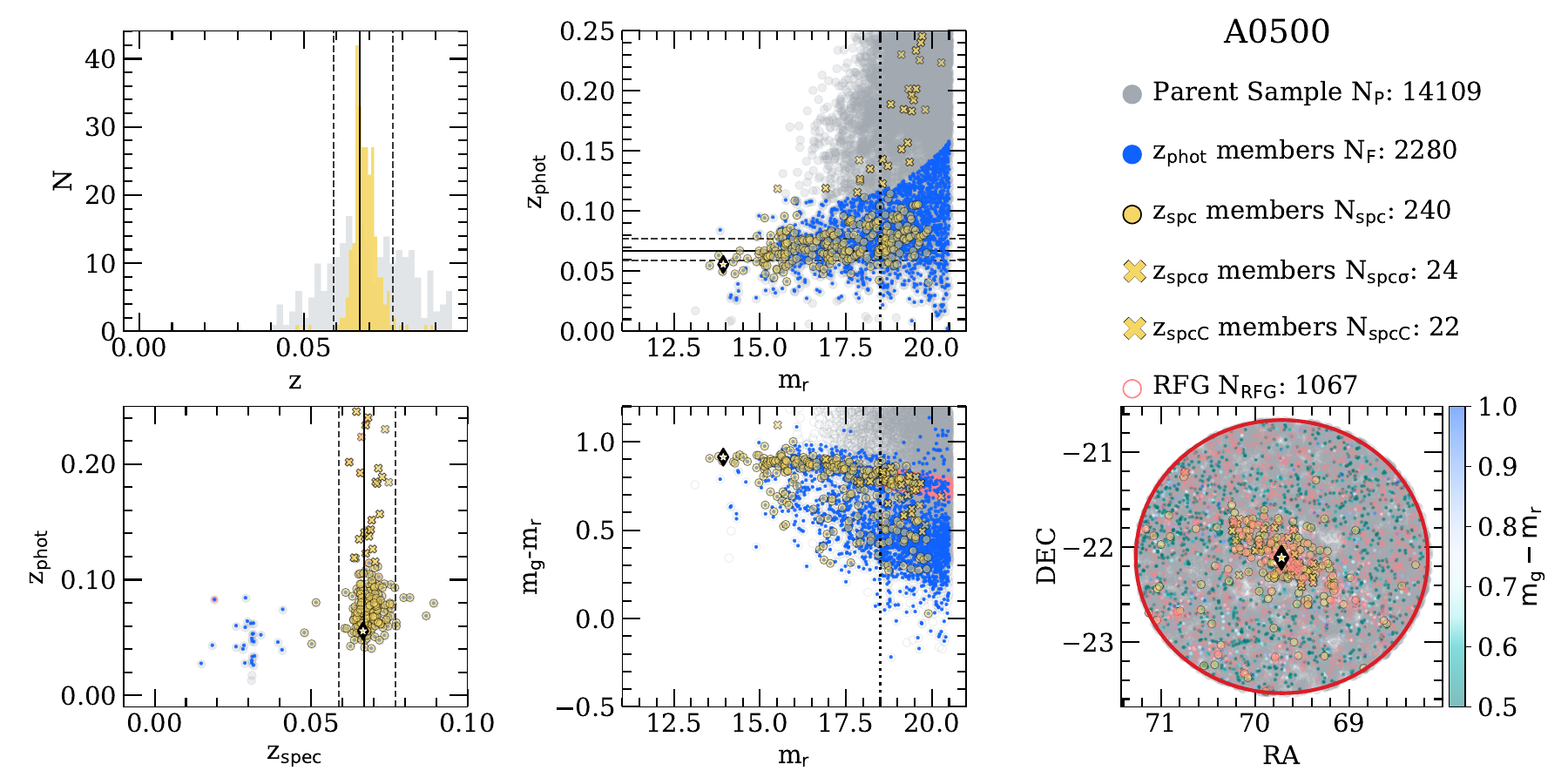}
    \caption{Photometric target selection of Abell 500, using CBPF $\rm{z_
     {phot}}$, including the red-sequence objects. The red-faint objects from
     S1506 completely fill the under-dense region in the faint-red-end and
     are homogenously distributed out to $\rm 5\times R_{200}$. Similar as
     Figure \ref{fig:ExBright} symbols represent the parent sample,
     spectroscopic objects and the photometric selected Abell 500 galaxy
     cluster candidate members in grey, yellow and blue respectively. We now
     include in magenta symbols the faint-red objects added through the
     fain-red-sequence fit (see Section \ref{sec:faint} for details). The
     upper panels show from left to right, the spectroscopic redshift
     distribution($\rm{z_{spec}}$), the $\rm{z_{phot}-m_{r}}$ diagram and
     relevant cluster information including the parent sample, total number
     of galaxy cluster selected objects, the confirmed spectroscopic cluster
     members (open circles), and the spectroscopically confirmed cluster
     members and spectroscopic objects within the cluster redshift range
     (yellow crosses) that were not selected by our method. The lower panels
     show from left to right the $\rm{z_{spec}}-\rm{z_
     {phot}}$, colour-magnitude and spatial distribution diagrams. The
     red-sequence-fit selected objects successfully fill-in the
     faint-end-under-dense region which are homogenously distributed within
     $\rm5\times R_{200}$.}
    \label{fig:ExFaintRef}
\end{figure*}

Figure \ref{fig:ExFaintRef} shows the photometric target selection of Abell
500 for the faint-end (S1505) and faint-end + red-sequence selected objects
(S1506). The objects selected from the red-sequence-fit described above are
shown in magenta symbols, where can be easily identified in the CMD panel
filling the under-dense region partially filled by our CBPF $\rm z_
{phot}$ estimations and that are homogeneously distributed within the $\rm
5\times R_{200}$.

The low-z faint sub-survey (S1505), includes faint targets with an efficient
selection, it posseses a high completeness criteria, and was compiled with a
target selection criterion focused on selecting the most likely galaxy
cluster candidate members. On the other hand, the low-z faint supplementary
sub-survey (S1506), samples all possible targets with the most complete
selection. It has a relaxed target selection criterion to make sure that all
putative member galaxies have a chance to be observed. In this way, S1506 can
be used to infer the statistics of rare galaxies based on the actual sampling
rate, as although the S1506-$\rm z_{spec}$ follow-up is low, it is not
weighting down any galaxy SED. Moreover, S1506 contains S1505 as a way to
test the $\rm z_{phot}$ errors by assessing whether a sampling targeting all
galaxies is achieved or not.

\subsection{Fornax target selection.}\label{sec:fornax}

The Fornax cluster proved the most challenging system in the low-z sample for
the target selection. It is the nearest cluster in the Southern hemisphere at
a distance of just 20\,Mpc ($z=0.005$). Even though it is the least massive
cluster in our sample, this proximity results in a virial radius of 1.8
degrees, 4--5 times larger than the median R$_{200}$ value in our low-z
sample. Similarly to Antlia, $\rm z_{phot}$ could not be used to select
Fornax galaxy members, as they are unreliable at this redshift
($z\sim0.005$). Thus, a CMD approach for selecting their galaxy members was
adopted making use of the \mbox{LS-DR10} photometric information 
\citep{LimaDias25}. 

Applying the same target selection criteria to the Fornax cluster for the
S1505 faint sub-survey resulted in an excessively large number of targets,
most of which we expect to be background sources, and so we implemented
additional cuts to bring the number of targets down to a reasonable level.
First we excluded objects with $\rm m_{g}-m_{r}$ colours more than $1\sigma$
above the cluster red sequence and also tightened the photometric redshift
criteria, retaining only galaxies with $z_{phot}{<}0.14$. However, the key
difference for the Fornax cluster sample is that we reduced the magnitude
limit from $\rm m_{r}{=}20.4$ to $\rm m_{r}{=}20.0$.

\section{Validation of the CHANCES-Low-z target catalogues for galaxy evolution studies: Tracing global and local galaxy environment.}\label{sec:Validation}

As an example of how these target catalogues can be used for different
scientific projects, we present here one of its applications to derive the
global and local galaxy environment of one of the CHANCES clusters. 

The photometric catalogues from the CHANCES target selection strategy
described in this work can be used to characterize the large-scale structure
(LSS)  around the clusters and superclusters (see \citealt{BaierSoto25}), 
as well as the local environment and substructures(\textit
{Piraino-Cerda et al. in prep.}), in order to determine the influence of the
environment at different scales on galaxy properties as well as galaxy
clusters properties.

To demonstrate that the cluster memberships based in photometric members are
good enough to trace large and smaller-scale structure, we have performed a
characterization of the cluster A3376, identifying cosmic filaments and
substructures within a region of $\rm 5\times R_{200}$ around the cluster
centre. For the identification of cosmic filaments we used the Discrete
Persistence Structures Extractor \citep[DisPerSE\footnote{\url
{https://www2.iap.fr/users/sousbie/disperse.html}};][]
{sousbie2011persistent}. We implemented DisPerSE  in two dimensions,
considering as input the photometric members coordinates. We assumed a
persistence threshold of $3\sigma$. For the computation of the local galaxy
density, we used the classification technique k-nearest neighbours (KNN). For
the computation of the local galaxy density $\Sigma_{k}$, we used the
K-Nearest Neighbor (KNN) method. \citet{Muldrew2012} showed that this method
is more efficient to evaluate local environments, using an appropriate small
value such as $k=5$. In order to correct for the lack of completeness and
Malmquist bias, we implemented a selection function $\psi(z)$, which depends
on the redshift distribution \citep{Lopes25}. We
ran KNN as a function of the projected positions of the galaxies and their
photo-z's, and spec-z's for those available. For the identification of
substructures we used the Hierarchical Density-Based Spatial Clustering of
Applications with Noise (hereafter HDBSCAN) clustering algorithm technique,
which is based on densities as HDBSCAN \citep[]{Campello2015}. We implemented
HDBSCAN in two dimensions using projected positions assuming a minimum
substructures members of 5. 

\begin{figure*}
    \centering
    \subfloat[S-PLUS]{\includegraphics[width = .5\linewidth]{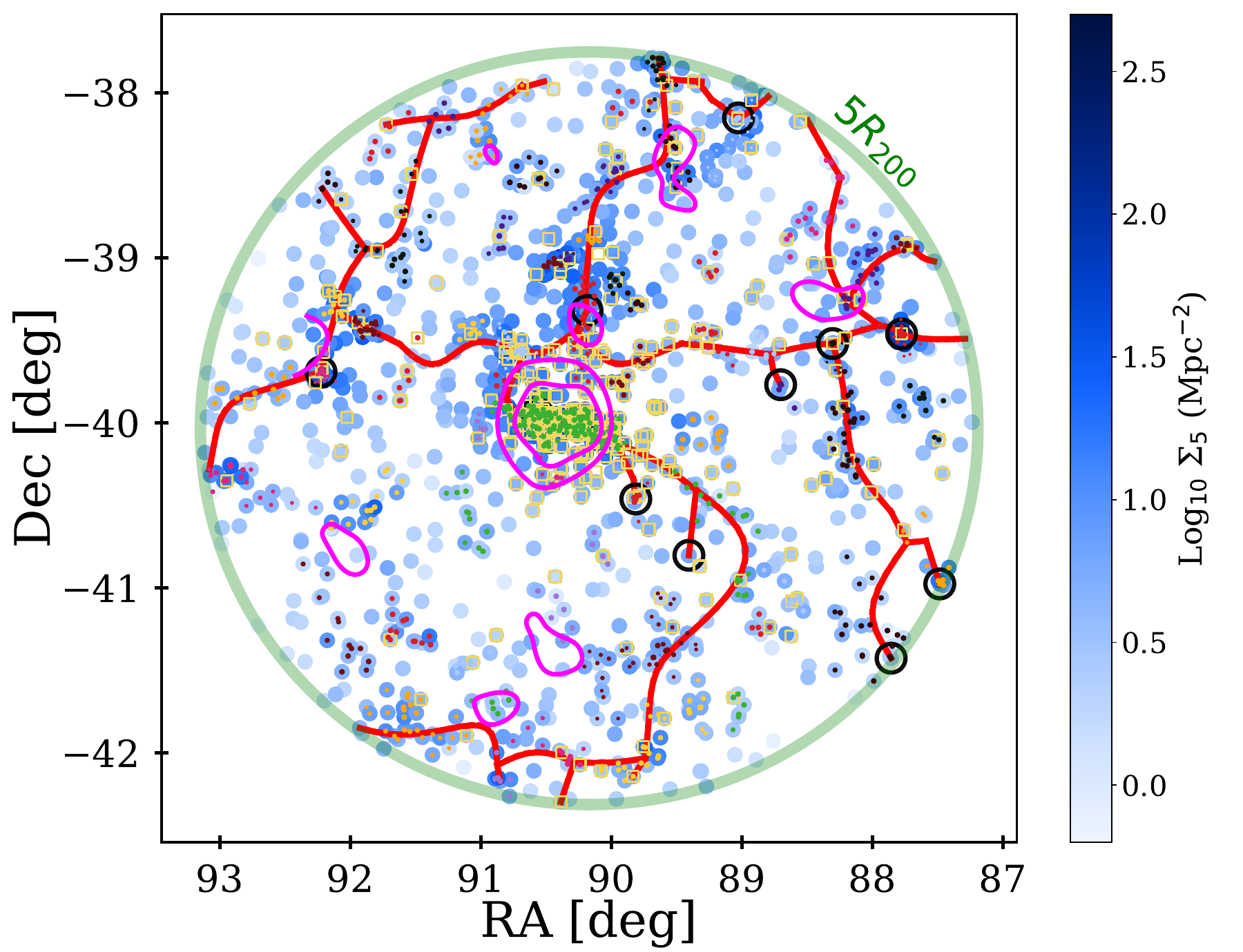}} 
    \subfloat[LS-DR9]{\includegraphics[width = .43\linewidth]{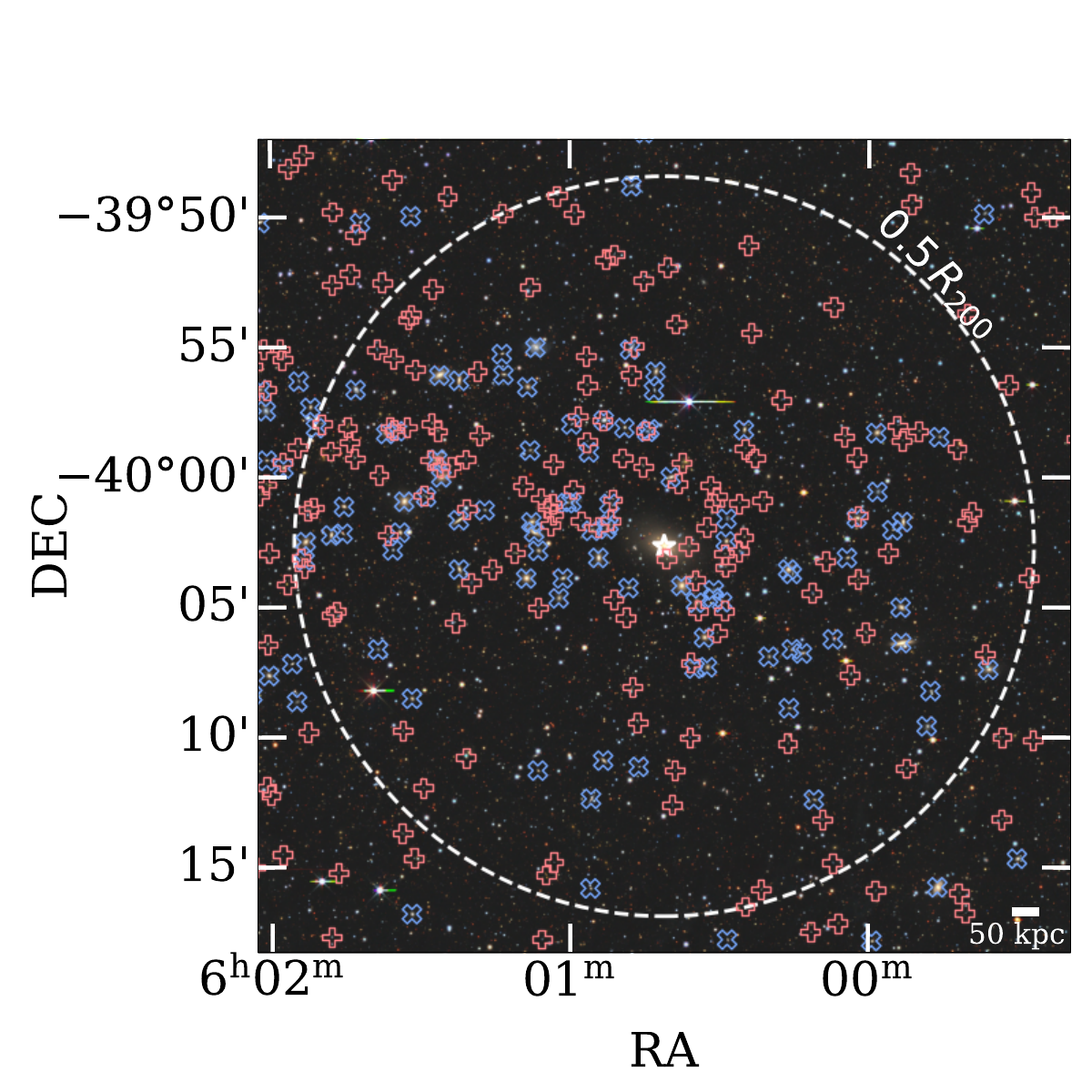}}
    \caption{Left panel: LSS and substructures identification in A3376. Blue
     colour bar shows the local galaxy density with KNN of photometric
     members that are part of the cluster. Yellow squares are the spectral
     members. The identified filaments are represented as red lines, where
     nodes are indicated as black dots. Green circumference is $\rm 5\times
     R_{200}$. Magenta contours are eROSITA large-scale X-ray detections in
     the cluster area. The identified substructures are plotted with
     different colours. Right panel: RGB cutout of A3376 the central region
     ($\rm0.5\times R_{200}$, dashed circle ), overlaid their corresponding
     galaxy cluster candidate members from S1501 (blue open plus symbols),
     S1505 (red open cross symbols) and the brightest cluster galaxy
     (BCG, white star) can be found.}
    \label{fig:LSS}
\end{figure*}

Figure \ref{fig:LSS} shows the cosmic filaments and substructures detected in
Abell 3376, as well as the X-ray emission observed by the extended ROentgen
Survey with an Imaging Telescope Array (eROSITA  \citealt{Merloni24}). 

\section{Summary and conclusions.}\label{sec:Conclusions}

In this paper, we have presented the target selection methodology for the
CHANCES Low-z sub-survey, which will obtain spectroscopic observations of
galaxies in and around 50 galaxy clusters and two superclusters at $z < 0.07$
using the 4MOST instrument. The CHANCES Low-z sub-survey is part of the
CHileAN Cluster galaxy Evolution Survey, a  4MOST community survey designed
to uncover the relationship between the formation and evolution of galaxies
and hierarchical structure formation as it happens, through deep and wide
multi-object spectroscopy. The CHANCES-Evolution sub-survey will observe 50
clusters at $0.07<z<0.45$ with the same magnitude limit ($\rm m_
{r}\leqslant20.4$) and coverage out to $\rm 5\times R_{200}$, while the
CHANCES-CGM sub-survey focuses on QSOs lying behind massive galaxy clusters
at z>0.35, to detect any Mg\,{\sc ii} absorption associated with galaxy
clusters and their surroundings.

We use  photometric data from DESI Legacy Imaging Surveys (\mbox
{LS-DR9}, \mbox{LS-DR10}) and \mbox{T80S/S-PLUS} to identify the most likely
galaxy cluster candidate members across a wide sky area ($\rm r<5\times R_
{200}$) and to depths of $\rm m_{r} < 20.4$. Our selection approach combines
photometric redshift estimates with magnitude-dependent redshift uncertainty
($\sigma_{\mathrm{NMAD}}$) criteria. By testing different magnitude limits
and $\rm N \times \sigma_{NMAD}$ combinations, we found that adopting $\rm
N \times \sigma_{NMAD} = 3.5$ for \mbox{T80S/S-PLUS} and $\rm
N \times \sigma_{NMAD} = 3.0$ for LS datasets provides an optimal balance
between completeness and purity.

To improve completeness at the faint red end of the galaxy population—where
public photo-zs tend to underperform—we used the LS-DR10-CBPF photo-z model
and supplemented the faint galaxy sample with red-sequence-selected galaxies.

Our final target catalogues are organized into three sub-surveys:
\begin{itemize}
    \item Low-z Bright: bright galaxies ($\rm m_{r} < 18.5$) with
     high photo-z reliability from \mbox{T80S/S-PLUS}, \mbox{LS-DR9}, \mbox
     {LS-DR10} and \mbox{LS-DR10-CBPF},
    \item Low-z Faint: faint galaxies ($18.5 \leq \rm m_{r} < 20.4$)
     selected using \mbox{T80S/S-PLUS} and \mbox{LS-DR10-CBPF} photo-zs,
    \item Low-z Faint supplementary: the Low-z Faint sample plus a
     supplementary sample of red-sequence galaxies to enhance completeness of
     faint red galaxy cluster candidate members.
\end{itemize}

Completeness and purity assessments using available spectroscopic data show
that our combined selection strategy recovers over 93\% of spectroscopically
confirmed members in the bright sample and approximately 89\% in the faint
sample, across clusters with good photometric and spectroscopic coverage.

Finally, we demonstrated that these photometric catalogues can be used to
trace both local and large-scale environments, enabling a wide range of
studies on the environmental impact on galaxy evolution.
These photometric catalogues will serve as primary input for the 
spectroscopic follow-up from 4MOST through CHANCES, which will provide $\sim$500,000 
spectra across clusters, groups, and filaments. This combination will enable 
a robust calibration of the DisPerSE-identified structures, turning projected 
filaments and substructures into 3D environments traced by redshift-confirmed members. 
In this way, our catalogues serve as the photometric backbone for the CHANCES 
spectroscopic programme, while also being compatible with other 4MOST spectroscopic 
surveys based on Legacy Survey imaging. This ensures that our low-z catalogues 
can be seamlessly combined with forthcoming wide-area spectroscopic efforts 
(e.g. DESI, Euclid, Rubin/LSST follow-up), thereby maximising their long-term 
legacy value for studies of galaxy evolution across the cosmic web.

These CHANCES Low-z target catalogues lay a solid foundation for the
spectroscopic survey starting soon with 4MOST, which will serve as a legacy
resource for environmental and evolutionary studies across a broad range of
cluster environments.

\begin{acknowledgements}
HMH acknowledges support from the Agencia Nacional de Investigaci\'on y Desarrollo (ANID)
through Fondecyt project 3230176 and the fruitful discussions with KMGS. 
HMH, YLJ, CPH, RBS, FPC, EI, VMS and DP gratefully acknowledge financial support 
from ANID – MILENIO–NCN2024-112. 
CL-D and AM acknowledges a grant from the ESO Comite Mixto ORP037/2022 and support from 
the Agencia Nacional de Investigaci\'on y Desarrollo (ANID) through Fondecyt project 3250511.
AM acknowledges support from the ANID FONDECYT Regular grant 1251882 and funding from the 
HORIZON-MSCA-2021-SE-01 Research and Innovation Programme under the Marie Sklodowska-Curie grant agreement number 101086388. 
HMH, AM, YLJ, CS, RD gratefully acknowledge support from the ANID BASAL project FB210003.  
YLJ and POV acknowledge support from FONDECYT Regular projects 1241426 and 1230441. 
CPH acknowledges support from ANID through FONDECYT Regular project 1252233.
RBS acknowledges support from the Agencia Nacional de Investigación y Desarrollo
(ANID)/Subdirección de Capital Humano/Doctorado Nacional/2023-21231017.
AVSC, ARL and RFH acknowledge financial support from CONICET, Agencia I+D+i 
(PICT 2019-03299) and Universidad Nacional de La Plata (Argentina).
LSJ acknowledges the support from CNPq (308994/2021-3) and FAPESP (2011/51680-6).
STF acknowledges the financial support of DIDULS/ULS through the funding ADI2553855.
MAF acknowledges financial support by the Emergia program (EMERGIA20\_38888) 
from the Junta de Andalucía and University of Granada.
EI gratefully acknowledge financial support from ANID FONDECYT Regular 1221846.
VHLS thanks the support of Coordination for the Improvement of Higher Education 
Personnel (CAPES).
SL and NT  acknowledge support by FONDECYT grant 1231187.
VMS acknowledge financial support from ESO ORP026/2021.
DP acknowledges financial support from ANID through FONDECYT Postdoctorado Project 3230379.
FAF acknowledges support from FAPESP grants 2024/00822-5 and 2024/22842-8
MSC acknowledges funding from São Paulo Research Foundation (FAPESP) grant 2023/10774-5.
CC acknowledges NSFC grant No. 11803044 and 12173045, the China Manned Space Program 
with grant no. CMS-CSST-2025-A07 and the Chinese Academy of Sciences South America Center
for Astronomy (CASSACA) Key Research Project E52H540301.
FRH acknowledges support from FAPESP grants 2018/21661-9 and 2021/11345-5.
CMO acknowledges support from FAPESP grant 2019/26492-3.
GBOS and GBOS acknowledges FAPESP funding through the TT5 fellowship under process 
number 2023/03688-5.
MJS acknowledges financial support from FAPESP grant 2022/00996-89.
LAG-S acknowledges funding for this work from CONICET
TSS acknowledge financial support from the São Paulo Research Foundation (FAPESP) 
through grant 2023/02762-7.
BFR acknowledges support from MNiSW grant DIR/WK/2018/12 and grant pl0201-01 at the 
Pozna\'n Supercomputing and Networking Center.
This work has been undertaken in the framework of the 4MOST collaboration
(\url{https://www.4most.eu/cms/home/}).
This work is based on observations collected with the T80-South telescope at CTIO, Chile, 
under the programme allocated by the Chilean Telescope Allocation Committee (CNTAC), 
no: CN2020B-30 and CN2022B-77.
This work is partially based on data from eROSITA, the soft X-ray instrument aboard SRG, 
a joint Russian-German science mission supported by the Russian Space Agency (Roskosmos), 
in the interests of the Russian Academy of Sciences represented by its Space Research 
Institute (IKI), and the Deutsches Zentrum für Luft- und Raumfahrt (DLR). The SRG 
spacecraft was built by Lavochkin Association (NPOL) and its subcontractors and is operated 
by NPOL with support from the Max Planck Institute for Extraterrestrial Physics (MPE). 
The development and construction of the eROSITA X-ray instrument was led by MPE, with 
contributions from the Dr. Karl Remeis Observatory Bamberg \& ECAP (FAU Erlangen-Nuernberg),
the University of Hamburg Observatory, the Leibniz Institute for Astrophysics Potsdam (AIP),
and the Institute for Astronomy and Astrophysics of the University of Tübingen, with the 
support of DLR and the Max Planck Society.

The Photometric Redshifts for the Legacy Surveys (PRLS) catalogue used in this paper was 
produced thanks to funding from the U.S. Department of Energy Office of Science, 
Office of High Energy Physics via grant DE-SC0007914.

The \mbox{S-PLUS} project, including the T80-South robotic telescope and the \mbox{S-PLUS}
scientific survey, was founded as a partnership between the Funda\c{c}ão de Amparo à Pesquisa 
do Estado de São Paulo (FAPESP), the Observatório Nacional (ON), the Federal University of 
Sergipe (UFS), and the Federal University of Santa Catarina (UFSC), with important financial 
and practical contributions from other collaborating institutes in Brazil, Chile (Universidad 
de La Serena), and Spain (Centro de Estudios de Física del Cosmos de Aragón, CEFCA). We further 
acknowledge financial support from the São Paulo Research Foundation (FAPESP) grant 
2019/263492-3, the Brazilian National Research Council (CNPq), the Coordination for the 
Improvement of Higher Education Personnel (CAPES), the Carlos Chagas Filho Rio de Janeiro 
State Research Foundation (FAPERJ), and the Brazilian Innovation Agency (FINEP). 
The \mbox{S-PLUS} collaboration members are grateful for the contributions from CTIO staff 
in helping in the construction, commissioning, and maintenance of the T80-South telescope 
and camera. We are also indebted to Rene Laporte, INPE, and Keith Taylor for their essential
contributions to the project. From CEFCA, we particularly would like to thank Antonio 
Marín-Franch for his invaluable contributions in the early phases of the project, David
Cristóbal-Hornillos and his team for their help with the installation of the data reduction 
package JYPE version 0.9.9, César Íñiguez for providing 2D measurements of the filter 
transmissions, and all other staff members for their support with various aspects of the
project. This work made use of Astropy: a community-developed core Python package and an
ecosystem of tools and resources for astronomy (\citealt{astropy_13,astropy_18,astropy_22} 
\url{http://www.astropy.org}). This work has benefited from open-source software including 
Matplotlib (Hunter 2007, \url{https://matplotlib.org/}), NumPy (\citealt{NumPy20}, 
\url{https://numpy.org/}), SciPy (\citealt{SciPy20}, \url{https://scipy.org/}), 
and Pandas (\citealt{Pandas20}, \url{https://pandas.pydata.org/}).

\end{acknowledgements}

\bibliographystyle{aa} 
\bibliography{References.bib} 
\clearpage
\begin{appendix}\label{sec:appendix}

\section{Photo-z CBPF appendix}
\begin{table}[ht]
\centering
\caption{Overall and faint low-z performance metrics for CBPF-$z$ and \mbox{LS-DR10} photo-$z$ estimates.}
\begin{tabular}{lcccc}
\hline
Metric & Selection & CBPF-$z$ & LS-DR10 \\
\hline
Mean Bias     & $\rm m_{r} < 20.5$ & 0.00032  & 0.05869  \\
Median Bias   & & $-0.0021$ & 0.0013  \\
$\sigma_{\rm NMAD}$         & & 0.01531  & 0.01534  \\
Outlier Fraction (\%) & & 1.4 & 9.0     \\
Mean error  & & 0.035 & 0.105    \\
\hline
Mean Bias     & $18.5 <\rm m_{r} < 20.5$;  & 0.028  & 0.146  \\
Median Bias   & $z_{\rm spec} < 0.3$ & 0.0078  & 0.0136  \\
$\sigma_{\rm NMAD}$ & & 0.028  & 0.032  \\
Outlier Fraction (\%) & & 5.6     & 19.6    \\
Mean error  & & 0.062  & 0.189    \\
\hline
Mean Bias     & $18.5 < \rm m_{r} < 20.5$;  &0.07  & 0.19  \\
Median Bias   & $z_{\rm spec} < 0.07$ & 0.021  &  0.031  \\
$\sigma_{\rm NMAD}$ & & 0.031  & 0.042  \\
Outlier Fraction (\%) & &  16.4     & 21.0    \\
Mean error  & & 0.106  & 0.234     \\
\hline
\end{tabular}
\label{tab:photoz_metrics}
\end{table}

\begin{figure}[ht]
    \centering
    \includegraphics[width=0.9\linewidth]{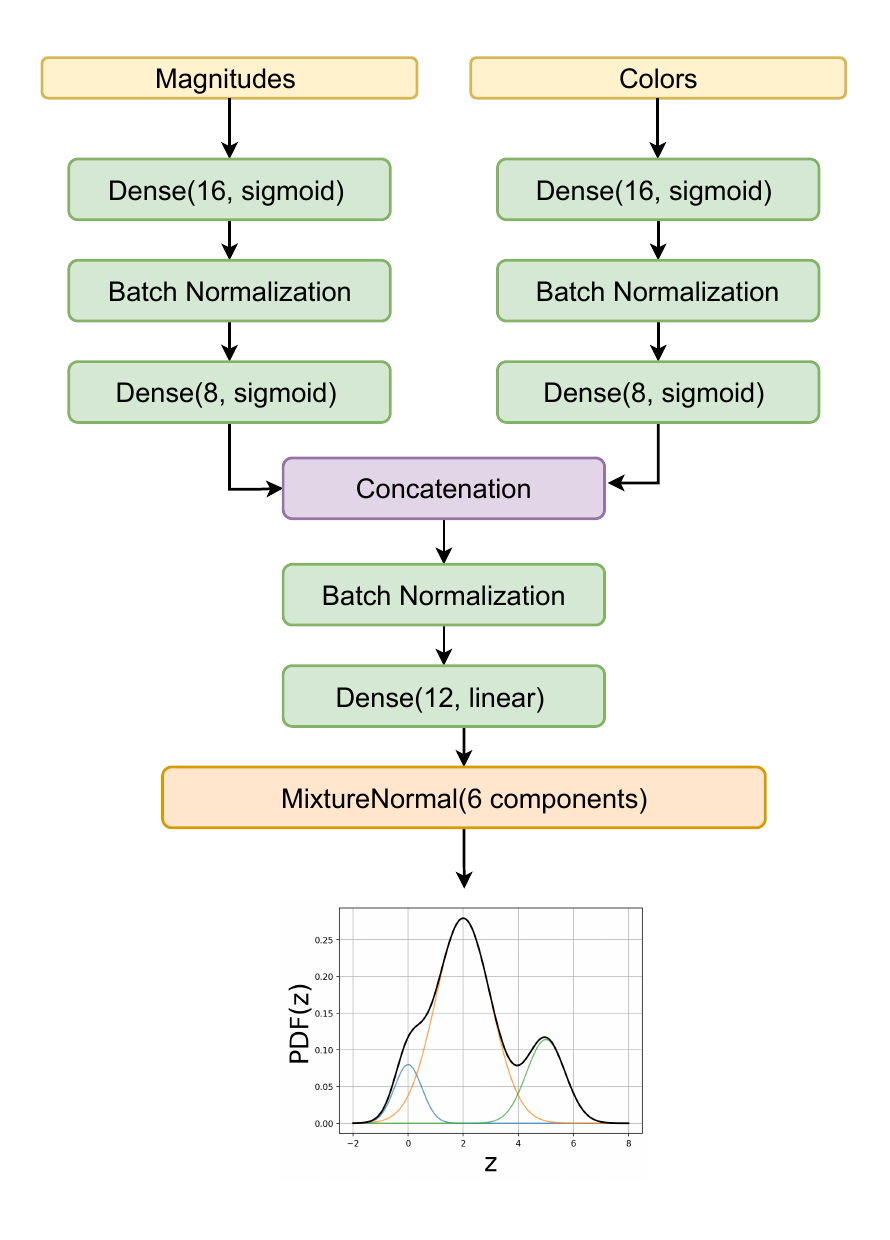}
    \caption{Architecture of the CBPF-$z$ model used for photometric redshift estimation. The network processes magnitudes and colours through separate branches, each composed of two dense layers with sigmoid activation and batch normalization. The outputs are concatenated and passed through additional dense and batch normalization layers, followed by a Gaussian Mixture output layer that models the redshift PDF.}
    \label{fig:mdn-CBPF}
\end{figure}

\section{$\sigma_{\rm NMAD}$ polynomial fit.}
\begin{table}[ht]
\centering
\caption{$\sigma_{\rm NMAD}$ polynomial best fit coefficients.}
\begin{tabular}{lccccc}
\hline
Survey & a & b & c & d\\
\hline
S-PLUS       & 1.77E-04 & -8.07E-03 & 1.24E-01 & -6.34E-01 \\
LS-DR9       & 2.60E-05 & -8.89E-04 & 9.81E-03 & -2.75E-02 \\
LS-DR10      & 1.77E-04 & -8.07E-03 & 1.24E-01 & -6.34E-01 \\
LS-DR10-CBPF & 1.48E-10 & -6.67E-09 & 9.73E-08 & -4.54E-07 \\
\end{tabular}
\label{tab:fitvals}
\end{table}

\clearpage

\section{Target selection figures and tables.}
\begin{table}[ht]
    \centering
    \caption{\mbox{S-PLUS} coverage. List of the 34 CHANCES low-z clusters covered by \mbox{S-PLUS}.}
    \begin{tabular}{ccc}
    \multirow{2}{*}{Cluster} & Spatial coverage & Area coverage\\
     & $\rm n \times \mathrm{R_{200}}$ & \% \\
    \hline \hline
A0085$\star$$\dag$           & 	3	& 85 		\\
A0133 		                 & 	5 	& 100 		\\
A0147$\star$ 		         & 	2 	& 86 		\\
A0151 		                 & 	- 	& 5 		\\
A0168 		                 & 	3 	& 91 		\\
A0496$\star$ 		         & 	2 	& 60 		\\
A0500$\star$$\dag$ 		     & 	5	& 100 		\\
A0548$\star$ 		         & 	2 	& 82 		\\
A0754$\star$ 		         & 	3 	& 71 		\\
A0780$\star$$\dag$ 		     & 	3.5 & 95 		\\
A0957$\dag$                  & 	5	& 100		\\
A1069  		                 & 	0.5	& 73 		\\
A1520 		                 & 	2.5	& 91 		\\
A1631$\dag$ 		         & 	5	& 100 		\\
A1644$\dag$                  & 	2	& 91 		\\
A2399$\dag$ 		         & 	4.5 & 98 		\\
A2415$\star$$\dag$     	     & 	2	& 87 		\\
A2717$\dag$ 	     	     & 	5	& 100 		\\
A2734        		         & 	- 	& 5 		\\
A2870        		         & 	1.5	& 53 		\\
A3223$\star$$\dag$ 		     & 	5	& 100		\\
A3266$\star$$\dag$	         & 	3	& 87 		\\
A3376$\star$$\dag$	         & 	3.5 & 82 		\\
A3490$\star$$\dag$ 		     & 	4	& 95 		\\
A3497$\star$ 		         & 	2	& 86 		\\
A3651$\star$ 		         & 	2 	& 92 		\\
A3667$\star$$\dag$           & 	3	& 81 		\\
A3716$\star$$\dag$ 		     & 	4	& 83 		\\
A3809$\dag$ 		         & 	5 	& 100		\\
AS560$\star$ 		         & 	5	& 100		\\
MKW4 		                 & 	2	& 79 		\\
MKW8 		                 & 	1	& 76 		\\
Hydra	                     & 	5	& 100		\\
Shapley  & 	\multirow{2}{*}{-}	& \multirow{2}{*}{30} \\
supercluster$\star$ (SSC)& & \\
Horologium-Reticulum & 	\multirow{2}{*}{-}	& \multirow{2}{*}{40} \\
supercluster$\star$ (HRS) & & \\
\\
\end{tabular}
\tablefoot{The second columns indicate the maximum radius covered by \mbox{S-PLUS} in terms of 
    $\rm n \times \mathrm{R_{200}}$, while the third column indicates an approximate area 
    covered by the \mbox{S-PLUS} data. \\
    $\star$ clusters from our dedicated CHANCES T80S observations. \\
    $\dag$ clusters with a good spatial and spectral (down to $\rm{m_{r}}=20.4$) coverages.}
\label{tab:S-PLUSCoverage}
\end{table}

\begin{figure*}
    \centering
    \includegraphics[width=0.9\linewidth]{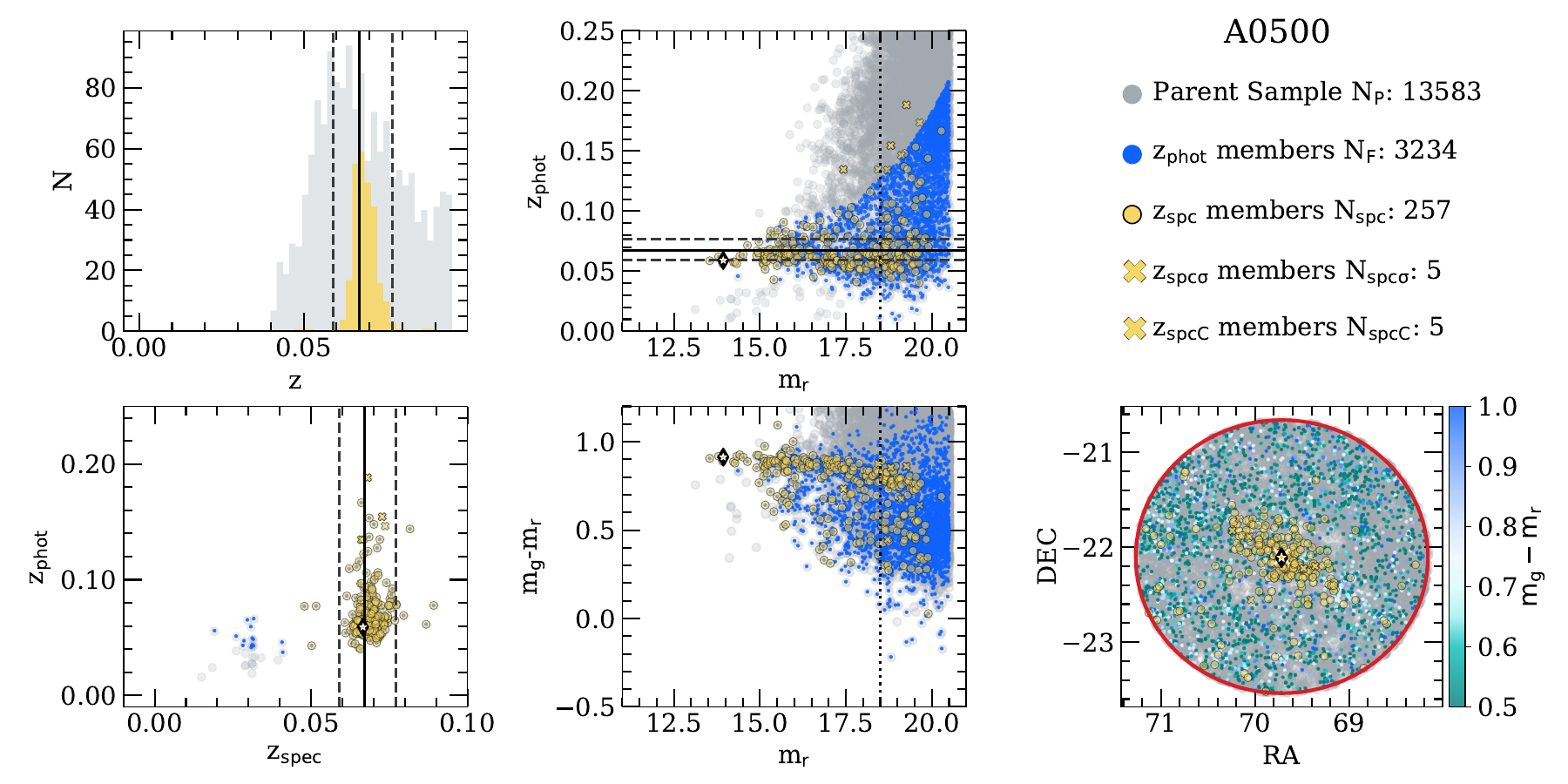}
    \caption{Similar to Figure \ref{fig:ExBright} but for \mbox{S-PLUS}  $\rm z_{phot}$ estimations including the whole
    $\rm m_{r}$ range and assuming $\rm N \times\sigma_{NMAD}=3.5$.}
    \label{fig:ExBrightS-PLUS}
\end{figure*}

\begin{figure*}
    \centering
    \includegraphics[width=0.9\linewidth]{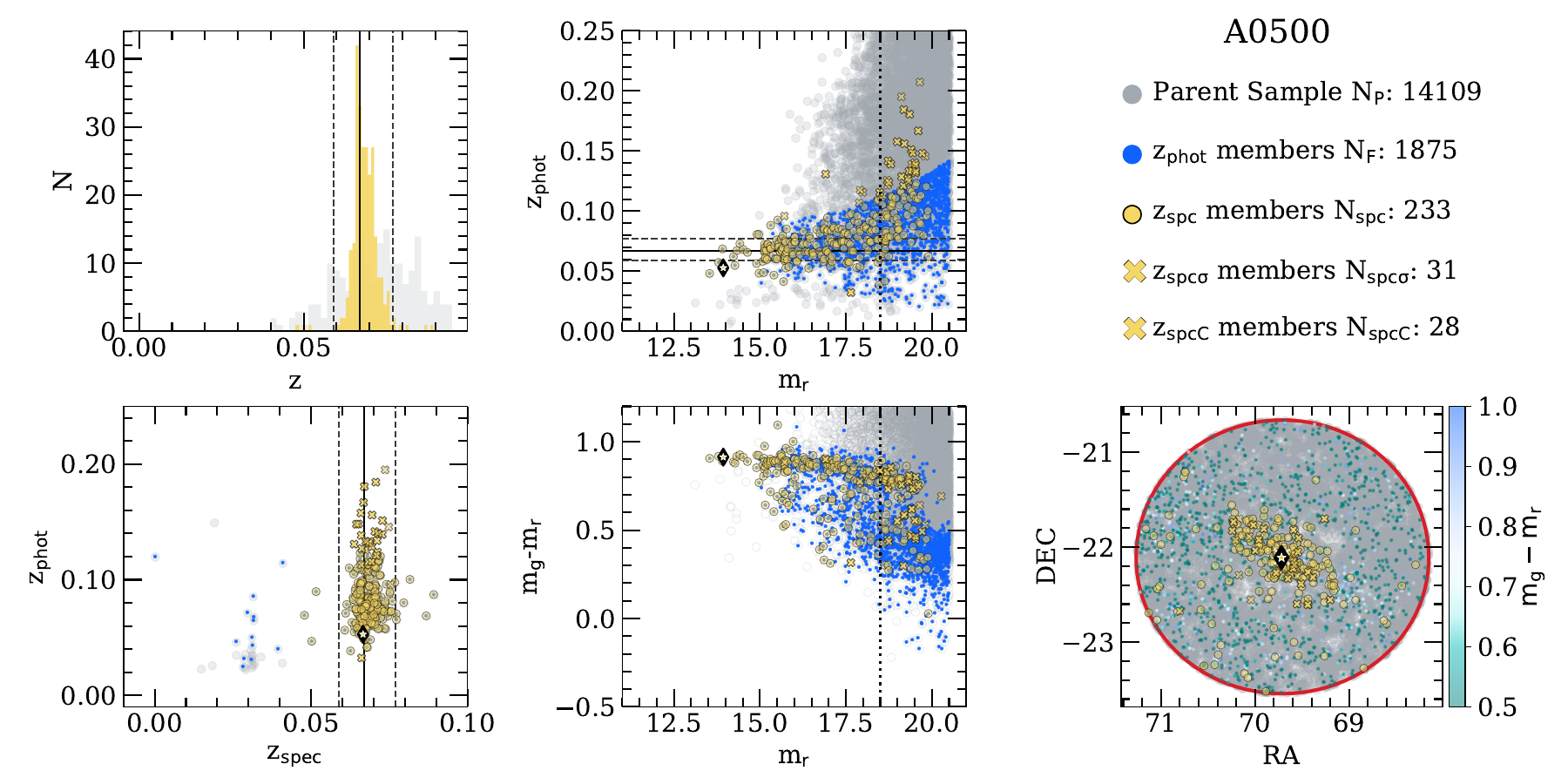}
    \caption{Similar to Figure \ref{fig:ExBright} but for \mbox{LS-DR9}  $\rm z_{phot}$ estimations including the whole
    $\rm m_{r}$ range and assuming $\rm N \times\sigma_{NMAD}=3$.}
    \label{fig:ExBrightDR9}
\end{figure*}

\begin{figure*}
    \centering
    \includegraphics[width=0.9\linewidth]{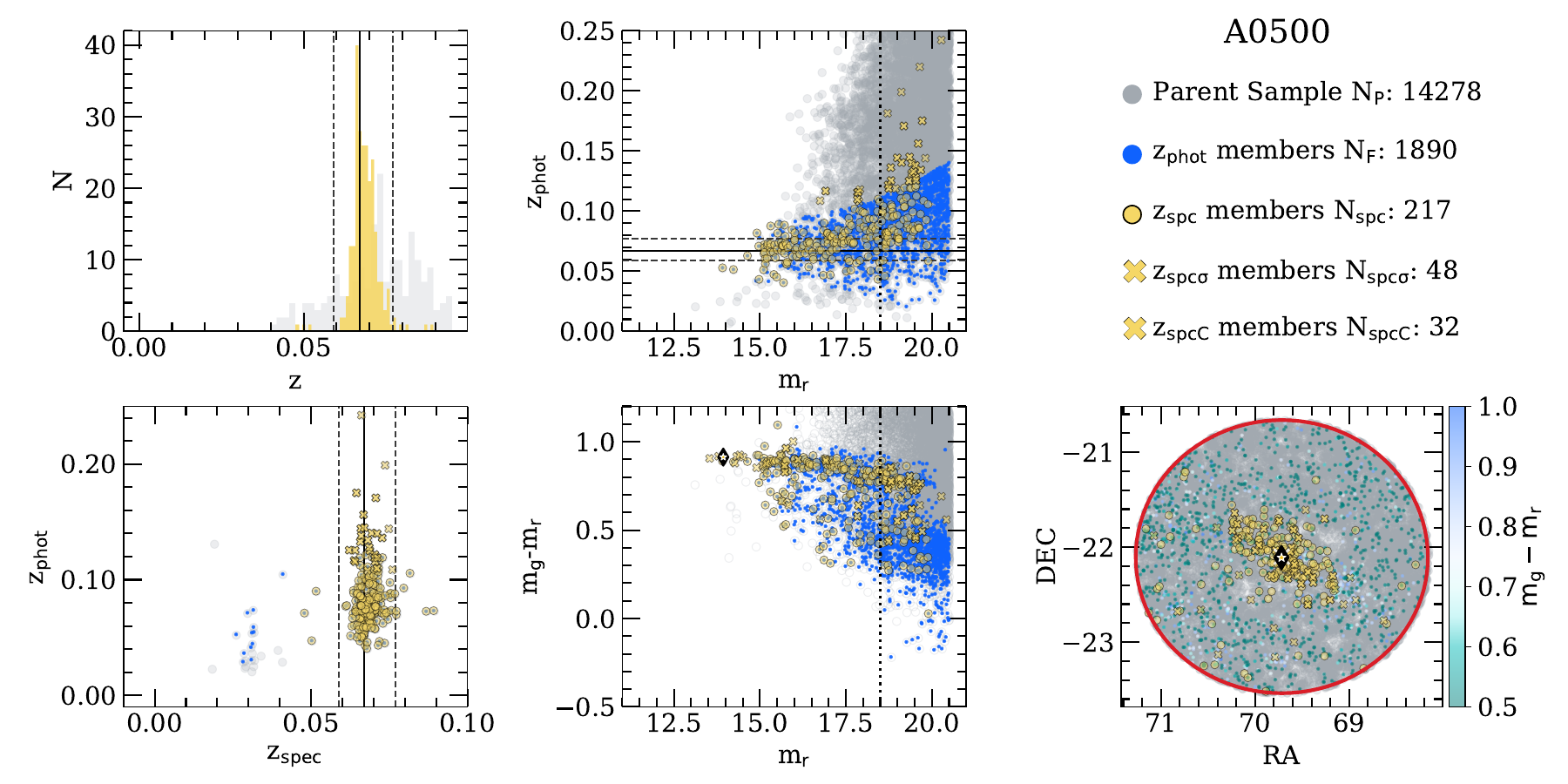}
    \caption{Same as Figure \ref{fig:ExBright} but for \mbox{LS-DR10}  $\rm z_{phot}$ estimations including the whole
    $\rm m_{r}$ range and assuming $\rm N \times\sigma_{NMAD}=3$.}
    \label{fig:ExBrightDR10}
\end{figure*}

\begin{figure*}
    \centering
    \includegraphics[width=0.9\linewidth]{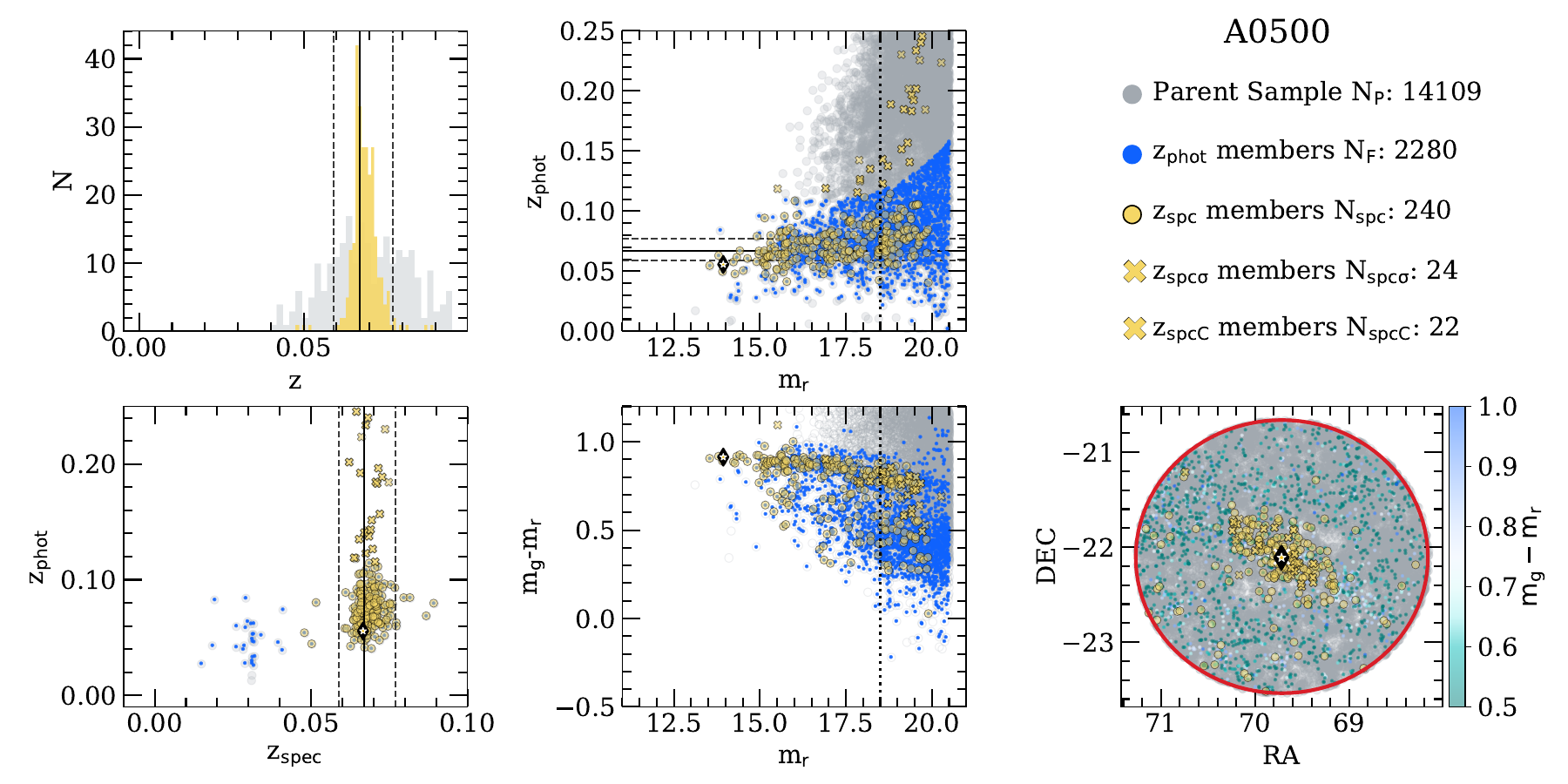}
    \caption{Similar to  Figure \ref{fig:ExBright} but for CBPF $\rm z_{phot}$ estimations including the whole
    $\rm m_{r}$ range and assuming $\rm N \times\sigma_{NMAD}=3$.}
    \label{fig:ExBrightCBPF}
\end{figure*}

\begin{figure*}
    \centering
    \includegraphics[width=0.9\linewidth]{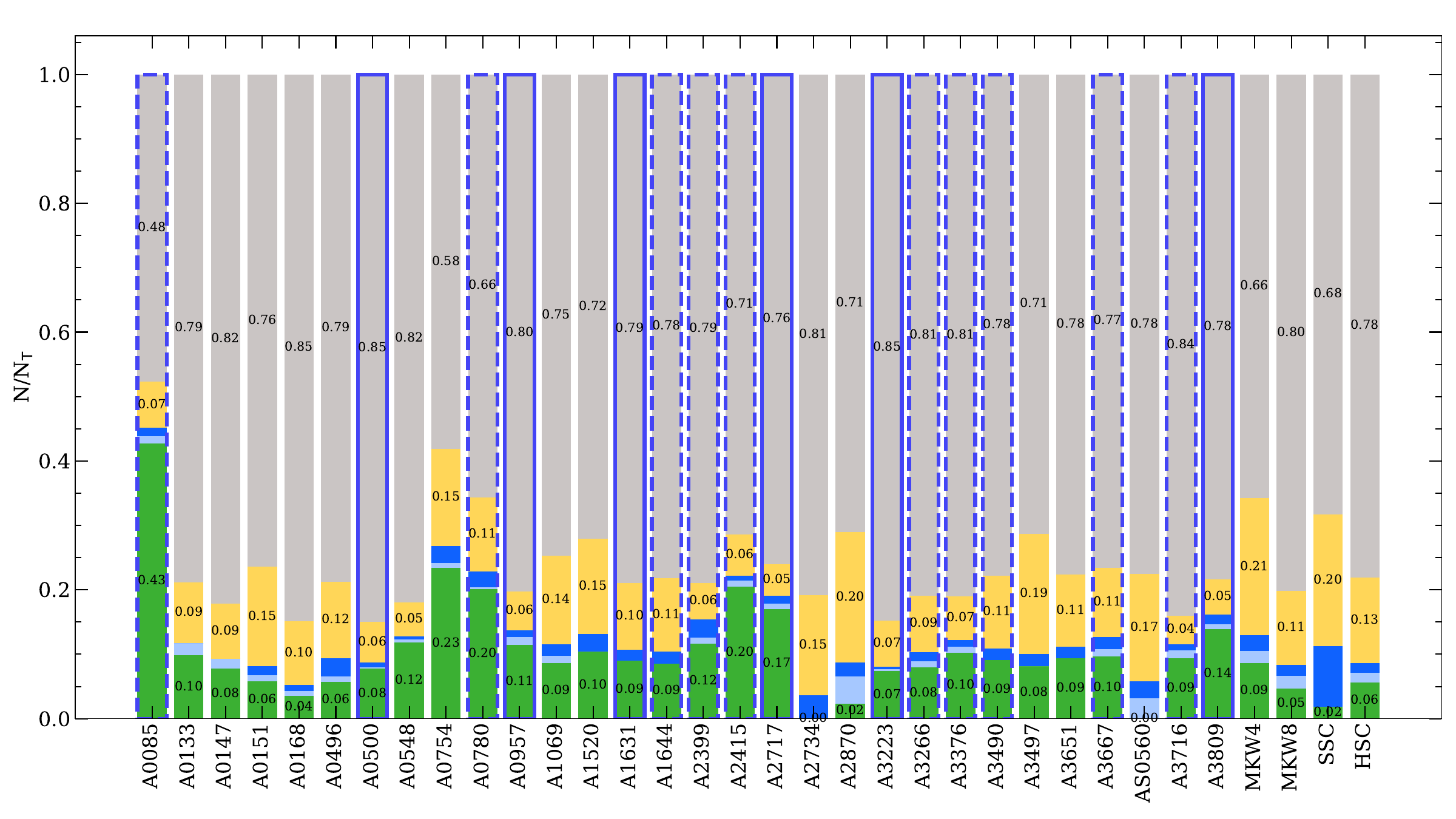}
    \caption{S1501 final individual target selection decomposition from the photometric 
    redshift selection according to their parent survey catalogue:
    \mbox{S-PLUS} (green), \mbox{LS-DR9} (light-blue), \mbox{LS-DR10} (blue) , \mbox{LS-DR10-CBPF} (yellow)
    and common objects (grey). Blue rectangles indicate
    the 6 clusters with good spatial and spectroscopic coverage down to $\rm m_{r}=20.4$,
    while dashed-rectangles indicate clusters with \mbox{S-PLUS} spatial coverage >80\%.}
    \label{fig:S1501_Decomp}
\end{figure*}

\begin{figure*}
    \centering
    \includegraphics[width=0.9\linewidth]{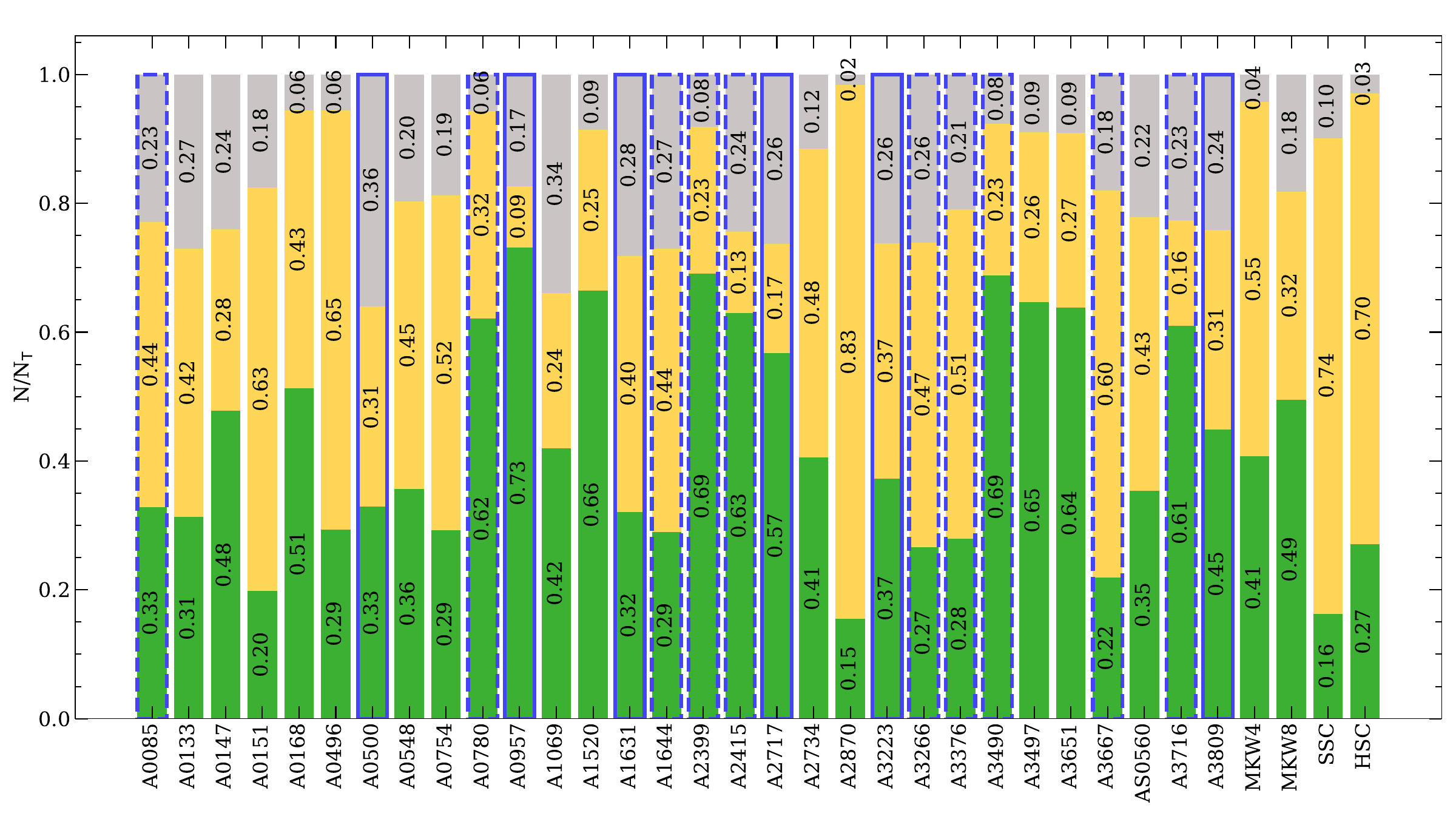}
    \caption{S1505 final individual target selection decomposition from the photometric 
    redshift selection according to their parent survey catalogue:
    \mbox{S-PLUS} (green), \mbox{LS-DR10-CBPF} (yellow) and common objects (grey). Blue rectangles indicate
    the 6 clusters with good spatial and spectroscopic coverage down to $\rm m_{r}=20.4$,
    while dashed-rectangles indicate clusters with \mbox{S-PLUS} spatial coverage >80\%.}
    \label{fig:S1505_Decomp}
\end{figure*}

\end{appendix}

\end{document}